\documentstyle[11pt,amssymb]{article}

\makeatletter
\@addtoreset{equation}{section}
\makeatother

\def\dalemb#1#2{{\vbox{\hrule height .#2pt
        \hbox{\vrule width.#2pt height#1pt \kern#1pt
                \vrule width.#2pt}
        \hrule height.#2pt}}}
\def\square{\mathord{\dalemb{6.8}{7}\hbox{\hskip1pt}}}
\def\0{{\sst{(0)}}}
\def\1{{\sst{(1)}}}
\def\2{{\sst{(2)}}}
\def\3{{\sst{(3)}}}
\def\4{{\sst{(4)}}}
\def\5{{\sst{(5)}}}
\def\6{{\sst{(6)}}}
\def\7{{\sst{(7)}}}

\def\R{\rlap{\rm I}\mkern3mu{\rm R}}

\def\td{\tilde}
\def\wtd{\widetilde}

\let\a=\alpha \let\b=\beta   
    
 \let\m=\mu

\def\nn{\nonumber} \def\bd{\begin{document}} \def\ed{\end{document}}
\def\ds{\documentstyle} \let\fr=\frac \let\bl=\bigl \let\br=\bigr
\let\Br=\Bigr \let\Bl=\Bigl 
\let\bm=\bibitem
\let\na=\nabla
\let\pa=\partial \let\ov=\overline 
\newcommand{\be}{\begin{equation}} 
\newcommand{\ee}{\end{equation}} 
\def\ba{\begin{array}}
\def\ea{\end{array}}
\def\ft#1#2{{\textstyle{{\scriptstyle #1}\over {\scriptstyle #2}}}}
\def\fft#1#2{{#1 \over #2}}
\def\del{\partial}
\def\sst#1{{\scriptscriptstyle #1}}
\def\oneone{\rlap 1\mkern4mu{\rm l}}
\def\ie{{\it i.e.\ }}
\def\via{{\it via}}
\def\semi{{\ltimes}}
\def\v{{\cal V}}

\newcommand{\ho}[1]{$\, ^{#1}$}
\newcommand{\hoch}[1]{$\, ^{#1}$}
\newcommand{\bea}{\begin{eqnarray}} 
\newcommand{\eea}{\end{eqnarray}} 
\newcommand{\ra}{\rightarrow}
\newcommand{\lra}{\longrightarrow}
\newcommand{\Lra}{\Leftrightarrow}
\newcommand{\ap}{\alpha^\prime}
\newcommand{\bp}{\tilde \beta^\prime}
\newcommand{\tr}{{\rm tr} }
\newcommand{\Tr}{{\rm Tr} } 
\newcommand{\NP}{Nucl. Phys. }
\newcommand{\tamphys}{\it Center for Theoretical Physics,
Texas A\&M University, College Station, Texas 77843\\
and SISSA, Via Beirut No. 2-4, 34013 Trieste, Italy\hoch{2}}
\newcommand{\ens}{\it Laboratoire de Physique Th\'eorique de l'\'Ecole
Normale Sup\'erieure\hoch{2,3}\\
24 Rue Lhomond - 75231 Paris CEDEX 05}

\newcommand{\auth}{E. Cremmer\hoch{\dagger}, B. Julia\hoch{\dagger}, 
H. L\"u\hoch{\dagger} and C.N. Pope\hoch{\ddagger1}}

\begin{document}
\begin{flushright}
\hfill{CTP TAMU-27/97}\\
\hfill{LPTENS-97/27}\\
\hfill{SISSARef-132/97/EP}\\
\hfill{hep-th/9710119}\\
\hfill{October 1997}\\
\end{flushright}


\begin{center}
{ \large {\bf Dualisation of Dualities. I.}}

\vspace{10pt}
\auth

\vspace{10pt}

{\hoch{\dagger}\ens}

\vspace{10pt}
{\hoch{\ddagger}\tamphys}

\vspace{10pt}

\underline{ABSTRACT}
\end{center}

      We analyse the global (rigid) symmetries that are realised on
the bosonic fields of the various supergravity actions obtained from
eleven-dimensional supergravity by toroidal compactification followed
by the dualisation of some subset of fields.  In particular, we show
how the global symmetries of the action can be affected by the choice
of this subset.  This phenomenon occurs even with the global
symmetries of the equations of motion.  A striking regularity is
exhibited by the series of theories obtained respectively without any
dualisation, with the dualisation of only the Ramond-Ramond fields of
the type IIA theory, with full dualisation to lowest degree forms, and
finally for certain inverse dualisations (increasing the degrees of
some forms) to give the type IIB series.  These theories may be called
the $GL_A$, $D$, $E$ and $GL_B$ series respectively.  It turns out
that the scalar Lagrangians of the $E$ series are sigma models on the
symmetric spaces $K(E_{11-D})\backslash E_{11-D}$ (where $K(G)$ is the
maximal compact subgroup of $G$) and the other three series lead to
models on homogeneous spaces $K(G) \backslash G\semi \R^s$. These can
be understood from the $E$ series in terms of the deletion of positive
roots associated with the dualised scalars, which implies a group
contraction.  We also propose a constrained Lagrangian version of the
even dimensional theories exhibiting the full duality symmetry and
begin a systematic analysis of abelian duality subalgebras.

{\vfill\leftline{}\vfill
\footnoterule
{\footnotesize	\hoch{1} Research supported in part by DOE 
Grant DE-FG03-95ER40917 \vskip	-12pt} \vskip 14pt
{\footnotesize \hoch{2} Research supported in part by EC under TMR
contract ERBFMRX-CT96-0045 \vskip -12pt} \vskip 14pt
{\footnotesize
        \hoch{3} Unit\'e Propre du Centre National de la Recherche
Scientifique, associ\'ee \`a l'\'Ecole Normale Sup\'erieure \vskip -12pt}
                       \vskip 10pt
{\footnotesize \hoch{\phantom{3}} et \`a l'Universit\'e de Paris-Sud 
\vskip -12pt}} 

\pagebreak
\setcounter{page}{1}

\section{Introduction}

     Eleven-dimensional supergravity \cite{cjs} occupies the
distinguished position of being the highest-dimensional supergravity
theory.  It provides a window on the elusive M-theory, which would
describe the strong coupling limits of ten-dimensional string theories
\cite{w1}.  The fact that M-theory compactified on $S^1$ gives rise to
the type IIA string can be seen both at the level of supergravity
\cite{cjs,cw,hn}, and in the sigma-model action \cite{bst,dhis}.  In
this paper we shall consider the classical, internal, global symmetry
groups of the bosonic sectors of the various maximal supergravities in
dimensions $D\le 11$, which can be obtained from eleven-dimensional
supergravity by toroidal compactification \cite{ss1,cs}.  These Lie
groups, discretised after quantisation, are conjectured to become the
duality symmetry groups of the toroidally-compactified type II quantum
string theories \cite{ht}.  As is well known, there exists a
formulation of each of these lower-dimensional theories, namely with
the canonical (maximal) choice of field dualisations, in which there
is a global $E_{(11-D)(11-D)}$ internal symmetry
\cite{cj3,cj2}.\footnote{In this paper, the exceptional groups $E_n$
will always be in their maximally non-compact form $E_{n(n)}$.  For
brevity, we shall write them simply as $E_n$.  For $n\le 5$ we have
$E_0$ trivial, $E_1=\R$, $E_2=GL(2,\R)$, $E_3= SL(3,\R)\times
SL(2,\R)$, $E_4= SL(5,\R)$ and $E_5= O(5,5)$.}  Specifically, these
symmetries are realised in the theories that are obtained by
performing the toroidal reduction to $D$ dimensions and then dualising
any field strength whose degree exceeds $\ft12 D$.  Thus when $D$ is
odd, this $E_{11-D}$ symmetry is realised on the gauge potentials and
is an invariance of the Lagrangian.  In fact the $E_{11-D}$ symmetry
for odd dimensions in this canonical choice of fields does not involve
any electric/magnetic type of duality at all; the name duality
symmetries is nevertheless widely used.

         However, the story is different in even dimensions.  In what
follows we shall use the term ``strict duality'' to mean a continuous
symmetry at the level of the equations of motion, whose Lie algebra
generators mix a set of field strengths with their Hodge duals (or
with additional field strengths of the dual degrees).  By extension,
the duality group has come to mean the full global internal symmetry
even when there is no strict duality at all.  We shall reserve the
name dualisation to a discrete Hodge-like duality that exchanges forms
of complementary degrees which appear in two dual Lagrangians with
locally equivalent equations of motion. Actually we propose the name
inverse dualisation for this operation when the degree of the form
increases.  One of the main questions will be to investigate the
effect of dualisations on the strict and the not so strict duality
symmetries.  In fact when $D$ is even, the field strengths of degree
$\ft12 D$ and their magnetic duals form a single irreducible
representation of $E_{11-D}$, and so for these fields, the strict
duality symmetries can only be implemented locally on the field
strengths, rather than on their gauge potentials.  Furthermore, only
the set of field equations plus Bianchi identities, rather than the
conventional Lagrangians, are invariant (recall the example of
electric/magnetic duality in $D=4$).  A local implementation of
duality on the potentials can only be achieved at the level of
equations of motion by the introduction of additional dual potentials.
Typically the equations then take the form of a twisted self-duality
condition \cite{cj3}; in this case, the subgroup of symmetries of the
Lagrangian is the parity-even subgroup. We may remark that the strict
dualities use the metric and thus are not really internal symmetries
in the usual sense of commuting with spacetime transformations; there
is no absolute Lorentz subgroup of diffeomorphisms in curved space.

     It is natural now to ask whether the process of dualising all the
field strengths whose degrees exceed $\ft12 D$ was crucial for
obtaining the $E_{11-D}$ global symmetry.  It was observed in
\cite{lptdual} that the global symmetries can change, depending on
whether or not certain dualisations are performed.  Indeed when $D$ is
odd, and the symmetry is realised at the level of the Lagrangian, it
is manifest that the dualisations are necessary for the symmetry to
act locally on the gauge potentials, since we cannot assemble two sets
of gauge potentials of dual degrees into an irreducible multiplet,
unless non-local symmetry transformations are allowed.  One might
think that this problem could always be circumvented at the level of
the equations of motion, since one's experience in simple examples
such as electric/magnetic duality in $D=4$ is that only the field
strengths, and not their bare gauge potentials, appear there.  Indeed,
if only the field strengths appear in the equations of motion and the
Bianchi identities, then one could view these field strengths as the
fundamental physical quantities on which the true symmetry
transformations should be defined. Then, any possible dualisation (or
inverse dualisation) that continues to allow the equations of motion
and Bianchi identities for a multiplet of field strengths to be
written purely in terms of the (now dualised) field strengths will
leave the global symmetry of the equations of motion unaffected, since
the transformations can be implemented as well on the field strengths
of the dualised reformulation.  On the other hand, if the result of
(inversely) dualising some members of an irreducible multiplet of
equal-degree field strengths is to cause the unavoidable appearance of
bare potentials (\ie not in the combination of field strengths) for
some of the remaining members of this multiplet in the equations of
motion and Bianchi identities, then the original global symmetry will
be modified.  Some examples of this phenomenon are discussed in
section 6.3.

      In the special case where scalar fields (0-form potentials) are
being (inversely) dualised, we shall presently show that the first
loss of global symmetry follows the loss of these scalars, because
their constant shift symmetries disappear too, or rather, the
corresponding group action becomes trivial in that sector: it is no
longer faithful. The action of the global symmetry group is a
nonlinear realisation on a homogeneous space but after the dualisation
the rigid symmetry is partially transmuted to a local gauge type
symmetry. As the dualisable scalar potentials appear only through
their field strengths and even though these do not mix with bare
potentials under the symmetry, the global invariance is already
reduced prior to dualisation if one looks only at its action on the
(1-form) field strengths.

          Note that the mere fact that the equations of motion and
Bianchi identities involve a particular field only via its field
strength is no guarantee that this field can be dualised.  A classic
example of this is the 4-form field strength in eleven-dimensional
supergravity, which apparently cannot be covariantly dualised to a
7-form.\footnote{In terms of the 4-form $F_\4=dA_\3$, the field
equation is $d\ast F_\4 = F_\4\wedge F_\4$ and the Bianchi identity is
$dF_\4=0$.  To rewrite the field equation as a Bianchi identity we
must define $F_\7= \ast F_\4 - A_\3\wedge F_\4$, giving $dF_\7=0$.
However its field equation is $d\ast F_\7 =-d\ast(A_\3 \wedge F_\4)$,
which cannot be recast into a local equation involving only $F_\7$.  A
more naive approach would be simply to define $F_\4=\ast F_\7$, giving
$d\ast F_\7=0$ and $d F_\7= (\ast F_\7)\wedge (\ast F_\7)$.  This does
not work either, since the latter equation cannot be interpreted as a
Bianchi identity that is solved in terms of a gauge potential $A_\6$
by writing $F_\7=dA_\6+\cdots$.}  In fact the question of whether or
not a particular field can be dualised must be studied at the level of
the Lagrangian; the possibilities for dualisation are not enlarged by
looking instead at the equations of motion.  At each stage of the
dualisation process, a sufficient condition for dualisability of a
{\it given} field is that it should appear in the action purely
through its field strength. (Here, and generally in these discussions,
when we say that a field or rather a collection of fields appear via
their field strengths, we mean that this can be achieved after some
field redefinitions and/or integrations by parts.)  The above
considerations lead to the following observations on how the
dualisation of fields can affect the global symmetry.  If an
irreducible multiplet of fields appear in the Lagrangian purely via
their field strengths, then the dualisation of any subset (proper or
improper) of these fields is possible, and it will not affect the
global symmetry of the corresponding equations of motion.  The
criterion for dualisability of a subset of fields in an irreducible
multiplet becomes more complicated if some of the fields in the
multiplet require the appearance of bare potentials in the Lagrangian,
and we shall not attempt an exhaustive discussion of this issue here.
In any case, the general statement about the global symmetry is that
if it is to be implemented with a finite number of derivatives on the
potentials for the fields of an irreducible multiplet, then all the
potentials must have the same degree, in other words dual degrees are
forbidden.  If instead the symmetry is to be implemented only on the
field strengths (necessarily in the equations of motion), then all the
members of the multiplet must appear in the field equations and
Bianchi identities only through their field strengths.  If the result
of dualisations is to make it that neither of these conditions is
satisfied, then the original global symmetry prior to the dualisations
will be broken.

     The fact that the global symmetry can depend on the choice of
dualisation \cite{lptdual}, and the fact that not all dualisations are
possible, are both consequences of the occurrence of non-linear terms
in the $D$-dimensional Lagrangian.  These terms have two origins,
namely the $F_\4\wedge F_\4\wedge A_\3$ term in the original
eleven-dimensional Lagrangian, and the non-linearity of the
eleven-dimensional Einstein-Hilbert action.  The latter implies that
modifications to the field strengths in lower dimensions will arise in
the Kaluza-Klein reduction process.  These are sometimes called
``Chern-Simons modifications,'' but the term is really a misnomer
since they actually come from the separation between the gauge
transformations originating from diffeomorphisms along the
compactified directions and the other gauge symmetries. In this paper
they will be called Kaluza-Klein modifications.  In order to
investigate these issues in more detail it is convenient to divide the
discussion into two parts, namely for the subsector comprising the
scalar fields, and then the remaining sectors involving the
higher-degree field strengths.

     In any dimension $D\ge6$, the scalar sector of the
$D$-dimensional theory that is obtained by dimensional reduction from
$D=11$ is unambiguous, since no dualisations of the higher-degree
forms can give rise to additional scalars.\footnote{Later though we
shall consider the inverse possibility of dualising existing scalars
to $(D-2)$-form potentials; this we call an inverse dualisation.}  In
these cases, the scalar sector of the Lagrangian has an $E_{11-D}$
symmetry.  In $D=5,4,3$, on the other hand, the field content of the
scalar manifold depends upon which 4-form, 3-form or 2-form field
strengths respectively one chooses to dualise, since these will give
additional contributions to the scalar sector.  The $E_{11-D}$ global
symmetries are achieved in these dimensions if one dualises all such
higher-degree fields, so as to maximise the total number of scalars.
If the full set of dualisations is not performed, then the global
symmetry of the scalar sector is altered.  This is because the
$E_{11-D}$ symmetry can only be expressed as transformations on the
scalars themselves, and not on their ``1-form field strengths.''
(Some of the $E_{11-D}$ transformations would act through non-local
functions of higher-degree fields if these were not dualised to
scalars, and some would simply disappear together with the
axions.\footnote{In this context, we are defining an axion to be any
scalar field other than the dilatons which come from the dimensional
reduction of the diagonal components of the metric. The dilatonic
scalars are the moduli parametrising the size of the compactifying
space.})
 
     In the Kaluza-Klein reduction of a generic higher-dimensional
theory, the global symmetry of the scalar sector may not necessarily
extend to the higher-degree sectors of the reduced theory.  In fact it
is only because of special features of eleven-dimensional supergravity
that its dimensional reductions allow the global symmetries of the
scalar manifolds to be extended to the full dimensionally-reduced
theories including the higher-degree fields.  For example, omitting
the $F_\4\wedge F_\4\wedge A_\3$ term in $D=11$ (or even just changing
its coefficient) would not affect the $E_{11-D}$ symmetry of the
scalar sector in $D\ge6$, but it would prevent its extension to the
higher-degree fields.  Even for $D=11$ supergravity itself, the entire
dimensionally-reduced theories may only exhibit the global symmetries
of their scalar sectors if appropriate dualisations of higher-degree
field strengths are also performed.  For example, in $D=6$ the
$E_5=O(5,5)$ symmetry of the scalar sector only extends to the entire
theory if the 3-form gauge potential is dualised to give an additional
vector, which, together with the 15 that are already present, can form
a 16-dimensional spinor representation of $O(5,5)$.  This is an
example where bare Kaluza-Klein vector potentials inevitably appear in
the original undualised formulation obtained by direct dimensional
reduction, even at the level of the equations of motion (see section
6.3).  Consequently, only by dualising the 3-form potential can the
$O(5,5)$ symmetry be realised in terms of transformations that involve
purely local functions of fields, namely on the 16 vector potentials.
Thus it should be emphasised that in this example, even at the level
of the equations of motion, the $O(5,5)$ symmetry cannot apparently be
realised unless the dualisation of the 3-form potential has been
performed.  We should however point out that the dualisation can be
effected in two ways: either by adding a Lagrange multiplier for the
Bianchi identity (this indeed requires that no bare potential appears
in the Lagrangian) or else by using a first order formalism of the
classical type, with field and potential considered as independent
variables, and integrating out the potential first.  Note that the
reverse dualisation exchanges the two types of procedures, and that
for instance the Freedman-Townsend dualisation from 2-form potentials
to scalars, leading to a sigma model in four dimensions, can be
effected despite the presence of bare 2-forms \cite{frto}.

     There are also examples where dualisations of higher-degree
fields are not obligatory in order for the global symmetry of the
scalar manifold to extend to the entire theory, at least at the level
of the equations of motion.  For example, the global $E_4=SL(5,\R)$
symmetry of the scalar sector in $D=7$ can be realised at the level of
the equations of motion in the entire theory regardless of whether or
not the 3-form potential is dualised to a 2-form potential.  This is
because the 3-form potential, together with the four 2-form potentials
of the original undualised theory, can all be made to appear in the
Lagrangian only via their field strengths.  Consequently, there will
never be bare potentials in the equations of motion or Bianchi
identities, and the set of $1+4$ field strengths will transform as a 5
of $SL(5,\R)$, regardless of whether or not the dualisation has been
performed.

           In this paper we shall analyse the rigid symmetries that
are realized on the bosonic fields of the various lower-dimensional
maximal supergravities.  Each such theory is obtained from
eleven-dimensional supergravity by toroidal compactification to $D$
dimensions, with the possible subsequent dualisation of some subset of
the fields.  A striking regularity is exhibited by the series of
theories obtained respectively by making no dualisations; with
dualisations of the so-called Ramond-Ramond fields of the type IIA
theory; with full dualisation to lower the degrees of all forms; 
and from there finally by inverse dualisations (raising the degrees of
certain forms) to the type IIB series.  We shall call these theories
the $GL_A$, $D$, $E$ and $GL_B$ series respectively.  This is only a
subset of the large number of classical forms of the theory. It turns
out that the scalar Lagrangians of the $E$ series are sigma models on
the symmetric spaces $K(E_{11-D})\backslash E_{11-D}$ (where $K(G)$ is
the maximal compact subgroup of $G$), while the other three series
lead to models on homogeneous spaces $K(G)\backslash G\semi\R^s$,
where $s$ is the dimension of a certain linear representation of
$G$. In fact, the $E$ series can be used as a means of generating the
other three series by performing appropriate inverse dualisations of
some of its axionic scalar fields.  The reason for this is that, as
will be shown in section 4, the axionic scalars in the fully-dualised
supergravities are in one-to-one correspondence with the positive
roots of the $E_{11-D}$ algebra \cite{lpsweyl}.  In fact we exploit
this to give a simple (triangular or Borel) parameterisation of the
$K(E_{11-D})\backslash E_{11-D}$ cosets for the scalar manifolds, in
which the axionic scalars are the parameters in the exponentiation of
the positive roots, while the dilatonic scalars are the parameters in
the exponentiation of the Cartan generators.  Let us recall three
equivalent formulations of symmetric space sigma models.  One
possibility is to work in a fixed gauge for the subgroup $K(G)$ and
use a triangular representative of each coset (permitted by the
Iwasawa decomposition); this amounts to using group elements in a
Borel subgroup (morally the upper triangular part of the group). The
Borel subgroup itself contains the Cartan subgroup times the strictly
upper triangular subgroup called below the group of positive
roots. The second possibility is to restore the $K(G)$ local gauge
invariance; this form is manifestly invariant under the full global
$G$ and not only under its Borel subgroup; the scalar fields (physical
and gauge) parameterise, before gauge fixing, the full group
$G$. Finally if we recall the analogy with the moving frames of
General Relativity \cite{cj3}, we may use the (local Lorentz
invariant) metric instead of the frames and then preserve manifest
$GL(4,\R)$ invariance without introducing the Lorentz gauge
invariance. Analogously, we shall use an internal metric ${\cal M}$
instead of an element of the group $G$; this will be the third
formulation of the symmetric space sigma models.
 
    The inverse dualisation of some of the axions appearing in the $E$
series now has the effect of removing the associated positive-root
generators from the parameterisation of the coset.  For the $GL_A$ and
$D$ series, these dualisations involve the subsets of positive roots
at the second level according to the grading of the root space of the
$E_{11-D}$ Lie algebra along the appropriate simple root (except in
$D=3$, where the scalars associated with both the third and second
level positive roots must be dualised for the $E$ to $GL_A$
contraction).  For the $GL_B$ series, the construction involves
dualising some scalars associated with commuting positive root vectors
selected by an appropriate double grading along two of the simple
roots.  A similar inverse dualisation for a fourth simple root relates
$ E_{11-D}$ and $ E_{10-D}$.  The generators corresponding to the
highest level that has not been inversely dualised will consequently
now commute. (This means that they could in turn be inversely
dualised; we discussed this further in a second paper \cite{cjlp2}).  The
commutativity of a set of generators and an Iwasawa type formula lead
to the property that the corresponding scalar fields can be
simultaneously covered with derivatives in the sigma model scalar
Lagrangian, and hence can be dualised.

      As stated earlier, in order to specify the classical theory
under consideration one should not restrict oneself to the scalar
sector alone but one must also specify the dualisations implemented on
higher-degree fields, which exchange degree $p\ge 2$ field strengths
with those of degree $D-p$. Once again, the Coxeter-Dynkin diagrams of
the $E$ series seem to contain all the information on the four
specific nodes corresponding to the simple roots alluded to above (in
any dimension for our four series).  The general situation is now more
difficult to summarise than in the above discussion of abelianisation
in the scalar sector.  The possibility of dualisation involves the
simultaneous existence and use of involutions of the Dynkin diagram of
$SL(11-D,\R)$ and of that of, for instance, $E_{11-D}$ for the $E$
series; they respectively exchange covectors ({\it i.e.}\ 1-forms) and
vectors ({\it i.e.}\ $(D-1)$-forms) and the corresponding
representations of the internal symmetry, and similarly for forms of
arbitrary degrees.  We shall explain these features in the second
paper of this series.

      The paper is organised as follows. In section 2, we obtain the
bosonic sector of the $D$-dimensional supergravity following 
directly from the dimensional reduction of eleven-dimensional
supergravity.  We show that these non-dualised theories have global
$GL(11-D,\R)\semi \R^q$ global symmetries, where $q=\ft16
(11-D)(10-D)(9-D)$.  In section 3, we study the cases $D=5,4$ and 3
where the full dualisations of all $(D-2)$-form potentials are
performed, so as to obtain the maximal numbers of scalars.  We show
that the symmetry of the scalar Lagrangian is changed by this
dualisation.  In particular, we show that the symmetry group contains
the Borel subgroup of $E_{11-D}$ and that the dimensions of the
maximal abelian symmetries are in each case reduced by the
dualisation.  In section 4, we study the coset structures of the
scalar Lagrangians, and their symmetries.  We show that the scalar
Lagrangians for the fully-dualised $E$ series, where the number of
scalars is maximised, have $E_{11-D}$ global symmetries for $3\le D
\le 10$.  When the dimensional reduction in section 2 is performed by
iteratively repeating the $D+1$ to $D$ dimensional reduction, the
scalars are precisely the parameters of the generators of the Borel
subgroups.  In other words, the triangular gauge is always the
simplest, and the Borel invariance is the most obvious symmetry.

    The symmetric space will be replaced by a double coset in the
cases where certain axions associated with some abelian positive roots
are undualised or simply have not been manufactured by dualisation.
This provides some group theoretical understanding of the dualisation
procedure involving scalar fields, and is discussed in section
5. Actually this leads to a situation where the scalar fields take
their values in a double coset space on which the normaliser of the
suppressed generators still acts transitively.

    In section 6, we show that the symmetries of the scalar sectors of
the maximal supergravities can be extended to the entire bosonic
theories including the higher-degree field strengths, which form
linear representations of the symmetry groups.  We discuss this in
detail in the fully-dualised $E$ series, and show that the
toroidally-compactified eleven-dimensional supergravities have
$E_{11-D}$ global symmetries after the full dualisation. Our
discussion will be simplified, and the full details are postponed to
our second paper in this series.  We shall also discuss in this
section how dualisations of higher-degree field strengths can affect
the symmetry group of the full Lagrangian.  In section 7, we study a
particular case where all the R-R fields are dualised to lower degrees
while the NS-NS fields remain intact.  In section 8, we study the
abelian global symmetries in the various versions of the
supergravities.  We show how the abelian constant shift symmetries can
be grouped into maximal abelian subsets of abelian $\R$ symmetries of
the positive-root systems for the theories.  In section 9, we study
type IIB supergravity and its dimensional reduction.  In particular,
we are interested in the versions where no dualisations are performed.
We conclude our paper in section 10.  We also present the full bosonic
Lagrangian following from the direct reduction of eleven-dimensional
supergravity in Appendix A.  In Appendix B, we present a scalar
Lagrangian with $SL(2,\R)$ global symmetry and study how the symmetry
is affected by dualisation.  Appendix C contains a discussion of
scalar Lagrangians with $O(n,n)$ global symmetries, and their
application in supergravity theories.\footnote{This analysis as well
as that of sections 6.2.2 and 6.3 was inspired by an ongoing research
program of one of us (B.J.) on the important differences between $4k$
and $4k+2$ dimensional spacetime.  The next most obvious one beyond
the duality properties, namely the change of sign in the
Schwinger-Zwanziger quantisation formula, has since then been studied
in detail in \cite{blps,dght}.}

\section{Direct reduction of $D=11$ supergravity and symmetries}

     In section 2.1, we shall discuss the toroidal dimensional
reduction of eleven-dimensional supergravity to $D$ dimensions.  In
the cases where none of the $D$-dimensional fields are dualised, we
shall show in section 2.2 that there is a global $ GL(11-D,\R) \semi
\R^q$ symmetry, where $q=\ft16(11-D)(10-D)(9-D)
=\{0,0,0,1,4,10,20,35,56\}$ in $D=\{11,10,9,8,7,6,5,4,3\}$, and the
$\semi$ symbol denotes a semi-direct product.

\subsection{Dimensional reduction of 11-dimensional supergravity}

          The bosonic sector of eleven-dimensional supergravity
contains the metric and a 4-form field strength $F_\4=dA_\3$. 
The Lagrangian is given by \cite{cjs}
\be
{\cal L} = e R -\ft1{48} e F_\4^2 + \ft16 *(F_\4\wedge F_\4 \wedge A_\3)
\ .\label{d11lag}
\ee
The subscripts on the potential $A_\3$ and its field strength
$F_\4=dA_\3$ indicate the degrees of the differential forms, and the
normalisation is that of \cite{dkl}.  Note that the relative
coefficient $\ft16$ of the $FFA$ term is inert under compatible
rescalings of the gauge potential and the metric, in the sense that
rescalings that preserve the ratio of the coefficients of the
Einstein-Hilbert and gauge-field kinetic terms also keep the
coefficient of $FFA$ in the same ratio.
  
  This rigid rescaling, which changes the entire action homogeneously,
is given by
\be
g_{\sst{MN}}\longrightarrow \lambda^2\, g_{\sst{MN}}\ ,
\qquad A_{\sst{MNP}}\longrightarrow \lambda^3\, A_{\sst{MNP}}\ .
\label{trombone}
\ee
Since it gives a homogeneous rescaling of the action, it is a symmetry
of the equations of motion.  It can alternatively be viewed as an
engineering scale invariance of the eleven-dimensional classical
equations, as a consequence of the fact that there is just one
overall dimensionful coupling constant, which sits in front of the
entire eleven-dimensional action.
 
     We shall reduce the theory to $D$ dimensions in a succession of
1-step compactifications on circles.  At each stage in the reduction,
say from $(D+1)$ to $D$ dimensions, the metric is reduced according to
the standard Kaluza-Klein prescription
\be
ds_{\sst D+1}^2 = e^{2\a\varphi} \, ds_{\sst D}^2 + e^{-2(D-2)\a\varphi}\,
(dz+{\cal A}_\1)^2\ ,\label{metred}
\ee
where the $D$ dimensional metric, the Kaluza-Klein vector potential
${\cal A}_\1={\cal A}_{\sst M}\, dx^{\sst M}$ with $M=0,1,...,D$ and
the dilatonic scalar $\varphi$ are taken to be independent of the
ignorable coordinate $z$ on the compactifying circle.  The constant
$\a$ is given by $\a^{-2} = 2(D-1)(D-2)$, and the parameterisation of
the metric is such that a pure Einstein action is reduced again to a
pure Einstein action together with a canonically-normalised kinetic
term for $\varphi$ and a dilated kinetic term for ${\cal F}_\2=d{\cal
A}_\1$:
\be
e\, R \longrightarrow e\, R -\ft14 e\,
e^{-2(D-1)\a\varphi}\, {\cal F}_\2^2 -\ft12 e\, (\del\varphi)^2\ .
\label{einstred}
\ee

    Gauge potentials reduce according to $A_{\sst{(n)}}(x,z)=
A_{\sst{(n)}}(x) + A_{\sst{(n-1)}}(x)\wedge dz$, implying that a
kinetic term for an $n$-form field strength $F_{\sst{(n)}}$ reduces
according to the rule:
\be
-\fft1{2 \, n!}\, e \, F_{\sst{(n)}}^2 \longrightarrow 
-\fft1{2 \, n!}\, e\,
e^{-2(n-1)\a\varphi}\,  F_{\sst{(n)}}^2  -\fft1{2\, (n-1)!} \, e\,
e^{2(D-n)\a\varphi}\,  F_{\sst{(n-1)}}^2\ .\label{fred}
\ee

There is a subtlety here in the definition of the
dimensionally-reduced field strength $F_{\sst{(n)}}$, which is most
easily seen by working with the adapted (``triangular") vielbein of
\cite{cj3} and tangent space (flat) indices, since this facilitates
the computation of the inner products in the kinetic terms and in fact
uses the bundle principal connection.  From the ansatz for the
reduction of the gauge potential we have
\be
F_{\sst{(n)}}\longrightarrow dA_{\sst{(n-1)}} + dA_{\sst{(n-2)}}\wedge 
dz = dA_{\sst{(n-1)}} -
dA_{\sst{(n-2)}}\wedge {\cal A}_\1 + 
dA_{\sst{(n-2)}}\wedge (dz+{\cal A}_\1)\ .\label{cs}
\ee
Thus while it is natural to define the dimensionally-reduced field
strength $F_{\sst{(n-1)}}$ by $F_{\sst{(n-1)}}=dA_{\sst{(n-2)}}$, we
shall define $F_{\sst{(n)}}$ by $F_{\sst{(n)}}=dA_{\sst{(n-1)}} -
dA_{\sst{(n-2)}}\wedge {\cal A}_\1$; it is this gauge-invariant field
strength that appears on the right-hand side of (\ref{fred}).  Note
that this makes the meaning of the symbol $F$ dimension-dependent. In
Appendix A the plain exterior derivative of a potential $A$ is called
$\td F$, tildes are also used here and there for different purposes
when there is no ambiguity.  Similar non-linear Kaluza-Klein
modifications to the lower-dimensional field strengths become
progressively more complicated as the descent through the dimensions
continues. These definitions are analysed further in Appendix A.

     It is not too difficult now to apply the above reduction
procedure iteratively \cite{lpsol}, to construct the $D$-dimensional
toroidally-compactified theory from the eleven-dimensional starting
point.  It is easy to see that the original eleven-dimensional fields
$g_{\sst{MN}}$ and $A_{\sst{MNP}}$ will give rise to the following
fields in $D$ dimensions,
\bea
g_{\sst{MN}} &\longrightarrow & g_{\sst{MN}}\ ,\qquad \vec\phi\ ,\qquad
{\cal A}_\1^{i}\ ,\qquad {\cal A}_\0^i{}_j \ ,\nn\\
A_\3 &\longrightarrow & A_\3\ ,\qquad A_{\2i}\ , \qquad A_{\1ij}\ ,
\qquad A_{\0ijk}\ ,\label{dfields}
\eea
where the indices $i, j, k$ run over the $11-D$ internal
toroidally-compactified dimensions, starting from $i=1$ for the 
step from $D=11$ to $D=10$.  The potentials $A_{\1ij}$ and $A_{\0ijk}$
are automatically antisymmetric in their internal indices, whereas the
0-form potentials ${\cal A}_\0^i{}_j$ that come from the subsequent
dimensional reductions of the Kaluza-Klein vector potentials ${\cal
A}^i_{\1}$ are defined only for $j>i$.  (Note that in the standard
notation, the set of potentials $(A_{\2i}, A_{\1ij}, A_{\0ijk})$
correspond to $(A_{\mu\nu i}, A_{\mu ij}, A_{ijk})$.  The quantity
$\vec\phi$ denotes the $(11-D)$-vector of dilatonic scalar fields
coming from the diagonal components of the internal metric.

           The detailed expression for the Lagrangian for the bosonic
sector of the $D$-dimensional toroidal compactification of
eleven-dimensional supergravity is presented in Appendix A.  Note that
at this stage the Lagrangian is simply the one obtained directly from
dimensional reduction, without performing any dualisations.  In the
next subsection, we show that this Lagrangian has a $GL(11-D,\R)\semi
\R^q$ global symmetry.

\subsection{No dualisation and $GL(N,\R)\semi \R^q$}

      The $SL(N,\R)$ part of the global symmetry is a completely
general consequence of the dimensional reduction to $D$ dimensions of
any $(D+N)$-dimensional theory that includes gravity \cite{cj3}. In
order to implement an internal $\R$ symmetry from the last generator
of $GL(N,\R)$, $D$ must be strictly larger than 2.  There is a
subtlety here, namely the effect of the Weyl rescalings that are
needed in order to go to the so-called Einstein frame. The $\R^q$ part
of the symmetry, on the other hand, comes from the local abelian gauge
symmetry of an antisymmetric tensor field strength in the original
$(D+N)$ dimensions.  Specifically, it describes the global shift
symmetries of the axionic scalars that are the potentials for 1-form
field strengths coming from the dimensional reduction. This
$GL(N,\R)\semi \R^q$ symmetry can be discussed in any dimension.  Let
us consider a theory in $(D+N)$ dimensions, containing a metric, a
dilaton $\phi$ and a degree $n$ antisymmetric tensor field strength
$F_n=dA_{n-1}$.  This theory is invariant under the general coordinate
transformations
\bea
\delta x^{\sst M} &=& - \xi^{\sst M} (x)\ ,\qquad
\delta \phi = \xi^{\sst M} \del_{\sst M} \phi\ ,\nonumber\\
\delta A_{\sst{M}_1\cdots\sst{M}_{n-1}} &=&
\xi^{\sst M} \del_{\sst M} A_{\sst{M}_1\cdots\sst{M}_{n-1}} +
(n-1)\, \del_{[\sst{M}_1} \xi^{\sst M}
A_{|\sst{M}|\sst{M}_2\cdots\sst{M}_{n-1}]}\ .\label{gctdpn}
\eea
Now, we compactify the theory to $D$ dimensions, splitting the index
$\sst M$ into a $D$-dimensional index $\mu$ and an $N$-dimensional
internal index $i$, with coordinates $x^\mu$ and $y^i$ respectively,
{\it i.e.}\ $x^\sst{M} = (x^\mu, y^i)$.  We then impose the toroidal
Kaluza-Klein condition that all the $D$-dimensional fields are
independent of the compactifying coordinates $y^i$, namely
$\del_i\phi=0=\del_i A_{\sst{M}_1,\ldots,\sst{M}_{n-1}}=0$.  Note that
the Kaluza-Klein ansatz requires that the transformed fields should
also be independent of $y^i$, implying that $\del_i \xi^\mu=0$,
$\del_i\del_\mu \xi^j=0$ and $\del_i\del_j \xi^k=0$.  These equations
have the solution
\be
\xi^\mu = \xi^\mu(x^\nu)\ ,\qquad
\xi^i=\Lambda^{i}{}_{j}\, y^j + \xi^i(x^\nu)\ ,
\ee
where the $\Lambda^i{}_j$ are constants. The resulting transformations
imply the following symmetries in $D$ dimensions:

\be
\cases{ \delta x^\mu = -\xi^\mu(x)\ , & reparameterisation
invariance in $D$ dimensions\ ,\cr
\delta y^i = -\xi^i(x)\ , &  local $\R^N$ invariance\ ,\cr
\delta y^i = -\Lambda^i{}_j\, y^j\ , & global $GL(N,\R)$ invariance\ .\cr
}
\ee
\medskip

In the sector with ignorable internal coordinates we cannot
distinguish the compactification torus from $\R^N$, but the massive
excitations would not transform under $GL(N,\R)$. Note that the naive
$GL(N,\R)\sim\R\times SL(N,\R)$ is to be combined with the rigid
rescaling (\ref{trombone}) to become an internal symmetry (\ie one
that leaves the metric invariant).  Indeed the plain $\R$ symmetry
rescales the volume of the compactifying space, and must be combined
with the rescaling (\ref{trombone}). This defines a new internal
scaling symmetry which we call $\R_s$.  The $SL(N,\R)$ however leaves
the volume fixed; this corresponds to the restriction $\sum_i
\Lambda^i{}_i=0$.  In particular, from (\ref{gctdpn}) we find that
$SL(N,\R)$ acts on internal world indices on fields according to the
rules
\be
\delta A_i = \Lambda^j{}_i\, A_j\ ,\qquad \delta V^i = -\Lambda^i{}_j\,
V^j \ .\label{glnud}
\ee 
 
        In the above discussion, we showed that the internal part of
the $(D+N)$-dimensional reparameterisation invariance describes a
$SL(N,\R)$ global symmetry from the $D$-dimensional point of view.
There is in addition a local gauge symmetry in $(D+N)$ dimensions,
namely
\be
\delta A_{\sst{M}_1\cdots \sst{M}_{n-1}} = (n-1)\, \del_{[\sst{M}_1}
\lambda_{\sst{M}_2\cdots \sst{M}_{n-1}]}\ .\label{dpng}
\ee
This gives rise to local gauge symmetries for the
dimensionally-reduced gauge potentials $A_{\sst{M}_1\cdots
\sst{M}_{n-2}}$ with one or more $D$-dimensional spacetime indices. In
the case of the $N!/((n-1)!(N-n+1)!)$ 0-form potentials, or axionic
scalars, however, only global shift symmetries remain.  To see this,
we note from (\ref{dpng}) that the transformation rules for the axions
are given by
\be
\delta A_{i_1\cdots i_{n-1}} =
(n-1)\, \del_{[i_1} \lambda_{i_2\cdots i_{n-1}]}\ .
\ee
In order for these variations to be nonzero and for (\ref{dpng}) to be
independent of the internal coordinates $y^i$, we must have
$\lambda_{i_2\cdots i_{n-1}} = c_{i_2\cdots i_{n-1} j}\, y^j$, where
$c_{i_2\cdots i_{n-1} j}$ is any constant antisymmetric tensor, giving
\be
\delta A_{i_1\cdots i_{n-1}} = c_{i_1\cdots i_{n-1}}\ .
\label{rps}
\ee
 
Thus the $D$-dimensional theory has also an $\R^q$ symmetry, with
$q=N!/((n-1)!(N-n+1)!)$.  Clearly these $\R$ symmetries commute with each 
other, since they are derived directly from the abelian gauge symmetry 
in $(D+N)$ dimensions.   However they do not all commute with the
$GL(N,\R)$ symmetries, since the axions $A_{i_1\cdots i_{n-1}}$ also
transform (covariantly) under $GL(N,\R)$.   This can be seen from the
commutation relations
\bea
[\delta_{c}, \delta_{\sst \Lambda}]=\delta_{\tilde c}\ ,\qquad
\tilde c_{i_1\cdots i_{n-1}}= 
(n-1) \Lambda^i{}_{[i_1}\, c_{|i|i_2\cdots i_{n-1}]}\ ,
\label{cdeltac}
\eea
where the $GL(N,\R)$ transformations parameterised by $\Lambda^i{}_j$ are
given by
\be
[\delta_{\sst{\Lambda}}, \delta_{\sst{\Lambda}'}]
= \delta_{\tilde\sst{\Lambda}}\ ,\qquad \tilde\Lambda^i{}_j = 
\Lambda^i{}_k\,  {\Lambda '}^k{}_j -{\Lambda '}^i{}_k \, \Lambda^k{}_j\ .
\ee
(Examples of fields with upstairs, contravariant world indices have
not arisen yet in our discussion, but we shall encounter them later.)
As a matter of fact note that, from the general coordinate invariance
of the 1-forms $\tilde\gamma^i{}_j \, (dy^j+\hat{\cal A}_\1^{j}) \,
\equiv dy^i + {\cal A}_\0^i{}_j\, dy^j +{\cal A}_\1^i $, we deduce
that the axions ${\cal A}_\0^i{}_j$ must transform inhomogeneously as
\be
\delta {\cal A}_\0^i{}_j = \Lambda^i{}_j + \Lambda^k{}_j \, {\cal A}_\0^i{}_k
\label{glnax}
\ee
under $SL(N,\R)$, while the 1-forms ${\cal A}_\1^i$ are inert. In
Appendix A this rule is derived from a careful distinction between
tangent and spacetime (internal) indices. $\hat{\cal A}_\1^{j}$ and
$\gamma^j{}_i$ (the inverse of $\tilde\gamma$) do transform as vectors
under $SL(N,\R)$.

    Let us now apply the above discussion to the dimensional reduction
of eleven-dimensional supergravity, for which we have $n=4$ and
$N=11-D$.  Thus the $D$-dimensional theory (without any dualisation)
has a global symmetry $GL(11-D,\R)\semi \R^q$, with $q=\ft16
(11-D)(10-D)(9-D)$.\footnote{Note that for all dimensions $D\ge3$, we
could (inversely) dualise all the $p$ axions $A_{\0ijk}$ to
$(D-2)$-form gauge potentials, since these axions may be all
simultaneously covered by derivatives everywhere in the Lagrangian.
The resulting theory would then have only a $GL(11-D,\R)$ global
symmetry.  In this and the next sections, we shall concentrate only on
those $D$-dimensional Lagrangians that are direct dimensional
reductions of eleven-dimensional supergravity and the ones obtained
 from these by the dualisation of the 4-form, 3-form or 2-form field
strengths to axions in $D=5$, 4 or 3 respectively.  We shall also
discuss the dimensional reduction of IIB supergravity in $D=10$,
without performing any dualisation.}  It should be emphasised that
$GL(11-D,\R)\semi \R^q$ is a symmetry at the level of the Lagrangian
that is derived from direct dimensional reduction without any
dualisation.  In the process, we also make use of Weyl rescalings so
that all the lower dimensional Lagrangians are written in the Einstein
frame.  This rescaling modifies the $\R$ part of $GL(11-D,\R)=\R\times
SL(11-D,\R)$, which becomes as a result an internal symmetry ($\R_s$).
This is the first instance of a hidden symmetry. It can be traced back
to the eleven-dimensional action, which is invariant up to a factor
under engineering rescalings, as well as (and equivalently) under Weyl
rescalings of the metric coupled to appropriate multiplicative
redefinitions of the 3-form. The next hidden symmetry arises in $D=8$,
for which the $\R_s$ factor becomes $SL(2,\R)$.  In dimension $D=7$
and below, the obvious and the hidden internal symmetries combine to
form a simple group.
 
    In this paper, we obtain lower-dimensional supergravities by
iteratively applying the $D+1$ to $D$ dimensional reduction.  This has
the effect that the manifest $SL(11-D,\R)$ symmetry is reduced to the
Borel subgroup, generated by the positive-root generators of the
group, with infinitesimal transformation parameters $\Lambda^i{}_j$
that are non-zero only for $i <j$, {\it i.e.} they are upper
triangular matrices.

         As we shall show in section 4, in the cases $D\ge 6$ the
global symmetry of the theory can be extended to $E_{11-D}$, provided
that certain higher-degree fields are dualised appropriately. In fact
the scalar Lagrangian in $D\ge6$ already has the full $E_{11-D}$
global symmetry, and the $GL(11-D,\R)\semi \R^q$ symmetry described
above is a subgroup of it. The fact that the extra scalars are
internal 3-forms is reflected by the fact that the extra root of the
$E_{11-D}$ group is above the third root of the $GL(11-D,\R)$ subgroup
that corresponds to the highest weight of that particular internal
$GL(11-D,\R)$ representation.  Note that in $D\ge6$ the number
$q=\{0,0,1,4,10\}$ of $\R$ symmetries for $D=\{10,9,8,7,6\}$ never
exceeds the dimension of the maximal abelian subalgebra of the group
$E_{11-D}$ corresponding to the fully-dualised theory, namely $\{1,2,
3, 6, 10\}$ \cite{ma}.

        The situation is different when $D\le 5$.  In these lower
dimensions, the theory contains $(D-2)$-form gauge potentials which
can be dualised to give rise to additional axionic scalars.  Before
any such dualisation is performed the theory has a $GL(11-D,\R)\semi
\R^q$ global symmetry, which can be enlarged to $E_{11-D}$ only after
performing certain necessary dualisations.  At first sight this is
rather counter-intuitive, since one might expect that at the level of
the equations of motion dualisation should have no effect on the
global symmetry of the theory.  Indeed this is true for the
dualisations of fields that appear in the Lagrangian only via their
field strengths, since, as we discussed in the Introduction, in such
cases the field strengths rather than their potentials can be treated
as the fundamental fields in the equations of motion.  However, in the
case of scalar potentials some transformations act on the scalars
themselves and not on their derivatives, and so the global symmetry
will be altered if any of the scalars are dualised to $(D-2)$-form
potentials.  An explicit example of a scalar Lagrangian with
$SL(2,\R)$ symmetry is discussed in Appendix B.1, to illustrate how
the dualisation of the axion alters the symmetry.  In the case of
supergravities in $3\le D \le 5$, the global symmetry analysis is more
complicated.  One way to see that the undualised theories do not have
the global $E_{11-D}$ symmetry is to look at the maximal abelian
$\R^q$ symmetries in the two versions.  In the undualised versions we
have certainly at least $\R^{20}$, $\R^{35}$ and $\R^{56}$ abelian
symmetries in $D=5$, 4 and 3, whose dimensions exceed that of the
maximal abelian subalgebras of $E_6$, $E_7$ and $E_8$, namely 16, 27
and 36 respectively. (The latter dimensions have been determined in
\cite{ma} by educated inspection.)  In the next section we shall show
explicitly how the $\R^q$ symmetry changes under the dualisation, in
each of the dimensions $D=5$, $4$ and $3$.

\section{Dualisation and the maximal scalar manifolds in $D=5,4,3$}

         In the previous section, we showed that the $D$-dimensional
supergravities coming from direct dimensional reduction of
eleven-dimensional supergravity without any dualisation have
$GL(11-D,\R)\semi \R^q$ global symmetries, where $q=\ft16 (11-D)
(10-D) (9-D)$.  The transformations are realised on the gauge
potentials, and thus correspond to symmetries of the Lagrangian (hence
of the equations of motion).  In fact in $D\ge6$, the Lagrangian for
the scalar sector actually has an $E_{11-D}$ global symmetry (see
section 4), which contains $GL(11-D,\R)\semi \R^q$ as a proper
subgroup in $6\le D \le 8$.  (In $D\ge 9$, the groups $GL(11-D,\R)$
and $E_{11-D}$ coincide.)  It is possible to extend the $E_{11-D}$
symmetry of the scalar sector to the full theory, when appropriate
dualisations of higher-rank fields are performed. In dimensions 10 and
9 the only subtle symmetry is the modified $\R_s$ generator coming
 from $GL(N,\R)$, but from dimension 8 on we encounter the Ehlers
miracle of an extra $SL(2,\R)$ factor (the name comes from the
analogous phenomenon in ordinary General Relativity reduced to 3
dimensions). In both cases this extra symmetry involves dualisation of
gauge forms. We shall return to this in section 6.

     In $D=5,4,3$, the scalar Lagrangian coming directly from the
dimensional reduction of the eleven-dimensional theory has only the
$GL(11-D,\R)\semi \R^q$ symmetry described above.  This is also true
at the level of the equations of motion.  The extension to $E_{11-D}$
is possible only when all the 3-form, 2-form or 1-form potentials
respectively are dualised to give additional scalars.  This is because
the $E_{11-D}$ symmetry of the dualised formulation must be
implemented on the scalars (0-form potentials) themselves, rather than
on their derivatives.  We shall use the term maximal scalar manifold
to refer to this situation where the number of scalar fields is made
as large as possible.  Recall that in $3\le D\le5$ the $GL(11-D,\R)
\semi \R^q$ symmetry is not contained in $E_{11-D}$ as we have seen by
observing that the $\R^q=\{\R^{20},\R^{35},\R^{56}\}$ abelian factors
are larger than the maximal abelian subalgebras
$\{\R^{16},\R^{27},\R^{36} \}$ of $\{ E_6,E_7,E_8\}$.  Thus the
fully-dualised and undualised formulations of the theories have
inequivalent global symmetries, neither of which encompasses the
other.

       We shall now explicitly perform the dualisations of the
$(D-2)$-form potentials to give rise to additional scalars in each of
the dimensions $D=5$, 4 and 3, and show how the global symmetries are
altered. Note that when we dualise all the $(D-2)$-form potentials,
the $GL(11-D,\R)$ symmetry is preserved in each case, now becoming a
subgroup of the full enlarged $E_{11-D}$ symmetry.  There are also
other possibilities, in which we may choose to dualise only a subset
of the $(D-2)$-forms.  For example, we might dualise only those fields
which, from the type IIA string point of view, are associated with the
Ramond-Ramond sector \cite{lptdual}.  In this case, the $GL(11-D,\R)$
symmetry is broken.  We shall discuss this in detail in section 7.

\subsection{Dualisation in $D=5$ supergravity}

        In $D=5$, the 3-form gauge potential $A_{\mu\nu\rho}$, which
comes from the dimensional reduction of the 3-form in $D=11$, can be
dualised to give a scalar field.  As in the case of the $SL(2,\R)$
example in Appendix B, an additional $\R$ symmetry is then created,
corresponding to a global shift symmetry of the new scalar field,
since it can be covered by a derivative everywhere in the Lagrangian.
Naively, one would expect, as in the case of the $SL(2,\R)$ example,
that the dualisation procedure would always increase the dimension of
the commuting $\R$ symmetries, since we have obtained new scalars that
can be covered by derivatives everywhere.  However, in the present
case there is an additional term in the scalar Lagrangian, coming from
the dimensional reduction of ${\cal L}_{FFA}$ in $D=11$. Without this
additional topological term, the analysis would indeed be analogous to
the $SL(2,\R)$ example in Appendix B, and we would have a dualised
theory in which the $\R^{20}$ symmetry of the original 0-form gauge
potentials of the undualised theory was enlarged to $\R^{20+1}$.  Let
us look in detail at the Lagrangian involving $F_\4=dA_\3 + \cdots$ in
$D=5$, given by
\be
{\cal L}(F_\4) = -\ft1{48} e e^{\vec a\cdot \vec\phi}\, F_\4^2 -\ft1{1728}e\, 
\epsilon^{ijklmn}\, A_{\0ijk}\,  \del_\mu A_{\0\ell mn}\, 
F_{\nu\rho\sigma\lambda}\, \epsilon^{\mu\nu\rho\sigma\lambda}\ ,
\label{d5f4lag}
\ee
where the second term comes from the ${\cal L}_{FFA}$ terms
(\ref{ffaterms}) and $\vec a$ is defined in Appendix A.  If no
dualisation of $A_\3$ were to be performed, we could add a topological
surface term so that all the 20 axions $A_{\0ijk}$ would be
simultaneously covered by derivatives, implying an abelian $\R^{20}$
symmetry.  But in order to dualise $A_\3$, we first introduce a
Lagrange multiplier $\chi$ to impose the Bianchi identity $dF_\4=0 +
\cdots$, by adding the term $\chi d F_\4$ to the Lagrangian (in order
to construct the scalar Lagrangian we may neglect the difference
$F_\4-dA_\3 $).  This leads to the first-order Lagrangian
\be
{\cal L}(F_\4) = -\ft1{48} e e^{\vec a\cdot\vec\phi}\, 
F_\4^2 -\ft1{1728}e\,
\epsilon^{ijklmn}\, A_{\0ijk} \del_\mu A_{\0\ell mn} 
F_{\nu\rho\sigma\lambda}\, \epsilon^{\mu\nu\rho\sigma\lambda} +
\ft1{24}e\, \chi\, \epsilon^{\mu\nu\rho\sigma\lambda} \,\del_\mu
F_{\nu\rho\sigma\lambda}\ .\label{d5folag}
\ee
Integrating by parts, we have
\be
{\cal L}(F_\4) = -\ft1{48} e e^{\vec a\cdot\vec\phi}\, F_\4^2 +\ft1{24}e\,
\epsilon^{\mu\nu\rho\sigma\lambda}\, F_{\nu\rho\sigma\lambda}
\Big(\del_\mu \chi - A_{\0ijk}
\del_\mu A_{\0\ell mn}\, \ft1{72}\epsilon^{ijk\ell mn} \Big)\ ,
\label{d5folag1}
\ee
 from which we can solve algebraically for $F_\4$, giving 
\be
F_{\mu\nu\rho\sigma}= e^{-\vec a\cdot\vec\phi}\, 
\epsilon_{\mu\nu\rho\sigma\lambda}\, 
\Big(\del^\lambda \chi -\ft1{72}
A_{\0ijk}\del^\lambda A_{\0\ell mn}\,  \epsilon^{ijk\ell mn}\Big)\ .
\ee
Thus after the dualisation of $A_\3$, the Lagrangian (\ref{d5f4lag}) 
becomes
\be
{\cal L}= -\ft12 e e^{-\vec a\cdot \vec \phi} G_\1^2\ ,
\label{d5dual}
\ee
where
\be
G_\1=d\chi -\ft1{72}\, 
  A_{\0ijk}\, d A_{\0\ell mn}\, \epsilon^{ijk\ell mn} \label{d5g}
\ee
is the 1-form field strength dual to $F_\4$, {\it i.e.}\ $F_\4 =
e^{-\vec a\cdot\vec\phi}\, \ast G_\1$.  Note that the dualisation has
the effect of reversing the sign of the dilaton coupling $ \vec a$.
We see that the $\R^{20}$ symmetry $\delta A_{\0ijk} = c_{ijk}$ of the
original Lagrangian (\ref{d5f4lag}) (under which all the other
original axions were inert) becomes
\be
\delta A_{\0ijk} = c_{ijk}\ ,\qquad
\delta \chi = k +\ft1{72}
c_{ijk}\, A_{\0\ell mn}\, \epsilon^{ijk\ell mn} \ ,
\ee
where $k$ is the constant shift symmetry associated with $\chi$.
Under the rescaling symmetry $\R_s$, $\delta \vec \phi = \ft12
\mu\,\vec g$, where $\vec g$ is defined in Appendix A, and $\delta
A_{\0ijk} = -\ft12 \mu\, A_{\0ijk}$, and $\delta \chi = -\mu\, \chi$.
These transformations leave the Lagrangians (\ref{d5folag1}) and
(\ref{d5dual}) invariant.  Thus the original $\R^{20}$ symmetries no
longer all commute once we consider the transformations of the new
scalar $\chi$.  Indeed we have
\bea
&&{[}\delta_{c}, \delta_{c'}{]}= \delta_{k}\ , \qquad 
k=\ft1{36} {c'}_{ijk}\,
c_{\ell mn}\, \epsilon^{ijk\ell mn}\ ,\nn\\
&&{[}\delta_\mu, \delta_k{]} = \delta_{\td k}\ , \qquad
\td k = \mu\, k\ ,\label{d5ccom}\\
&&{[} \delta_\mu, \delta_c {]} = \delta_{c'}\ ,\qquad
c'_{ijk} = \ft12 \mu\, c_{ijk}\ ,\nn
\eea
The first line of (\ref{d5ccom}) implies that there are now only 10
commuting $\R$ symmetries. Note that the factor $1/36$ is purely
combinatorial and would correspond to a one if one were to order the
indices of $c_{ijk}$.  Without loss of generality, the $\R^{10}$
symmetry may be taken to correspond to the parameters $c_{\a\beta6}$,
where we have split the index $i$ into $i=(\alpha,6)$, with
$\a=1,\ldots 5$.  The original abelian global $\R^{20}$ symmetry of
the original Lagrangian (\ref{d5folag}) for $F_\4$ therefore becomes
non-abelian in general, with an abelian $\R^{11}$ left, after the
dualisation that replaces $F_\4$ by $G_\1$.  Note that the reduction
of the abelian symmetry $\R^{20} \rightarrow \R^{11}$ is already seen
in (\ref{d5folag1}) at the stage when the Lagrange multiplier $\chi$
is first introduced, even before the field $F_\4$ is eliminated using
its algebraic equation of motion.  In this first-order formulation,
$F_{\mu\nu\rho\sigma}$ is no longer viewed as a field strength; rather
it is an auxiliary field that can be integrated out to give rise to
the dualised Lagrangian (\ref{d5dual}), and this route to the dual
action forbids bare potentials $A_{\mu\nu\rho}$.

        So far we have restricted our discussion only to the
Lagrangian for the 4-form field strength.  As we discussed earlier,
before dualisation the symmetry group of the full scalar Lagrangian is
$GL(6,\R) \semi \R^{20}$.  There is in fact a maximal abelian $\R^9$
subalgebra in $SL(6,\R)$, but this symmetry does not commute with the
$\R^{20}$.  After dualisation, we see that the commuting $\R^{20}$
symmetry is reduced to $\R^{10}$, because of the ${\cal L}_{FFA}$
terms.  This raises the possibility that some of the $\R^9$ symmetry
in $SL(6,\R)$ might commute with the remaining $\R^{10}$ symmetry.  To
see that this indeed occurs, let us denote the $SL(6,\R)$
transformation parameters by $\Lambda^i{}_j$.  Since the axions
$A_{\0ijk}$ transform covariantly under $SL(6,\R)$, we have
\bea
{[}\delta_{\sst{\Lambda}}, \delta_{\sst{\Lambda}'}{]} &=&
\delta_{\tilde\sst{\Lambda}}\ ,\qquad\qquad
\tilde\Lambda^i{}_j = 
\Lambda^i{}_k\,  {\Lambda '}^k{}_j -{\Lambda '}^i{}_k \,
\Lambda^k{}_j\ ,\nn\\
{[\delta_c,\delta_{\sst{\Lambda}}]} &=&\delta_{\tilde c}\ ,\qquad\qquad
\tilde c_{ijk} = 3\, \Lambda^\ell{}_{[i} c_{jk]\ell}\ .
\label{d5com}
\eea
It is straightforward to verify that (\ref{d5ccom}) and (\ref{d5com})
generate the complete Borel subalgebra of $E_6$, (restricting the
algebra of $SL(6,R)$ to its Borel subalgebra.)  Note that the
transformations associated with $\Lambda^\a{}_6$ commute with
themselves, as well as with those of $c_{\beta\gamma6}$.  In addition,
the shift symmetry $k$ commutes with $SL(6,\R)$ since the
corresponding axion $\chi$ is a singlet under $SL(6,\R)$.  Thus in the
fully dualised $D=5$ supergravity, the theory contains $10+1+5=16$
commuting $\R$ symmetries, corresponding to the parameters
\be
c_{\a\beta 6}\ ,\qquad k\ , \qquad \Lambda^{\a}{}_6\ ,\label{d5p}
\ee
(These commuting $\R$ symmetries imply that the corresponding axions,
namely $A_{\0\a\beta6}$, $\chi$ and ${\cal A}_\0^\a{}_6$, can be
covered by derivatives simultaneously in the Lagrangian.)  This result
is not an accident.  In section 4, we shall show that the scalar
Lagrangian of the fully-dualised five-dimensional supergravity has an
$E_6$ global symmetry, and the commuting subset in (\ref{d5p})
precisely corresponds to the maximal abelian $\R^{16}$ subalgebra of
$E_6$ (recall that here, as in the rest of the paper, we are referring
to the maximally non-compact forms (also called split real forms) of
our Lie algebras, namely $E_{6(6)}$ in this case). We may insist that
maximal abelian subalgebra stands for maximal abelian subalgebra of
maximal dimension. Clearly we must stay away from the requirement of
ad-semisimplicity of its generators that characterise Cartan tori; in
a sense we are looking for maximally nilpotent generators instead.

\subsection{Dualisation in $D=4$ supergravity}

       In $D=4$ dimensions, there are seven 2-form gauge potentials
$A_{\2i}$, which can be dualised to give additional axionic scalar
fields.  In the undualised form, the theory possesses a $
GL(7,\R)\semi \R^{35}$ global symmetry, with the $\R^{35}$ realised by
the transformations $\delta A_{\0ijk} = c_{ijk}$.  We shall now show
how, owing to the presence of the ${\cal L}_{FFA}$ terms in the
Lagrangian, the $\R$ symmetries are modified by the dualisation. The
subsector of the Lagrangian involving the 2-form gauge potentials,
which eventually turn into axionic scalars in the dualised theory, is
given by (see Appendix A)
\be
{\cal L} = -\ft1{12} e \sum_{i=1}^{7} 
e^{\vec a_i\cdot \vec \phi} (F_{\3i})^2 -\ft1{72}
 A_{\0ijk}\, d A_{\0\ell mn}\wedge 
d A_{\2p} \, \epsilon^{ijk\ell mnp}\ ,
\label{d4f3lag}
\ee
where the associated field strengths $F_{\3i}$ have the
Kaluza-Klein non-linear modifications
\be
F_{\3i} = \gamma^j{}_i\, dA_{\2j}+\cdots \,\label{f3cs}
\ee
with $\gamma^i{}_j$ given by (\ref{gam}).  

    Multiplying (\ref{f3cs}) by $\tilde \gamma^i{}_k$, the inverse
matrix defined in (\ref{gaminv}), and taking an exterior derivative,
we see that the Bianchi identities for the field strengths $F_{\3i}$
are given by
\be
d(\td \gamma^i{}_k\, F_{\3i}) =0\ \label{f3bi}
\ee
modulo non-scalar terms.
They can be enforced by introducing seven
Lagrange multipliers $\chi^i$, leading to the first-order Lagrangian
\bea
{\cal L} &=& -\ft1{12} e \sum_{i=1}^7 
e^{\vec a_i\cdot \vec \phi} (F_{\3i})^2 -
\ft1{432}e\,\epsilon^{ijk\ell mnp}\,
A_{\0ijk}\, \del_\mu A_{\0\ell mn}\, 
\td \gamma^q{}_p\,F_{\nu\rho\sigma q}\, 
\epsilon^{\mu\nu\rho\sigma}\nn\\ &&\qquad\qquad - \ft16 e\, \del_\mu\chi^i\,
\td\gamma^j{}_i\, F_{\nu\rho\sigma j} \,
\epsilon^{\mu\nu\rho\sigma}\ .
\label{d4folag}
\eea
It is now easy to solve the algebraic equations of motion for
$F_{\3i}=e^{-\vec a_i\cdot\vec\phi}\,\ast G_\1^i$, 
and to substitute this back into the Lagrangian, giving
\be
{\cal L} = -\ft12 e\,  \sum_{i}^7 e^{-\vec a_i\cdot \vec \phi}
(G_\1^{(i)})^2\ ,\label{d4dual}
\ee
where
\be
G_\1^i=\td\gamma^i{}_j\, \Big(d \chi^j +\ft1{72}
       A_{\0k\ell m} \, d A_{\0npq}\, \epsilon^{jk\ell mnpq}
\Big)\ .\label{d4g}
\ee

          As in the previous five-dimensional case, here the
original $\R^{35}$ and the scaling $\R_s$ global symmetry of the 
undualised Lagrangian (\ref{d4f3lag}) become
\bea
\delta A_{\0ijk}& =& c_{ijk}-\ft13 \mu\, A_{\0ijk}\ ,\qquad
\delta \chi^{i} = k^{i} - \ft1{72}  c_{jk\ell}
\, A_{\0mnp}\, \epsilon^{ijk\ell mnp} -\ft23 \mu\, \chi^i\ ,\nn\\
\delta \vec \phi &=& \ft13 \mu \, \vec g\ .
\eea
The transformations associated with the $c_{ijk}$ no longer all
commute if one now includes the action on the 7 new scalars, since
\be
[\delta_{c}, \delta_{c'}] =\delta_k\ ,\qquad k^i=
\ft1{36}  {c}_{jk\ell}\, {c'}_{mnp}\, \epsilon^{ijk\ell mnp} 
\ .\label{d4com}
\ee
The maximal commuting subset is the $\R^{15}$ corresponding, for
example, to the parameters $c_{\a\beta7}$, where the index $i=(\a, 7)$
with $\a=1,2,\ldots, 6$.  Since $\delta_{k}$ commutes with $\delta_c$,
the scalar Lagrangian (\ref{d4dual}) now has an $\R^{15+7} = \R^ {22}$
global symmetry.

       Of course this is not the whole story.  The full undualised
Lagrangian has also a $SL(7,\R)$ global symmetry, which itself
contains a maximal abelian subalgebra $\R^{12}$. Prior to dualisation,
this $\R^{12}$ symmetry does not commute with the $\R^{35}$, since the
axions $A_{\0ijk}$ transform (covariantly) under $SL(7,\R)$, with
commutation relations of the form given in (\ref{d5com}).  After
dualisation, however, the abelian $\R^{35}$ symmetry becomes
non-abelian in general with abelian $\R^{15}$ left.  Moreover, the
$SL(7,\R)$ remains unscathed under this full dualisation, and so it
becomes possible that some of the maximal abelian $\R^{12}$ subalgebra
of $SL(7,\R)$ now commutes with $\R^{15}$.  Indeed, following the
analogue of the analysis we performed in $D=5$, we find that six of
the $SL(7,\R)$ abelian transformations, associated with parameters
$\Lambda^\a{}_7$, commute with the $\R^{15}$ symmetry described by the
parameters $c_{\a\beta 7}$: As we remarked above, the $\R^7$
symmetries of the seven new axionic scalars coming from the
dualisation commute with $\R^{15}$, but they do not all commute with
$\delta_{\Lambda^\a{}_7}$; only six of them do, namely the six
associated with the parameters $k^\a$.  To see this we need, in
addition to the previously-given commutators of transformations,
\bea
{[} \delta_c, \delta_{\sst \Lambda} {]} &=& \delta_{\td c}\ ,\qquad
\td c_{ijk} = 3 \Lambda_{[i}{}^\ell\, c_{jk]\ell}\ ,\nn\\
{[} \delta_\mu, \delta_c {]} &=& \delta_{c'}\ ,\qquad
c'_{ijk} = \ft13 \mu\, c_{ijk}\ ,\nn\\ 
{[}\delta_k,\delta_\Lambda{]} &=& \delta_{\tilde k}\ , \qquad \tilde k^i= -
\Lambda^i{}_j\, k^j\ ,\label{lamk}\\
{[}\delta_{\mu}, \delta_k{]} &=& \delta_{\hat k}\ , \qquad \hat k^i=
\ft23 \mu\, k^i \ .\nn
\eea
Thus in the full Lagrangian of
the dualised theory, we have abelian symmetries with dimensions
$15+6+6=27$, associated with the parameters
\be
c_{\a\beta7}\ ,\qquad k^\a\ ,\qquad \Lambda^\a{}_7\ .
\ee
Note that $\R^{27}$ is precisely the maximal abelian subalgebra of
$E_7$.  As in $D=5$, the associated axions, $A_{\0\a\beta 7}$,
$\chi^\a$ and ${\cal A}_\0^\a{}_7$, can be simultaneously covered by
derivatives everywhere in the Lagrangian.  As in the previous case,
the complete algebra of these transformations is a subalgebra of the
Lie algebra $E_{7}$ which contains the Borel subalgebra of $E_7$.

\subsection{Dualisation in $D=3$ supergravity}

     Again, we begin by considering the subsector of the scalar
Lagrangian coming from the dualisation of higher-degree field
strengths.  In this case, it is the 8 2-forms ${\cal F}_\2^{i}$ and 28
2-forms $F_{\2ij}$ that yield additional scalars after dualisation.
 From (\ref{dgenlag}) and (\ref{ffaterms}), we see that these appear in
the Lagrangian in the form
\be
{\cal L} = -\ft14 e\, \sum_i e^{\vec b_i\cdot \vec\phi}\, ({\cal
F}_\2^{i})^2 -\ft14e\, \sum_{i<j} e^{\vec a_{ij}\cdot \vec\phi}\,\
(F_{\2ij})^2 -\ft1{144}\td F_{\1ijk}\wedge \td
F_{\1\ell mn}\wedge A_{\1pq}\, \epsilon^{ijk\ell mnpq}
\ ,\label{d3lag1}
\ee
and we must now use the full non-linear Kaluza-Klein modifications, 
given by (\ref{A.6}), because all fields are scalar in this 
three-dimensional theory;
\bea
F_{\2ij} &=& \gamma^k{}_i\, \gamma^\ell{}_j\,\Big( d A_{\1 k\ell} -
\gamma^m{}_n\, d A_{\0k\ell m}\wedge {\cal A}_{\1}^n\Big)\ ,\label{b1}\\
{\cal F}_\2^i &=& d{\cal A}_\1^i -\gamma^j{}_k\, d{\cal
A}_\0^i{}_j \wedge {\cal A}_\1^k\ .\label{b2}
\eea

     The first task before carrying out the dualisations is to express
the Bianchi identities associated with ${\cal F}_\2^{i}$ and
$F_{\2ij}$, and the final cubic interaction term in (\ref{d3lag1}),
purely in terms of the Kaluza-Klein modified field strengths, so that
these may then be eliminated algebraically once the Lagrange
multipliers enforcing the Bianchi identities are introduced.  Using
(\ref{aredef}-\ref{hatfieldstr}), we see that multiplying (\ref{b2})
by $\gamma^j{}_i$ gives
\be
\gamma^j{}_i\, {\cal F}_\2^{i} = d(\gamma^j{}_i\, {\cal A}_\1^{i})\ ,
\label{key}
\ee
and hence 
\be
d(\gamma^j{}_i\, {\cal F}_\2^{i} ) =0\ .\label{bian1}
\ee
Multiplying (\ref{b1}) by $\td \gamma^i{}_p\, \td\gamma^j{}_q$, and taking
the exterior derivative, we obtain, after using (\ref{key}), 
\be
d(\td \gamma^i{}_p\, \td \gamma^j{}_q\, F_{\2ij} - A_{\0pqm}\,
\gamma^m{}_n\, {\cal F}_\2^{n})=0\ .\label{bian2}
\ee
Equations (\ref{bian1}) and (\ref{bian2}) are the relevant Bianchi
identities, expressed in terms of the Kaluza-Klein modified field
strengths.  For the final cubic interaction term ${\cal L}_{FFA}$ in
(\ref{d3lag1}), we first use (\ref{b1}), multiplied by two inverse
$\gamma$ matrices, to re-express it as
\be
{\cal L}_{FFA}= \ft1{144}\Big(dA_{\0ijk}\,
A_{\0\ell mn}\, \td
\gamma^r{}_p\, \td \gamma^s{}_q\wedge F_{\2rs}  +
dA_{\0ijk}\, A_{\0\ell mn}\wedge dA_{\0pqr}\wedge \gamma^r{}_s\,
{\cal A}_\1^{s}\Big)\epsilon^{ijk\ell mnpq}\ .
\ee
For the second term in this expression, we use the Schouten identity
that a total antisymmetrisation over the nine indices $ijk\ell mnpqr$
vanishes to show that we may write
\be
 dA_{\0ijk}\, A_{\0\ell mn}\wedge
dA_{\0pqr}\, \epsilon^{ijk\ell mnpq}  = 
-\ft13 d\Big( dA_{\0ijk}\,
A_{\0\ell mn}\, A_{\0pqr}\, \epsilon^{ijk\ell mnpq} \Big)\ .
\ee
This implies, after an integration by parts and using (\ref{key}), that
${\cal L}_{FFA}$ can also be written purely in terms of the
modified field strengths, as
\be
{\cal L}_{FFA}= \ft1{144} \Big( dA_{\0ijk}\,
A_{\0\ell mn}\, \td
\gamma^r{}_p\, \td \gamma^s{}_t\wedge F_{\2rs} +\ft13 
dA_{\0ijk}\, A_{\0\ell mn}\, A_{\0pqr}\gamma^r{}_s\wedge {\cal
F}_\2^{s}\Big) \epsilon^{ijk\ell mnpq}\ .
\ee

     It is now a straightforward matter, after introducing Lagrange
multipliers $\lambda_{i}$ and $\lambda^{ij}$ to enforce the Bianchi
identities (\ref{bian1}) and (\ref{bian2}) respectively, to show that
the Lagrangian (\ref{d3lag1}) becomes, upon elimination of the
original 2-form fields by means of their algebraic equation of motion,
\be
{\cal L}= -\ft12 e\, \sum_i e^{-\vec b_i\cdot\vec\phi}\, (G_{\1i})^2
-\ft12 e\, \sum_{i<j} e^{-\vec a_{ij}\cdot \vec\phi}\, (G_\1^{ij})^2
\ ,
\ee
where the dualised field strengths $G_{\1i}$ and $G_\1^{ij}$ are given
by
\bea
G_{\1i} &=& \gamma^j{}_i\, \Big(d\lambda_j -\ft12 A_{\0jk\ell}\,
d\lambda^{k\ell} -\ft1{432} dA_{\0k\ell m}
\, A_{\0npq}\, A_{\0rsj}\, \epsilon^{k\ell mnpqrs}\Big)
\ ,\label{g1eq}\\
G_\1^{ij} &=& \td\gamma^i{}_k\, \td\gamma^j{}_\ell\, \Big(d\lambda^{k\ell}
+\ft1{72}  dA_{\0mnp}\, A_{\0qrs}\, \epsilon^{k\ell mnpqrs}
\Big)\ .\label{gg1eq}
\eea
Thus the original commuting $\R^{56}$ and $\R_s$ symmetry $\delta A_{\0ijk} 
=c_{ijk}$ now become
\bea
\delta A_{\0ijk} &=& c_{ijk}-\ft16 \mu\, A_{\0ijk}\ ,\qquad
\delta \lambda^{ij} = \kappa^{ij} - \ft1{72}
A_{\0k\ell m}\,  c_{npq}\,  \epsilon^{ijk\ell mnpq}
-\ft13\mu\, \lambda^{ij} \ ,\nn\\
\delta \lambda_i &=& \epsilon_{i} + \ft12 c_{ijk} \, \lambda^{jk}-
\ft{1}{432} A_{\0ijk}\, A_{\0\ell mn}
\, c_{pqr}\,  \epsilon^{jk\ell mnpqr} -\ft12\mu \lambda_i\ ,\\
\delta \vec \phi &=& \ft16 \mu \,\vec g\ .\nn
\eea
These transformations have the following non-trivial commutation
relations
\bea
{[}\delta_{c}, \delta_{c'}{]} &=& \delta_\kappa\ ,\qquad\qquad
\kappa^{ij} = -\ft1{36}  c_{k\ell m}\, {c'}_{npq}\,\epsilon^{ijk\ell mnpq}
\ ,\nn\\
{[}\delta_{\kappa}, \delta_{c}{]} &=& \delta_{\epsilon}\ ,
\qquad \qquad \epsilon_i = \ft12 c_{ijk}\, \kappa^{jk}\ ,
\label{d3com}\\
{[}\delta_{\mu}, \delta_{\epsilon} {]} &=& \delta_{\epsilon'}\ ,
\qquad\qquad \epsilon_i'=\ft12\mu\, \epsilon_i\ ,\nn\\
{[}\delta_{\mu}, \delta_\kappa {]} &=& \delta_{\kappa'}\ ,\qquad\qquad
{\kappa'}^{ij} = \ft13 \mu\,  \kappa^{ij}\ ,\nn\\
{[} \delta_c, \delta_{\sst \Lambda} {]} &=& \delta_{\td c}\ ,\qquad
\td c_{ijk} = 3 \Lambda_{[i}{}^\ell\, c_{jk]\ell}\ ,\nn\\
{[} \delta_\mu, \delta_c {]} &=& \delta_{c'}\ ,\qquad
c'_{ijk} = \ft16 \mu\, c_{ijk}\ ,\nn
\eea
together with
the standard $SL(8,\R)$ transformations, since $A_{\0ijk}$,
$\lambda^{ij}$ and $\lambda_{i}$ all transform covariantly under 
$SL(8,\R)$.  We see that the transformations described by the parameters
$c_{ijk}$ no longer all commute, and the maximal set of commuting $\R$ 
symmetries is associated with the parameters
\be
\kappa^{ij}\ ,\qquad \epsilon_{i}\ ,
\ee
corresponding to the commuting shift symmetries of all the $36=28+8$
new axionic scalars coming from the dualisation.  Note that this
$\R^{36}$ symmetry is precisely the maximal abelian subalgebra of
$E_8$, which is unique up to conjugation. Once more the normalisations
have been chosen to include appropriate symmetry factors that
correspond to structure constants $\pm1$.

\section{Coset structure of scalar Lagrangians and their symmetries}

     In this section, we examine the complete scalar Lagrangians for
all maximal supergravities in $D\le 11$ dimensions.  These scalars
include the $(11-D)$ dilatons $\vec\phi=(\phi_1,\phi_2,\ldots,
\phi_{11-D})$, the $\ft12 (11-D)(10-D)$ axions ${\cal A}_\0^i{}_j$ and
the $\ft16 (11-D)(10-D)(9-D)$ axions $A_{\0ijk}$.  In addition, in
dimensions $D\le 5$ there is an option, as we saw in the previous
section, to include further axions obtained as the potentials for the
duals of $(D-1)$-form field strengths.

     The construction and the analysis of the symmetries of the scalar
Lagrangians turns out to be very simple in the field variables that we
are using here, which arise from the step-by-step Kaluza-Klein
reduction from $D=11$.  Let us first consider the dimensions $D\ge 6$,
where there are no complications coming from the possibility of
further axions arising from the dualisation of higher-degree field
strengths.  The key observation, which is shown by elementary
computation using the definitions given in Appendix A, is that in each
dimension $D\ge 6$ the set of dilaton vectors $\vec b_{ij}$ and $\vec
a_{ijk}$ for the corresponding axions ${\cal A}_\0^i{}_j$ and
$A_{\0ijk}$ are in one-to-one correspondence with the positive-root
vectors of the $E_{11-D}$ algebra.  In fact it is easy to see from
(\ref{dilatonvec}) and (\ref{fsum}) that one can take $\vec b_{i,i+1}$
and $\vec a_{123}$ (for $D\le 8$) to be the simple roots, with all
others generated as sums of these with non-negative integer
coefficients.  Furthermore, in the dimensions $D=5$, $D=4$ or $D=3$,
one can also easily see from (\ref{dilatonvec}) and (\ref{fsum}) that
if the dilaton vectors $-\vec a$, $-\vec a_i$, or $(-\vec b_i, -\vec
a_{ij})$ respectively, corresponding to the axions coming from the
dualisations discussed in the previous section, are included then the
entire sets of dilaton vectors for all the axions coincide with the
positive roots of $E_6$, $E_7$ and $E_8$ respectively.  Again, the
simple roots can be taken to be $\vec b_{i,i+1}$ and $\vec
a_{123}$. The former are the simple roots of $SL(11-D,\R)$, and the
latter, which arises after the appropriate Weyl rescaling, is the seed
of $E_{11-D}$. Thus in all dimensions we may summarise the information
about the dot products of the dilaton vectors for the full sets of
axions by the Dynkin diagram:
\bigskip\bigskip

\centerline{
\begin{tabular}{ccccccccccccc}\\
 $\vec b_{12}$& &$\vec b_{23}$& &$\vec b_{34}$& &$\vec b_{45}$
& &$\vec b_{56}$& &$\vec b_{67}$& &$\vec b_{78}$ \\
 o&---&o&---&o&---&o&---&o&---&o&---&o\\
 &   & &   &$|$&   & &   & &   & &   & \\
 &   & &   &o&   & &   & &   & &   & \\
 &   & &   &$\vec a_{123}$& & &   & &   & &   & \\
\end{tabular}}
\bigskip\bigskip

\centerline{Diagram 1: The dilaton vectors $\vec b_{i,i+1}$ with $i\le n-1$
and $\vec a_{123}$}
\centerline{ generate the $E_n$ Dynkin diagram\phantom{xxxxxxxxxx}}
\bigskip\bigskip
In each dimension $D$, the diagram is truncated to the part that
survives when only the simple roots with indices $i\le 11-D$ are
retained.  This defines the simple root that has to to be removed in
order to disintegrate $E_{11-D}$ to $E_{10-D}$. Note that the
undualised versions of supergravities discussed in section 2 have
symmetry $SL(11-D,\R)$, corresponding to removing the root $\vec
a_{123}$.  In section 7, we shall discuss the case of R-R dualisation,
where all R-R fields are dualised when this results in fields of lower
degrees.  In this case the symmetry group contains $O(10-D,10-D)$ as a
subgroup, corresponding to removing the simple root $\vec b_{12}$.  In
section 9, we shall discuss the direct dimensional reduction of the
type IIB theory, where no dualisations are performed.  These cases
correspond to removing both $\vec b_{23}$ and $\vec a_{123}$, and
hence they contain $SL(2,R)\times SL(9-D,R)$ as subgroups.  For future
convenience, we shall denote the simple roots $(\vec a_{123}, \vec
b_{12}, \vec b_{23}, \ldots, \vec b_{78})$ by $\vec r_i$ with $i=(1,2
\ldots, 8)$.

       In the next two subsections, we shall prove that the scalar
Lagrangians of toroidally-compactified eleven-dimensional
supergravities have $E_{11-D}$ global symmetries, after all the
$(D-2)$-form potentials are dualised to give rise to the maximal
number of scalar fields, and after the canonical Weyl rescaling.  Our
strategy will be to give $K(E_{11-D})\backslash E_{11-D}$ coset
constructions of scalar Lagrangians, and then to show that these
coincide precisely with the scalar Lagrangians of the fully-dualised
supergravities that we obtained in section 3.  The occurrence of the
global $E_{11-D}$ symmetries was conjectured in \cite{cj3,cj2}, based
on the structure of the scalar fields in each dimension.  In fact,
maximal supergravities with these symmetries have been obtained in all
dimensions, but in general by direct construction, rather than by
dimensional reduction from $D=11$ (see, for example,
\cite{book}). However, a complete proof that they can be obtained from
eleven-dimensional supergravity has been given only for the cases of
$D=9$ \cite{bho}, $D=4$ \cite{cj3} and $D=3$ \cite{jul,miz}.

\subsection{Scalar manifolds in $D\ge 6$}

     First let us consider the dimensions $D\ge6$, where we just have
the axions ${\cal A}_\0^i{}_j$ and $A_{\0ijk}$ associated with the
dilaton vectors $\vec b_{ij}$ and $\vec a_{ijk}$ respectively.  Since
these are given by (\ref{dilatonvec})
\be
\vec b_{ij} = -\vec f_i + \vec f_j\ ;\qquad \vec a_{ijk}= \vec f_i+\vec f_j 
+\vec f_k -\vec g\ ,\label{abdef}
\ee
it follows immediately that
\be
\vec b_{ij} +\vec b_{jk} = \vec b_{ik}\ ,\qquad
\vec a_{ijk} + \vec b_{i\ell} = \vec a_{\ell jk} \ .\label{abcom}
\ee
Defining the generators associated with the positive roots $\vec
b_{ij}$ and $\vec a_{ijk}$ as 
$E_i{}^j$ and $E^{ijk}$ respectively, we see from (\ref{abcom}) that they
will obey commutation relations of the form 
\bea
{[}E_i{}^j, E_k{}^\ell{]} &=& \delta^j_k\, E_i{}^\ell -
\delta_i^\ell\, E_k{}^j\ ,\label{eegl}\\
{[}  E_\ell{}^m, E^{ijk}, {]}&=&- 3\delta^{[i}_\ell\, E^{|m|jk]} \ 
,\label{ee}\\
{[} E^{ijk},E^{\ell mn} {]} &=& 0\ ,\label{eeaxion}
\eea
where it is understood that $E_i{}^j$ is defined only for $i<j$, while
$E^{ijk}$ is totally antisymmetric and is defined for any ordering of
its indices.  Note that the commutators (\ref{eegl}) and (\ref{ee})
arise in all dimensions: (\ref{eegl}) describes the positive-root
(nilpotent) subalgebra of $SL(11-D,\R)$, while (\ref{ee}) reflects the
fact that $E^{ijk}$ transforms covariantly under $SL(11-D,\R)$.  The
third commutator (\ref{eeaxion}) is dimension dependent.  For $D\ge
6$, which we are currently considering, the generators $E^{ijk}$
commute. For $D\le 5$, whether they commute or not depends on whether
we perform dualisations or not.  We shall discuss this later, case by
case.  Note also that if we include the $\R_s$ factor of $GL(11-D,\R)$
and write the set of Cartan generators as a vector $\vec H$, then we
will have
\be
{[}\vec H, E_i{}^j{]} = \vec b_{ij} \, E_i{}^j \ ,
\qquad
{[}\vec H, E^{ijk}{]} = \vec a_{ijk}\, E^{ijk}\qquad {\rm no\ sum} \ .
\label{cartond6}
\ee
In fact having defined the positive-root subalgebra, the summation
rules (\ref{abcom}) for the dilaton vectors come as no surprise, since
they are then a direct consequence of the Jacobi identities of this
subalgebra.

     Now define $\v=\v_1\, \v_2\, \v_3$, with
\bea
\v_1 &=& e^{\ft12 \vec\phi\cdot \vec H}\ ,\nn\\
\v_2 &=& \prod_{i<j} U_{ij} = \cdots U_{24}\, U_{23} \cdots U_{14}\,
U_{13}\, U_{12}\ ,\label{decomp}\\
\v_3 &=& e^{\sum_{i<j<k} A_{\0ijk}\, E^{ijk}} \ ,\nn
\eea
where 
\be
U_{ij} \equiv 
e^{{\cal A}_\0^i{}_j\, E_i{}^j} \qquad ({\rm no\ sum})\ ,
\ee
and the right-hand-side of $\v_2$ is defined with anti-lexical ordering, 
as indicated in (\ref{decomp}).  It follows that 
\be
d\v\, \v^{-1}= d\v_1\, \v_1^{-1} + \v_1\, (d\v_2\, \v_2^{-1})\, \v_1^{-1}  + 
\v_1\, \v_2\, (d\v_3\, \v_3^{-1})\, \v_2^{-1}\, \v_1^{-1}\ .\label{gmg}
\ee
One can now see from the commutation relations above that
\bea
d\v_1\, \v_1^{-1}&=& \ft12 d\vec\phi\cdot \vec H\, \nn\\
d\v_2\, \v_2^{-1}&=& \sum_{i<j} {\cal F}_\1^i{}_j\, E_i{}^j
\ ,\label{gdef}\\
d\v_3\, \v_3^{-1}&=& \sum_{i<j<k} dA_{\0ijk}\, E^{ijk}\   ,\nn
\eea 
where ${\cal F}_\1^i{}_j$ are the fully Kaluza-Klein modified field
strengths given by 
\be
{\cal F}_\1^i{}_j= \gamma^k{}_j\, d{\cal A}_\0^i{}_k\ ,
\ee
and $\gamma^i{}_j$ is given in (\ref{gam}).  (We have made use of the
relation $de^X\, e^{-X}=dX + \ft12 [X,dX] +\ft16 [X,[X,dX]]+\cdots$,
where, because of the nature of the parameterisation of $\v$, we need
only the first term in this series.  Also, we need $e^{X}\, Y\, e^{-X}
= Y + [X,Y] +\ft12 [X,[X,Y]]+ \cdots$.)  Note that in terms of the
fundamental representation of $E_i{}^j$, we have $(\v_2)^i{}_j =
\td\gamma^i{}_j$ and $(\v_2^{-1})^i{}_j = \gamma^i{}_j$.  The
anti-lexical ordering in $\v_2$ is crucial to the above derivation.
This in particular implies that $(\v_2)^i{}_j$ becomes linear in
${\cal A}_\0^i{}_j$ and makes the proof of (\ref{gdef})
straightforward.
 
     Conjugating the expression for $d\v_2\, \v_2^{-1}$ with $\v_1$,
we find
\be
\v_1\, (d\v_2\, \v_2^{-1})\, \v_1^{-1}= \sum_{i<j} 
e^{\ft12 \vec b_{ij}\cdot \vec\phi}\,  {\cal F}_\1^i{}_j\, E_i{}^j\ .
\ee
Also, we can see that
\be
\v_2\, (d\v_3\, \v_3^{-1}) \, \v_2^{-1} =\sum_{i<j<k} F_{\1ijk}\, E^{ijk}\ ,
\label{f123}
\ee
where $F_{\1ijk}$ are the fully modified field strengths
given by
\be
F_{\1ijk} = \gamma^\ell{}_i\, \gamma^m{}_j \,\gamma^n{}_k\, dA_{\0\ell mn}\ .
\ee
After conjugating (\ref{f123}) with $\v_1$, and adding together all the
terms in (\ref{gmg}), we obtain the result
\be
d\v\, \v^{-1} = \ft12 d\vec\phi\cdot\vec H +\sum_{i<j}e^{\ft12 \vec
b_{ij}\cdot \vec\phi}\,  {\cal F}_\1^i{}_j\, E_i{}^j
+\sum_{i<j<k} e^{\ft12
\vec  a_{ijk}\cdot \vec \phi}\, F_{\1ijk}\, E^{ijk}\ .\label{d6scal}
\ee
It follows from this that the entire scalar 
Lagrangian in $D\ge 6$ is expressible as 
\bea
{\cal L} &=& \ft14 e\,{\rm tr}\Big(\del {\cal M}^{-1}\, \del {\cal M})\Big)\ ,
\label{dmlag}\\
&=& -\ft12 e\, {\rm tr}\Big(\del \v \,\v^{-1}\,( \del \v\, 
\v^{-1})^{\rm T} \Big)
   -\ft12 e\, {\rm tr}\Big(\del \v \,\v^{-1}\, (\del \v\, \v^{-1}) \Big)\ .
\label{dglag}
\eea
where we have defined the internal metric 
\be {\cal M}:= \v^{\rm T}\, \v\ ,\label{calm}
\ee
and the superscript ``T'' denotes the transpose.
The first two  normalisations below are imposed by (\ref{dmlag}): 
\be
{\rm tr}(H_i\, H_j)=2\delta_{ij}\ ,\qquad {\rm tr}(E_i{}^j\, 
E^T_k{}^\ell) = \delta_{ik}\, \delta^{j\ell}\ ,
\qquad {\rm tr}(E^{ijk}\, E^{T\ell mn})=6\delta_{[\ell}^{i}\,
\delta_{m}^{j}\, \delta_{n]}^{k}\ ,\label{normK1}
\ee
where in the second expression it is understood that $i<j$ and
$k<\ell$. The last normalisation is at our disposal provided we
respect the $SL(11-D,\R)$ covariance but the relative normalisation of
the first two terms (which form the Casimir invariant in the defining
representation of $SL(11-D,\R)$) is dictated by the obvious invariance
under that group. The present choice of normalisation is canonical for
simply laced groups and leads to the most symmetric expressions. Also,
we have
\be
{\rm tr}(E_i{}^j\, E_k{}^\ell)=0\, \qquad 
{\rm tr}(E^{ijk}\, E^{\ell mn})=0\ ,
\ee
again for $i<j$, $k<\ell$.  Comparison shows that the Lagrangian
obtained by substituting (\ref{d6scal}) into (\ref{dglag}) is
identical to the scalar sector of the $D$-dimensional supergravity
Lagrangian given in (\ref{dgenlag}), when $D\ge6$. Using the manifest
$SL(11-D,\R)$ invariance one only has to adjust the coefficient of the
exponent in $\v_3$ to establish the invariance. The commutativity of
the extra generators (\ref{eeaxion}) allows for this arbitrary
rescaling. Of course one still makes use of the presence of the
antisymmetric third order tensor representation and the special Weyl
rescaling that modifies the $\R_s$ subgroup of $GL(11-D,\R)$ to
exhibit the $E_{11-D}$ symmetry in the scalar sector.

    Having written the scalar Lagrangian in the form (\ref{dglag}),
its global symmetry is now made manifest.  If $U$ is a constant matrix
in the global symmetry group, we can send
\be
\v\longrightarrow \v'' = \v\, U\ ,
\ee
which leaves $d\v\, \v^{-1}$ invariant.  This takes us out of the
``positive-root'' gauge, in that $\v''$ is no longer expressible in
the form (\ref{decomp}). We now define $\v'$ by
\be
 \v' = {\cal O}\, \v\, U\ ,
\ee
where $\v'$ {\it is} in the positive-root gauge (\ref{decomp}), and 
${\cal O}$ is
some appropriate compensating element of the maximal compact subgroup
(the denominator group), for which 
${\cal O}^{\rm T}\, {\cal O}=1$.  Then under the compensated 
transformation, we
have
\be
{\cal M}\longrightarrow {\cal M}'= U^{\rm T}\, \v^{\rm T}\, {\cal O}^{\rm T}\, 
{\cal O}\, \v\, 
U  =U^{\rm T}\, \v^{\rm T}\, \v \, U  = U^{\rm T}\, {\cal M}\, U\ ,
\ee
which is easily seen to leave (\ref{dmlag}) invariant.  Since in
$D\ge6$ the dilaton vectors have been established to correspond to the
positive roots of the $E_{11-D}$ algebra, it follows that this is the
global symmetry group of the scalar Lagrangian.  We may restore the
$K(E_{11-D})$ local gauge invariance to streamline some formulas
because the Lagrangian is built out of $K(E_{11-D})$ invariants.
 
       So far, we have obtained cosets $K(E_{11-D})\backslash
E_{11-D}$ for $D\ge 6$.  We constructed them using the Borel subgroups
of $E_{11-D}$, which are generated by the positive-root and Cartan
generators.  The scalars are the parameters for these generators, with
the axions associated with positive-root generators and dilatons
associated with the Cartan generators. (The Iwasawa decomposition and
the Borel subgroups of the global supergravity symmetry groups
$E_{11-D}$ were extensively studied in \cite{solv1,solv2,solv3}.)  We
showed that these scalar cosets are precisely the same as the scalar
manifolds obtained from eleven-dimensional supergravity by dimensional
reduction to $D$ dimensions.  The cosets have manifest $SL(11-D)$
symmetries, since all the generators carry $SL(11-D)$ indices.  In
Appendix C, we give an explicit construction of these scalar cosets
with manifest $E_{11-D}$ invariance, for $D\ge6$.

        Before we go to the next subsection to study the coset
structure of the scalar manifolds for $D\le 5$, there are some group
theoretical points that need to be addressed.  The structure of the
matrix ${\cal M}$ defined in (\ref{calm}) is correct for $D\ge 6$,
owing to the nature of the $E_{11-D}$ groups in these cases, but it is
not as it stands precisely applicable in lower dimensions.  For a
generic non-compact Lie group $G$, one has ${\cal M} = \tau(\v^{-1})\,
\v$, where $\tau$ is the Cartan involution whose fixed point set is
the maximal compact subgroup of $G$.  For the groups $SL(N,\R)$ or
$SO(N,N)$, where the maximal compact subgroups are $SO(N)$ and
$SO(N)\times SO(N)$ respectively, we simply have $\tau(\v^{-1}) =
\v^{\rm T}$.  That is why we could use ordinary transpose ``T'' to
discuss the coset structure for $D\ge 6$.  For $E_6$ in $D=5$, we have
in the fundamental representation: ${\cal M}= (\Omega \otimes
\Omega)\, \v^\dagger\, \v= \v^T (\Omega \otimes \Omega)\, \v $, where
following the second reference of \cite{cj2} $\Omega$ is the constant
invariant symplectic matrix in $U\!Sp(8)$, which is the maximal
compact subgroup of $E_6$.  Another example is $E_7$, with $SU(8)$ as
its maximal compact subgroup.  In this case we have $\tau(\v^{-1}) =
\v^\dagger$.  For convenience, we shall introduce a ``generalised
transpose'' $\#$, defined as $\v^\# = \tau (\v^{-1})$.  Then the
formulae obtained in this section become applicable to general
dimensions simply by replacing the transpose ``T'' by the generalised
transpose ``$\#$'' everywhere, and in particular the scalar Lagrangian
(\ref{dglag}) becomes
\be
{\cal L}= -\ft12 e\, {\rm tr}\, \Big(\del \v \,\v^{-1}\, 
(\del \v\, \v^{-1})^{\#} \Big)
   -\ft12 e\, {\rm tr}\, \Big(\del \v \,\v^{-1}\, 
(\del \v\, \v^{-1}) \Big)\ .\label{gtlag}
\ee
Acting on the generators $E_i{}^j$ and $E^{ijk}$ corresponding to the
positive roots, the Cartan involution has the effect of turning them
into minus their conjugate negative-root generators, \ $E^\#_i{}^j$
and $E^{\# ijk}$, and changes the sign of the Cartan generators.  (In
other words, $\tau$ sends $(E^+,E^-,\vec H) \rightarrow (-E^-, -E^+,
-\vec H)$, where $E^+$ and $E^-$ denote the sets of positive and
negative root generators.  This implies that all the compact
generators, $(E^+-E^-)$, are left unchanged, while all the non-compact
generators, $(E^++E^-)$ and $\vec H$, are reversed in sign.)  Note
that in this positive-root system, $E_i{}^j$, and hence also their
conjugate generators $E^\#_i{}^j$, are defined only for $i<j$.

It turns out that the maximal duality symmetries of supergravities are
real Lie groups in split form (maximally noncompact real form with
Cartan subalgebra that may be chosen along the noncompact directions).
For all these real forms the Cartan involutions follow exactly the
pattern we have just discussed.

\subsection{Maximal scalar manifolds in $D=5,4,3$}

        The above discussion was for the cases $D\ge6$, where there
are no extra axions coming from dualisations of higher-rank field
strengths.  In $D\le 5$, there can be additional scalars coming from
the dualisations of $(D-2)$-form potentials.  In this subsection, we
shall consider the cases where the maximal numbers of such
dualisations are performed, leading to scalar manifolds with global
$E_6$, $E_7$ and $E_8$ symmetries in $D=5,4$ and 3.  The essential
point is that again all the axions, including those obtained by
dualisation, have dilaton vectors corresponding to the positive roots
of the $E_{11-D}$ algebra, and all the dilatonic scalars are
associated with the Cartan generators.  In fact the manifestly
$SL(11-D,\R)$-covariant expressions that initially arise in the
theories are nothing but the decompositions of $E_{11-D}$ covariant
expressions with respect to the $SL(11-D,\R)$ subgroup.  Thus we again
have a very simple way of expressing the coset representatives in a
straightforward generalisation of the previous parameterisation
$\v=\v_1\, \v_2\, \v_3$ used above, leading eventually to an
expression for the scalar Lagrangian in the form (\ref{dglag}),
implying that it has a manifest $E_{11-D}$ global symmetry.

     First, consider $D=5$.  We have, in addition to the commutation
relations implied by (\ref{abcom}) that 
\be
\vec a_{ijk} + \vec a_{\ell m n} = -\vec a\ ,\label{d5vecsum}
\ee
when $ijk\ell mn$ are all different.  As we saw in the previous section, 
$-\vec a$ is indeed the dilaton vector associated with the axion $\chi$
obtained from the dualisation of $F_\4$. Introducing a new generator $D$,
associated with this extra axion, we see that it will satisfy the commutation
relations
\be
[E^{ijk},E^{\ell mn}] = -\epsilon^{ijk\ell mn}\, D\ , \qquad[\vec H,D]=-\vec
a\, D\ ,\qquad [E_i{}^j,D]=0\ , \qquad [E^{ijk},D]=0\ ,\label{d5com1}
\ee
while the remaining commutators are unchanged from their previous
form.  The normalisation of the extra generator $D$ is chosen to be
canonical, as in the case of the other positive-root generators,
namely ${\rm tr}(D^\# \, D)=1$.  The coefficient on the right-hand
side of the first commutator in (\ref{d5com1}) is the one that arises
in $E_6$ with the same canonical normalisations of the generators.
This can easily be seen by using the Weyl group of $E_6$ to relate
this commutator to a commutator of generators with already-known
normalisation.  For example, we may begin from the commutator for
$[E_1{}^2,E_2{}^3]= E_1{}^3$, which involves only the generators of
$SL(6,\R)$, and is given in (\ref{eegl}).  In terms of the associated
dilaton vectors, this corresponds to the sum rule $\vec b_{12} + \vec
b_{23} =\vec b_{13}$.  Since our positive roots all have length 2, the
Weyl reflection of a weight $\vec\gamma$ in the root $\vec\a$ is given
by $\vec\gamma\rightarrow \vec\gamma - (\vec\a\cdot\vec\gamma) \vec
\a$.  Elementary computations using the results in Appendix A show
that under a Weyl reflection in the root $\vec a_{145}$, the previous
dilaton sum is mapped to $\vec a_{245} + \vec b_{23} = \vec a_{345}$,
and that a second Weyl reflection in the root $\vec a_{126}$ then maps
this into $\vec a_{245} +\vec a_{136}= -\vec a$.  Since the
positive-root generators $X_a$ have all been canonically normalised so
that ${\rm tr}(X^\#_a\, X_b) =\delta_{ab}$, which is invariant under
Weyl reflections, this shows that the commutator $[E_1{}^2,E_2{}^3]=
E_1{}^3$, after the two Weyl reflections, becomes $[E^{245}, E^{136}]
=\pm D$, in agreement with the scale factor in the first commutator in
(\ref{d5com1}).\footnote{This argument fixes the magnitudes, but not
the signs, of the right-hand sides of the commutators.  The signs are
in general determinable by more subtle arguments \cite{tits}.  For our
present purposes, it suffices to note that any choice of signs that is
consistent with the $SL(11-D,\R)$ Jacobi identities represents a valid
reduction of $E_{11-D}$ to $SL(11-D,\R)$, with the different possible
such choices being related by trivial redefinitions of generators.}
Note that the result of the first Weyl reflection provides a
consistency check on the normalisation of the commutator (\ref{ee}),
since it implies that $[E^{245}, E_2{}^3] =\pm E^{345}$.  Note also
that in fact the form of the first commutation relation in
(\ref{d5com1}) can be seen already in (\ref{d5ccom}); it is dictated
by $SL(6,\R)$ covariance, and the summation rule (\ref{d5vecsum}) is
the direct consequence of the Jacobi identity for $\vec H$, $E^{ijk}$
and $E^{\ell mn}$. The internal dilation covariance dictates the
dimensions of the generators for the $\R$ factor of $GL(6,\R)$.

     We now extend the parameterisation of the previous
$D\ge6$ cosets, by introducing an extra factor $\v_4$, so that 
$\v=\v_1\, \v_2\, \v_3\, \v_4$, where
\be
\v_4 = e^{\chi\, D}\ ,
\ee
and $\v_1$, $\v_2$ and $\v_3$ are given by (\ref{decomp}) as before.
There will then be the following changes when we calculate $d\v\,
\v^{-1}$.  Firstly, we pick up an extra term $d\chi\, D$, from
$d\v_4\, \v_4^{-1}$.  In addition, the computation of $d\v_3\,
\v_3^{-1}$ will be modified, because of the non-zero commutators of
the form $[E^{ijk}, E^{\ell mn}]$.  This will cause non-linear
Kaluza-Klein modifications to $d\chi$ to be generated, producing
precisely the field strength $G_\1$ obtained in (\ref{d5g}) from the
dualisation of $F_\4$.  After conjugating this with $\v_1$, and
including the other contributions, we find that
\be
d\v\, \v^{-1}= \ft12 d\vec\phi\cdot\vec H +\sum_{i<j} e^{\ft12\vec b_{ij} 
\cdot\vec\phi}\, {\cal F}_\1^i{}_j\, E_i{}^j + \sum_{i<j<k} e^{\ft12\vec
a_{ijk}\cdot\vec \phi}\, F_{\1ijk}\, E^{ijk} 
+e^{-\ft12\vec a\cdot\vec \phi}\, G_\1   \, D\ .\label{d5dg}
\ee
Substituting into (\ref{gtlag}), we get precisely the $D=5$ scalar
Lagrangian obtained in the previous section by dualising the 3-form
potential $A_\3$ to give an additional axionic scalar.  It is now
manifest from (\ref{d5dg}) that it will have an $E_6$ global symmetry.
A crucial feature here is that the normalisation of the first
commmutation relation in (\ref{d5com1}), which is dictated now by the
nonabelian structure of the $E_6$ algebra, is identical to the
normalisation that was needed above for the coset construction of the
$D=5$ Lagrangian.

     Now consider $D=4$.  In this case, we have 3-form field strengths
$F_{\3i}$ that are dualised to 1-forms, and hence give the further
seven axions $\chi^i$, $i=1,\ldots, 7$ that we discussed in section
3.2.  The associated change in the commutation relations is reflected
in the relation
\be
\vec a_{ijk} +\vec a_{\ell mn} = -\vec a_p\ ,\label{d4vecsum}
\ee
where $ijk\ell mnp$ are all different.  This follows from
(\ref{abdef}), together with the fact that the dilaton vector for
$F_{\3i}$ is $\vec a_i = \vec f_i - \vec g$, and that the dualisation
reverses the sign of the dilaton vector, as we saw previously.
Introducing generators $D_i$ associated with the extra axions
$\chi^{i}$, we see that they will obey the commutation relations
\bea
{[}E^{ijk}, E^{\ell mn}{]} &=& \epsilon^{ijk\ell mnp}\, D_p\ , \nn\\
{[}E_j{}^k, D_i{]} &=& \delta_i{}^k\, D_j\ , \qquad
\qquad [D_i,E^{jk\ell}]=0\ ,\label{d4com17}\\ 
{[}\vec H, D_i {]} &=& -\vec a_i\, D_i\ ,\qquad\qquad {\rm no\ sum}\ .
\nn
\eea
Note that the non-trivial commutator in $D=4$, namely the one for
$E^{ijk}$ with $E^{\ell mn}$, corresponds to (\ref{d4com}), and that
(\ref{d4vecsum}) can be derived from the Jacobi identity for $\vec H$,
$E^{ijk}$ and $E^{\ell mn}$. As usual, we choose the canonical
normalisation for the extra generators, so that ${\rm tr}(D^\#_i\,
D_j)=\delta_{ij}$.  The normalisation of the first commutator in
(\ref{d4com17}) is the one dictated by decomposing $E_7$ commutation
relations under $SL(7,\R)$, and can be established by the same
technique that we used in $D=5$, namely by making $E_7$ Weyl
reflections to relate it to an already-known commutator.  In fact
successive Weyl reflections in the roots $\vec a_{145}$ and $\vec
a_{126}$ map the commutator $[E_1{}^2,E_2{}^3]= E_1{}^3$ into
$[E^{245}, E^{136} ]= \pm D_7$.

    We now write the coset representative as $\v_1\, \v_2\, \v_3\,
\v_4$, where $\v_1$, $\v_2$ and $\v_3$ are given by (\ref{decomp}) as
before, and here
\be
\v_4= e^{\sum_i\chi^{i}\, D_i}\ .
\ee
 From this, and the commutation relations, we see that
\bea
d\v\, \v^{-1}&=& d\v_1\, \v_1^{-1}+ \v_1\, d\v_2\, \v_2^{-1}\, \v_1 +
\v_1\, \v_2\, d\v_3\, \v_3^{-1}\, \v_2^{-1}\, \v_1^{-1} +
\v_1\, \v_2\, d\v_4\, \v_4^{-1}\, \v_2^{-1}\, \v_1^{-1} \nn\\
&=& \ft12 d\vec \phi\cdot\vec H+\sum_{i<j<k}e^{\ft12\vec a_{ijk}
          \cdot\vec\phi}\, F_{\1ijk}\, E^{ijk} \nn\\
&&+\sum_{i<j} e^{\ft12\vec b_{ij}\cdot\vec\phi}\, {\cal F}_\1^i{}_j \,
E_i{}^j +\sum_i e^{-\ft12 \vec a_i\cdot \vec\phi}\, G_\1^{i}\, D_i\ ,
\eea
where $G_\1^{i}$ are the 1-forms dual to $F_{\3i}$, defined in
(\ref{d4g}).  Substituting this into (\ref{gtlag}), we obtain
precisely the $D=4$ scalar Lagrangian of section 3.2, with its $E_7$
global symmetry made manifest.  Again, it should be emphasised that
the normalisations of the commutators in (\ref{d4com17}), which are
dictated by the structure of the $E_7$ algebra, are exactly such as to
give the correct expressions for the kinetic terms for the new axions
arising from dualisation.  The $E_7$ symmetry of the maximal
4-dimensional supergravity coming from compactification of
eleven-dimensional supergravity with full dualisation was obtained in
\cite{cj3}.

     Finally, we turn to the coset construction for the scalar
manifold of 3-dimensional maximal supergravity.  In this case, the
dualisation of the 2-form field strengths ${\cal F}_\2^{i}$ and
$F_{\2ij}$ gives rise to $8+28=36$ additional axions $\lambda_i$ and
$\lambda^{ij}$, with the associated dilaton vectors $-\vec b_i$ and
$-\vec a_{ij}$, as we showed in the previous section.  From
(\ref{dilatonvec}), one easily sees that the following relations hold,
\be
\vec a_{ijk} + \vec a_{\ell mn} = -\vec a_{pq}\ ,\qquad \vec a_{ijk} +
(-\vec a_{jk}) = -\vec b_i\ ,\label{d3vecsum}
\ee
where in the first equation, $ijk\ell mnpq$ are all different.  These
relations amongst the positive roots of $E_8$ result from the
following commutation relations for the extra generators $D^i$ and
$D_{ij}$ associated with $\vec b_i$ and $\vec a_{ij}$:
\bea
{[} E^{ijk}, E^{\ell mn}{]} &=& - \sum_{p<q} \epsilon^{ijk\ell mnpq}
\, D_{pq}\ ,\nn\\ 
{[} E^{ijk}, D_{\ell m} {]} &=&
-6\delta^{[i}{}_{[\ell}\delta^{j}{}_{m]}
\, D^{k]}\ ,\qquad [E^{ijk}, D^\ell] =0\ ,\label{e3}\\
{[} E_i{}^j, D_{k\ell}{]} &=& 2\delta^j_{[k}\, D_{|i|\ell]} 
\ ,\qquad\qquad\quad [ E_i{}^j,D^k] = -\delta_i^k\, D^j\ ,\nn\\
{[} \vec H, D^i{]} &=& -\vec b_i\, D^i\ ,\qquad\qquad\quad\quad 
[\vec H, D_{ij} ] = -\vec a_{ij}\, D_{ij}\ .\qquad
 \nn
\eea
The first three commutators, characteristic of the $D=3$ case, were
encountered in section 3.3, and from these we can use the Jacobi
identities to derive the dilaton-vector summation rules
(\ref{d3vecsum}). As in the previous cases, we are taking the extra
generators to be canonically normalised, namely ${\rm tr}(D^{\#i}\,
D^j)=\delta^{ij}$ and ${\rm tr} (D^\#_{ij}\, D_{k\ell}) = 2
\delta^{i}_{[k}\, \delta^j_{\ell ]}$. The normalisations of the
commutators that produce $D^i$ and $D_{ij}$ in (\ref{e3}) are those
dictated by the $E_8$ algebra, and can be established, as in the
previous cases, by relating them to already-known commutators by means
of Weyl reflections.  For example, starting again from
$[E_1{}^2,E_2{}^3]= E_1{}^3$, and applying successive Weyl reflections
in the roots $\vec a_{245}$ and then $\vec a_{126}$ gives $[E^{245},
E^{136}] =\pm D_{78}$.  A third Weyl reflection in the root $\vec
a_{278}$ then gives $[E^{245}, D_{45}] = \pm D^2$.

The parameterisation of  the coset in this case is taken to be
\be
g = \v_1\, \v_2\, \v_3\, \v_4\, \v_5 ,
\ee
where $\v_1$, $\v_2$ and $\v_3$ are given by (\ref{decomp}) as usual, and 
\be
\v_4= e^{\sum_i
\lambda_{i}\, D^i}\qquad \v_5 = e^{\sum_{i<j}\lambda^{ij}\, D_{ij}} \ .
\ee
A mechanical calculation gives the result that
\bea
d\v\, \v^{-1}&=& \ft12 d\vec\phi\cdot\vec H +\sum_{i<j} e^{\ft12 \vec
b_{ij}\cdot\vec\phi}\, {\cal F}_\1^i{}_j\, E_i{}^j +\sum_{i<j<k} e^{\ft12
\vec a_{ijk}\cdot\vec\phi}\, F_{\1ijk}\, E^{ijk}\nn \\
&& +\sum_i
e^{-\ft12\vec b_i\cdot
\vec\phi}\, G_{\1i}\, D^i +\sum_{i<j} e^{-\ft12\vec a_{ij}\cdot
\vec\phi}\, G_\1^{ij}\, D_{ij}\ ,\label{d3dg}
\eea
where in addition to the usual field strengths ${\cal F}_\1^{i}$ and
$F_{\1ijk}$ for the axions ${\cal A}_\0^i{}_j$ and $A_{\0ijk}$, the
field strengths $G_{\1i}$ and $G_\1^{ij}$ for the additional axions
$\lambda_i$ and $\lambda^{ij}$ are precisely the ones given in
(\ref{g1eq}) and (\ref{gg1eq}).  Substituting (\ref{d3dg}) into
(\ref{gtlag}) gives a manifestly $E_8$ invariant formulation for the
$D=3$ scalar Lagrangian obtained in section 2.1.  Again, the
commutator structures that are dictated by $E_8$ covariance are
precisely the ones needed to reproduce the kinetic terms of the
additional axions arising from dualisation.  Note that in $D=3$, this
scalar Lagrangian describes the entire bosonic sector of the theory.
The $E_8$ symmetry of the bosonic sector of $D=3$ maximal
supergravity, obtained from dimensional reduction of $D=4$ maximal
supergravity, was proved in \cite{jul}.  A rather different proof of
the $E_8$ symmetry in $D=3$ was given recently in \cite{miz}.

\section{Dualisation and double coset structure}

    In the previous section, we constructed the cosets that give the
scalar Lagrangians for the maximal supergravities (obtained from
dimensional reduction of eleven-dimensional supergravity) in all
dimensions $D\ge 3$, in the formulations where the numbers of scalars
are maximised by dualising all $(D-2)$-form potentials (in $D\le5$).
We also showed that they have global $E_{11-\sst D}$ symmetries.  If
no such dualisations were performed, the original scalar Lagrangians
in dimensions $3\le D\le 5$ would instead have $GL(11-D,\R)\semi \R^q$
global symmetries.  The easiest way to understand the relation between
the global symmetries of these two formulations, and indeed to
understand the symmetries for any other choice of dualisations, is to
take the fully-dualised versions with the $E_{11-D}$ symmetries as the
starting point.

    Let us illustrate this by examining the relation between the
global $E_{11-D}$ symmetries of the fully-dualised formulations and
the $GL(11-D,\R)\semi \R^q$ symmetries of the undualised formulations.
$SL(11-D,\R)$ has positive-root generators $E_i{}^j$, whose algebra is
given by (\ref{eegl}).  The $E^{ijk}$ form a $q$-dimensional linear
representation under $SL(11-D,\R)$.  Curiously enough, in $D\ge6$ the 
positive-root generators $E_i{}^j$ of $SL(11-D,\R)$,
together with $E^{ijk}$, are precisely the positive-root generators
for $E_{11-\sst D}$.  In dimensions $3\le D\le 5$, on the other hand,
the situation is different. In these cases, the positive-root algebra
for $SL(11-D,\R)$ and its $\R^q$ representation can instead be obtained
as a subalgebra of a  contraction of $E_{11-\sst D}$ positive-root
algebra.  For example in $D=5$, $E_6$ has an additional generator in
its positive-root algebra, namely $D$, which is a singlet under
$SL(6,\R)$.  It is generated by the commutation of $E^{ijk}$ with
$E^{\ell mn}$:
\be
[E^{ijk}, E^{\ell mn}]= -\epsilon^{ijk\ell mn}\, D\ .\label{eed}
\ee
In a Wigner-In\"onu contraction, we would rescale $D\rightarrow \lambda\,
D$ and send $\lambda\rightarrow 0$, then all the generators $E_{ijk}$
would become commuting, giving rise to an $\R^{20}$ symmetry.
Furthermore, we can
then consistently truncate the generator $D$, to obtain a subalgebra.
Here both steps are done simultaneously by setting $D=0$.
The remaining positive-root generators are precisely those of
$SL(6,\R)$ and its $\R^{20}$ linear representation, which together
generate exactly the symmetry of the undualised scalar manifold,
namely $GL(6,\R)\semi \R^{20}$.  The necessity of the truncation of
the generator $D$ from the algebra is a reflection of the fact that
the corresponding scalar $\chi$ is now replaced by its dualised
version, namely a $(D-2)$-form gauge potential, and thus disappears
 from the scalar Lagrangian.  It should be emphasised that although we
can legitimately set the generator $D$ to zero in the positive-root
algebra, the corresponding scalar field $\chi$ cannot be consistently
set to zero if the remaining scalars are non-vanishing.  (In other
words, setting $\chi$ to zero would in general be inconsistent with
its equation of motion.)  This can be
seen from the non-linear Kaluza-Klein structure given in (\ref{d5g}).
Similarly, if we dualise $\chi$ to $A_\3$ then the field strength
$F_\4$, unlike other higher-rank fields, cannot be consistently set to
zero either.  However in this case, the terms involving $F_\4$ are no
longer part of the scalar Lagrangian.

       From the group theoretical point of view, the above truncation
of the generator $D$ corresponds to a double coset $U\!Sp(8)\backslash
E_6/N^{\vec r_1}_2$, since the generator $D$ corresponds to the
unique level-2 positive root with respect to the simple root $\vec r_1$, it
generates the Lie algebra of $N^{\vec r_1}_2$. $N^{\vec r_1}_2$ is
by definition the algebra of
the set of positive root vectorss of level 2 with respect to the simple root
$\vec r_1$.\footnote{Recall that any positive root can be written as a
sum $\sum_i \ell_i\, \vec r_i$ over the simple roots, where the
coefficients $\ell_i$ are non-negative integers which define the level
of the positive root with respect to each of the simple roots $\vec
r_i$. The total level of the root is $\ell=\sum_i \ell_i$.}  
The single coset construction, which was given in section 4, has the
effect of removing all the negative roots of the group. In the double
coset, additional higher-level positive roots are removed as well.
(Note that our notation for simple roots $\vec r_i$ of $E_{11-D}$ is
defined below Diagram 1, and in particular $\vec r_1$ denotes the simple
root $\vec a_{123}$.)  There are in total 36 positive roots in the
$E_6$ algebra, with only one that is at level 2 with respect to the
simple root $\vec r_1$ in $E_6$.  The remainder comprise 20 level 1
and 15 level 0 positive roots.  Since the levels of roots are additive
under commutation of the associated generators, it follows that the 20
level 1 generators become commuting, as indicated by (\ref{eed}), if
the would-be level 2 generator is contracted and truncated out.  As a
result, the double coset described above can then be re-expressed as
the single coset $SO(6)\backslash GL(6,\R)\semi \R^{20}$, and the
associated theory has a $GL(6,R)\semi \R^{20}$ global symmetry.
By general arguments, the double coset remains a coset for the
normalizer of the $N^{\vec r_1}_2$ subgroup acting on the
right. Clearly this normalizer is generated by the level 0 and 1
roots, {\it i.e.} it is precisely the group $GL(6,\R)\semi \R^{20}$.

      A similar analysis applies also in $D=4$ and $D=3$, where the
extra generators coming from the dualisations can again be rescaled,
allowing the positive-root algebra to be contracted.  We can then
extract a subalgebra, namely the positive-root algebra of
$GL(11-D,\R)$ and its $\R^q$ linear representation and verify the
symmetry $GL(11-D,\R)\semi \R^q$ of the undualised theory.  For $D=4$
and $D=3$, these dualisations correspond to the double cosets
$SU(8)\backslash E_7/N^{\vec r_1}_2$ and $SO(16) \backslash
E_8/N^{\vec r_1}_{2,3}$ respectively.  To be more specific, in $D=4$
the global $E_7$ symmetry algebra has 63 positive roots.  With respect
to the simple root $\vec r_1$, their gradings are: 7 level 2 roots, 35
level 1 roots and 21 level 0 roots.  The 7 level 2 roots correspond
precisely to the 7 axions that are generated by dualisation of the 7
2-form potentials.  When these 7 axions are inversely dualised back to
2-form potentials, the 7 level 2 roots are contracted and truncated
out, and as a consequence the 35 level 1 generators become commuting.
The resulting theory then has a global $GL(7,\R)\semi \R^{35}$
symmetry.  In both $D=5$ and $D=4$ the directly-reduced theories,
where no dualisations to give additional scalars are performed,
correspond to truncating out the level 2 roots of the $E_6$ or $E_7$
algebras respectively.  In $D=3$ the story is slightly different.  The
$E_8$ group has 120 positive roots, and with respect to the simple
root $\vec r_1$ there are 8 at level 3, 28 at level 2, 56 at level 1
and 28 at level 0.  The level 3 and level 2 positive roots are
commuting, corresponding to the axions that are generated by dualising
all the 36 vectors of the theory.  If these axions were absent because
we left undualised the corresponding vectors, so that the
corresponding level 3 and level 2 roots were omitted, then the 56
level 1 generators would become commuting on the remaining fields.
Thus the non-dualised theory in $D=3$ possesses a $GL(8,\R)\semi
\R^{56}$ global symmetry.

    There are many other possible choices of dualisations that could
be performed on the scalar sectors of the theories.  Some of these
will correspond to partial dualisations of a subset of the
$(D-2)$-form potentials of the original undualised theories, while
others can involve ``inverse'' dualisations of fields that were
originally scalars.  In all cases, the global symmetries can be
deduced by taking the fully-dualised theory as the starting point, and
studying the algebra of the subset of $E_{11-D}$ positive roots
corresponding to the axions that remain after the chosen dualisations.
For example in $D=3$, we can truncate out 28 level 2 or 8 level 3
generators, since they are commuting and thus are associated with
axions that can be dualised.  In fact, if we truncate out all of them,
then we get the non-dualised case with $GL(8,\R)\semi \R^{56}$, which
we discussed earlier.  If we dualise only the level 3 roots we again
end up with a coset, but if however we dualise the 28 level 2 roots
without dualising the higher 8, then the normaliser is only the level
0 $GL(8,\R)$ which does not act transitively and this is not a coset
situation.  In section 7 we shall study one particular class of
examples, in which only those original $(D-2)$-form potentials that
lie in the Ramond-Ramond sector (from the ten-dimensional type IIA
viewpoint) are dualised to give additional scalars.  This corresponds
to truncating out the highest-level positive roots with respect to the
simple root $\vec r_2=\vec b_{12}$, rather than the truncations with
respect to $\vec r_1=\vec a_{123}$ that we discussed above.  In
section 9, we discuss the symmetry group of the direct dimensional
reductions of type IIB supergravity in $D=10$, where no dualisations
are performed.  These correspond to truncating out sets of commuting
positive roots under a double grading with respect to the two simple
roots $\vec r_1=\vec a_{123}$ and $\vec r_3=\vec b_{23}$. For
completeness we may recall that the disintegration of $E_{11-D}$ to
$E_{10-D}$ amounts to omitting the roots of level 1 with respect to
the last simple root that appeared during the dimensional reduction,
except for the case $D=3$ where both level 1 and 2 roots must be
omitted.

         In all these cases, the symmetry can be understood from the
point of view of the double coset of the $E_{11-D}$ group.  In section
2, the dimensional reduction of eleven-dimensional supergravity was
performed by iteratively repeating the $D+1$ to $D$ dimensional
reduction.  It followed that the scalars are then precisely the
parameters of the generators of the Borel group modulo the maximal
compact subgroup. Thus the fully-dualised Lagrangian is already
naturally written with the coset $K(E_{11-D})\backslash E_{11-D}$
structure, where $K(G)$ denotes the maximal compact subgroup of $G$.
To understand the dualisation, we recall that in section 4 we saw that
the axions are in one-to-one correspondence with the positive roots,
while the dilatons are associated with the Cartan generators.  We
shall show in section 8 that the maximal abelian subalgebras of the
positive-root (nilpotent) algebras precisely describe the commuting
$\R$ symmetries of the sets of axions that can be simultaneously
covered by a derivative.  Thus inverse dualisation of scalars to
higher-forms can be understood as removals of commuting generators
 from the coset $K(G)\backslash G$, giving rise to a double coset.

\section{Global symmetries of higher-rank fields}

     In the previous sections, we studied the global symmetries of the
scalar sectors of the toroidally compactified supergravity
Lagrangians, showing in particular that they have global $E_{11-D}$
symmetries when the numbers of scalars are maximised by dualising all
$(D-2)$-form potentials.  We also showed that alternative choices of
dualisation change the scalar Lagrangian, and its symmetries.  In this
section we shall study how the global symmetries of the scalar sector
can also be realised on the higher-degree fields. We shall show that
in general it is necessary to perform appropriate dualisations in the
higher-degree sectors in order to preserve the global symmetry of the
scalar sector.  In particular, when the dualisation of $(D-2)$-form
potentials to scalars has been performed, the $E_{11-D}$ symmetry is
preserved for the higher-degree sectors of the Lagrangian provided
that all field strengths with degrees greater than $\ft12 D$ are
dualised.  In some cases, these dualisations are necessary in order to
preserve the $E_{11-D}$ symmetry, while in others, the symmetry may
still be realised (on the higher-rank field strengths rather than
their potentials) at the level of the equations of motion even in the
absence of the dualisations.  An example of the former is in $D=6$,
where the $E_5=O(5,5)$ global symmetry of the scalar Lagrangian is
broken (even at the level of the equations of motion) unless the
4-form field strength is dualised to give an additional 2-form.  An
example where the dualisation is optional is in $D=7$, where the
$E_4=SL(5,\R)$ symmetry of the scalar Lagrangian is preserved in the
full Lagrangian if the 4-form is dualised to give another 3-form field
strength, and it is also preserved at the level of the equations of
motion if the 4-form is left undualised.  We shall discuss these, and
other examples, below.

     To begin, let us consider the cases where all field strengths of
ranks greater than $\ft12 D$ are dualised, and the theories have the
$E_{11-D}$ global symmetries.  The discussion divides into two,
depending on whether $D$ is odd or even. In odd dimensions, the
symmetry acts on the potentials, and is realised at the level of the
Lagrangian.  In an even dimension, on the other hand, the field
strengths of rank $\ft12 D$ and their duals form a single irreducible
multiplet under the $E_{11-D}$ group, and thus the symmetry can be
implemented only at the level of the equations of motion, realised on
the field strengths rather than the potentials.  (The symmetry
transformations still act on the potentials for the lower-degree field
strengths.)

\subsection{Odd dimensions}

    In $D=9$, no dualisations are necessary, and the global $GL(2,\R)$
symmetry of the scalar manifold extends to the entire Lagrangian, as
we described in section 2.2.  The bosonic Lagrangian and its
$GL(2,\R)$ symmetry was proved in \cite{bho,lpsweyl}.  The situation
is more complicated in $D=7$ and $D=5$; we shall describe the former
in complete detail, and postpone a similar discussion of the latter to
a subsequent publication.  (In $D=3$, the fully-dualised theory
contains only scalar fields, and its $E_8$ symmetry was fully
discussed already in section 4.2.)

     We shall first consider $D=7$.  The scalar Lagrangian, which has
a manifest global $GL(4,\R)$ symmetry, in fact has a larger
$SL(5,\R)$ global symmetry.  This can be made manifest by defining
generators $E_\a{}^\beta$, where $\a=(i,5)$ is an $SL(5,\R)$ index,
and $E_i{}^5 = \ft16 \epsilon_{ijk\ell}\, E^{jk\ell}$.  The
commutation relations for $E_i{}^j$ and $E^{ijk}$ now become
\be
[ E_\a{}^\beta, E_\gamma{}^\delta] = \delta^\beta_\gamma\, E_\a{}^\delta -
\delta_\a^\delta\, E_\gamma{}^\beta\ .
\ee
The manifest $O(5) \backslash SL(5,R)$ coset is presented in Appendix
C.  

       At the level of the Lagrangian, where this symmetry acts on the
gauge potentials, it extends to the higher-degree sectors of the
theory only if we dualise $A_\3$ to give a further 2-form potential
$A_{\2}$ which, together with the four potentials $A_{\2i}$, (and
after some field redefinitions, which we shall discuss in detail
below) form an irreducible 5 under $SL(5,\R)$. The grouping of
$SL(4,\R)$ representations into $SL(5,\R)$ representations can already
be seen in the structure of the dilaton vectors.  Consider first the
3-form field strengths, including the dualisation of the 4-form.  The
associated dilaton vectors are $(\vec a_i,-\vec a)$.  As we saw in
section 4, the simple roots of the $SL(5,\R)$ algebra in $D=7$ are
given by $\vec b_{12}$, $\vec b_{23}$, $\vec b_{34}$ and $\vec
a_{123}$.  It is a simple matter to see, from the definitions in
Appendix A, that $-\vec a$ is the highest-weight vector of the
5-dimensional representation of $SL(5,\R)$, with the rest of the
multiplet filled out by acting with the negatives of the simple roots,
according to the scheme
\be
\vec a_4 = -\vec a -\vec a_{123}\ ,\qquad \vec a_3 = \vec a_4 -\vec b_{34}
\ ,\qquad \vec a_2=\vec a_3-\vec b_{23}\ ,\qquad \vec a_1 = \vec a_2 -\vec
b_{12}\ .
\ee
Similarly, the dilaton vectors for the 2-form field strengths, namely
$(\vec a_{ij}, \vec b_i)$, are the weight vectors of the $\bar{10}$
representation of $SL(5,\R)$, with $\vec a_{34}$ as the highest-weight
vector.

    To see the $SL(5,\R)$ symmetry explicitly at the level of the
Lagrangian, we begin by dualising the 3-form potential $A_\3$.  To do
this, we introduce a 2-form Lagrange multiplier $A_\2$, to enforce the
Bianchi identity for $F_\4$, which will now be treated as an auxiliary
field, adding an extra term $d A_\2\wedge \tilde F_\4$ to the
Lagrangian.  As usual, it is advantageous to replace $\tilde F_\4$
immediately by its Kaluza-Klein modified field strength $F_\4=\hat
F_4$ as given in (\ref{interm}), and likewise to replace all
occurrences of $\tilde F_\4$ in the Wess-Zumino terms given in
(\ref{ffaterms}) by $F_\4$.  This makes the algebraic equation for the
auxiliary field $F_\4$ very easy to solve, giving $F_\4= e^{-\vec
a\cdot\vec \phi}\, \ast G_\3$, where
\be
G_\3 = dA_\2 + \Big(\ft16 dA_{\2i}\, A_{\0jk\ell} 
+\ft18 A_{\1ij}\wedge dA_{\1k\ell} \Big)\, \epsilon^{ijk\ell} 
\ ,\label{dualg}
\ee
At this point, it is convenient to introduce some redefined potentials,
along the lines of those described in Appendix A.  In fact for the 1-forms
$A_{\1ij}$, we make exactly the redefinition given in (\ref{aredef}).  For
the 2-forms $A_{\2i}$, however, it turns out that the most appropriate
redefinition here is different from the hatted potentials given
 in (\ref{aredef}), and is instead given by
\be
\bar A_{\2i} = A_{\2i} +\ft12 A_{\1ij}\wedge\hat{\cal A}_\1^j
\ .\label{n2def}
\ee
At the same time, we make the following redefinition for the dualised
potential $A_\2$:
\be
\bar A_\2 = A_\2 + \ft1{12} \epsilon^{ijk\ell}\, A_{\0ijk} \,
A_{\1\ell m}\wedge \hat{\cal A}_\1^m\ .\label{lmdef}
\ee
Note that $\bar A_{\2i}$ and $\bar A_\2$ have the transformations
\be
\delta \bar A_{\2i} =0\ ,\qquad 
\delta \bar A_{\2} = \Lambda^i{}_5\, \bar A_{\2i}\ ,
\ee
under the variations $\delta A_{\0ijk}=c_{ijk}$, 
where $\Lambda^i{}_5 = \ft16 \epsilon^{ijk\ell}\, c_{jk\ell}$.

     We are now in a position to define $SL(5,\R)$-covariant potentials,
$B_\1^{\a\beta}$ and $B_{\2\a}$, related to those described above by
\bea
B_\1^{ij} &=& \ft12 \epsilon^{ijk\ell}\, \hat A_{\1k\ell}\ ,\qquad\quad\ \
B_\1^{i5}= \hat{\cal A}_\1^i\ ,\nn\\
B_{\2i} &=& \bar A_{\2i}\ ,\qquad\qquad\qquad B_{\2 5} = \bar A_\2\ .
\eea
We also define their field strengths $H_\2^{\a\beta}$ and $H_{\3\a}$, by
\be
H_\2^{\a\beta} = dB_\1^{\a\beta}\ ,\qquad 
H_{\3\a} = dB_{\2\a} + \ft18 \epsilon_{\a\beta\gamma\delta\sigma}\,
B_\1^{\beta\gamma}\wedge d B_\1^{\delta\sigma}\ .\label{hfs}
\ee
Note that while the gauge transformation for $B_{\2\a}$ is simply
$\delta B_{\2\a}=d\Lambda_{\1\a}$, the one for $B_\1^{\a\beta}$ must be
accompanied by a compensating transformation for $B_{\2\a}$, in order to
ensure the gauge invariance of $H_{\3\a}$:
\be
\delta B_\1^{\a\beta}=d\Lambda_\0^{\a\beta}\ ,\qquad 
\delta B_{\2\a} = -\ft18 \epsilon_{\a\beta\gamma\delta\sigma}\,
B_\1^{\beta\gamma}\wedge d\Lambda_\0^{\delta\sigma}\ .\label{comga}
\ee
We now find that all the remaining Wess-Zumino terms
given in (\ref{ffaterms}), together with the additional terms acquired from
the introduction of the Lagrange multiplier, turn out, after calculations
of not inconsiderable complexity,  to be expressible in
the simple $SL(5,\R)$-covariant form
\be
{\cal L}_{\rm WZ} = 
-\ft12 \Big(H_{\3\a} \wedge H_{\3\beta}\wedge
B_\1^{\a\beta} - 4 H_{\3\a}\wedge B_{\2\beta} \wedge dB_\1^{\a\beta}
\Big)\ .\label{d7wz}
\ee
Note that this is invariant, up to a total derivative, under the gauge 
transformations not only for $B_{\2\a}$ but also for $B_\1^{\a\beta}$,
given in (\ref{comga}).
In terms of $H_\2^{\a\beta}$ and $H_{\3\a}$, the field strengths appearing 
in the Lagrangian given in Appendix A take the form
\bea
\hat {\cal F}_\2^i &=& H_\2^{i5}\ ,\qquad \hat F_{\2ij} =\ft12
\epsilon_{ijk\ell}\, H_\2^{k\ell} + A_{\0ijk}\, H_\2^{k5}\ ,\nn\\ 
\hat F_{\3i} &=& H_{\3 i}\ ,\qquad G_\3 = H_{\3 5} -\ft16 
\epsilon^{ijk\ell}\, A_{\0ijk}\, H_{\3\ell}\ ,\label{ffs}
\eea
where $G_\3$ is defined by (\ref{dualg}). 

      The scalar Lagrangian for coset $O(5)\backslash SL(5,\R)$ is
discussed in Appendix C. Putting this together with the results for
the higher-rank fields obtained above, we can write down the
manifestly $SL(5,\R)$ invariant bosonic Lagrangian for $D=7$ maximal 
supergravity:
\bea
{\cal L} &=& e\, R + \ft14 e\, \tr(\del_\mu G^{-1}\, \del^\mu G) -
\ft1{12}e\, H_{\3\a} G^{\a\b} H_{\3\b} -\ft18e\,
H_{\2}^{\a\beta} G_{\a\gamma} G_{\b \delta} H_{\2}^{\gamma\delta}
\nn\\
&&-\ft12 \Big(H_{\3\a} \wedge H_{\3\beta}\wedge
B_\1^{\a\beta} - 4 H_{\3\a}\wedge B_{\2\beta} \wedge dB_\1^{\a\beta}
\Big)\ .
\eea
where $G^{\a\b}$ and $G_{\a\b}$ are given by (\ref{gsl5r}).

     The discussion for five dimensions proceeds in a similar way.
The supergravity Lagrangian in $D=5$ with manifest $E_6$ symmetry was
constructed in the second reference in \cite{cj2}.  In order to have
$E_6$ as the global symmetry, we must first of all dualise the 3-form
potential $A_\3$ to give another scalar $\chi$, as discussed in
section 4.  For the higher-form gauge potentials, it is necessary to
dualise the six 2-form potentials $A_{(2)i}$ to give additional
vectors, which together with the six Kaluza Klein vectors, and fifteen
vectors from anti-symmetric tensor in $D=11$ form the 27-dimensional
representation of $E_6$.  The detailed discussion of this example will
be presented in a forthcoming paper.
  
\subsection{Even dimensions}

     The discussion becomes more complicated in even dimensions only
because of the occurrence of field strengths whose degree is equal to
$\ft12 D$. For all the other higher-degree fields, the global
$E_{11-D}$ symmetry of the fully-dualised scalar manifold acts on the
potentials in the same manner as in the case of odd dimensions.
However, because the field strengths of degree $\ft12 D$ together with
their duals form a single irreducible representation of $E_{11-D}$,
the symmetry can only be realised on these field strengths themselves,
rather than on their potentials.  Consequently, the $E_{11-D}$
symmetry of the full theory can only be realised at the level of the
equations of motion in even dimensions.

     There is, however, a convenient way to reformulate the theory so
that the global $E_{11-D}$ symmetry can in fact be implemented at the
level of an auxiliary Lagrangian. We shall call it the doubled
Lagrangian.  This can be done by introducing an auxiliary set of field
strengths of degree $\ft12 D$, equal in number to the original set,
together with their associated potentials.  One can then construct a
Lagrangian whose full set of field equations can be consistently
truncated to give the equations of motion of the original theory.  By
this means, the global $E_{11-D}$ symmetry can be implemented on the
doubled set of potentials, whose total number is now equal to the
dimension of the relevant irreducible representation of $E_{11-D}$.
(A similar trick was recently used in order to construct a Lagrangian
for type IIB supergravity, by adding the anti-self-dual half of the
4-form potential, whose equations of motion could be consistently
truncated to those of type IIB supergravity \cite{bbo}.)  We shall now
use this technique to discuss the global $E_{11-D}$ symmetries of the
fully-dualised theories in $D=8$, 6 and 4.  We shall give the complete
construction of the $E_7$-invariant bosonic theory in $D=4$, while, in
the more complicated cases of $D=6$ and $D=8$, we shall just focus our
attention on the field strengths of degrees 3 and 4 respectively.

\bigskip
\noindent{$D=8$:}
\bigskip

     The scalar Lagrangian in this case has a global $SL(3,\R)\times
SL(2,\R)$ symmetry.  The 3-form potential $A_\3$ is a singlet under
$SL(3,\R)$, since it carries no internal indices and the $SL(3,\R)$ is
contained in the standard global $GL(3,\R)$ symmetry that results from
compactification on a 3-torus.  (The $SL(3,\R)$ has simple roots $\vec
b_{12}$ and $\vec b_{23}$, while the $SL(2,\R)$ has the simple root
$\vec a_{123}$.) However, the field strength of $A_\3$ is actually a
doublet under $SL(2,\R)$, which can be seen from the fact that $(\vec
a, -\vec a)$ form the weight vectors of the 2 of $SL(2,\R)$, with
$-\vec a$ as highest weight.  These two dilaton vectors are associated
with the 4-form field strength and its dual.  The 2-form potentials
$A_{\2i}$ form a singlet under $SL(2,\R)$, and a triplet under
$SL(3,\R)$.  This can be seen from their dilaton vectors $(\vec a_1,
\vec a_2, \vec a_3)$, which are the weight vectors of the 3 of
$SL(3,\R)$, with $\vec a_3$ as highest weight.  Finally, the vector
potentials $A_{\1ij}$ and ${\cal A}_\1^{i}$ have dilaton vectors
$(\vec a_{ij}, \vec b_i)$ that form the weights of the $(3,2)$
representation of $SL(3,\R)\times SL(2,\R)$, with $\vec a_{23}$ as
highest weight.

     The extension to the higher-degree fields of the $SL(3,\R)$
factor in the global symmetry of the scalar manifold is
straightforward, since it is contained in the $GL(3,\R)$ that was
described in section 2.  We shall instead concentrate on the
$SL(2,\R)$ factor, since this involves new issues associated with the
occurrence of the 4-form field strength, whose degree is half the
spacetime dimension.  In fact, we can consistently truncate out all
the fields that are non-singlets under $SL(3,\R)$, namely the 2-form
and 3-form field strengths and the scalars of the $SL(3,\R)$ factor in
the scalar manifold, since the remaining $SL(2,\R)$ scalars and the
4-form field strength, which is an $SL(3,\R)$ singlet, cannot act as a
source for them.  We may therefore simplify the discussion of the
$SL(2,\R)$ structure of the theory by performing this truncation.  We
then introduce a second 3-form potential $\tilde A_\3$, and define the
4-form field strengths
\be
H_\4 = \pmatrix{dA_\3 \cr d\td A_\3}\ ,
\ee
which form a doublet under $SL(2,\R)$.
The $SL(2,\R)$-invariant Lagrangian can then be
written in the form
\bea
{\cal L} &=& eR +\ft14 e\, {\tr}(\del_\mu {\cal M}^{-1}\, 
\del^\mu {\cal M}) -\ft1{48}e\, H_\4^{\rm T}\, {\cal M}\, H_\4\ ,
\nn\\
&=& eR -\ft12e\, (\del\phi)^2 -\ft12 e\, e^{2\phi}\, (\del\chi)^2 -\ft1{96}e\,
e^{-\phi}\, F_\4^2 -\ft1{96}e\,e^{\phi}\, \tilde F_\4^2\ ,\label{sllag}
\eea
where $F_\4=dA_\3$, $\tilde F_\4 = d\tilde A_\3 -\chi\, dA_\3$, and
$\phi=-\vec a\cdot \vec\phi$, $\chi=A_{\0123}$. The Bianchi identities
and equations of motion for $F_\4$ and $\tilde F_\4$ are therefore
\bea
dF_\4 &=& 0\ ,\qquad\qquad d(\tilde F_\4 +\chi\, F_\4)=0\ ,\nn\\
d\ast(e^\phi\, \tilde F_\4 ) &=&0\ , \qquad d\ast( e^{-\phi}\, F_\4 -\chi\,
e^\phi\, \tilde F_\4) =0\ .
\eea

     We see from these equations that it is consistent to impose the relation
\be
\widetilde F_\4 = e^{-\phi}\, \ast F_\4\ ,\label{trunc}
\ee
leading to the equations
\be
dF_\4=0\ ,\qquad\qquad d(e^{-\phi}\, \ast F_\4 + \chi\, F_\4)=0\ ,
\label{sleq}
\ee
which are precisely the Bianchi identity and equation of motion for
the 4-form field strength $F_\4$ that follow from the $D=8$ Lagrangian
in Appendix A.  In fact the latter, after performing the truncation to
the $SL(3,\R)$ singlets described above, takes the form
\be
{\cal L}= eR -\ft12 e(\del\phi)^2 -\ft12e\,  e^{2\phi}\, (\del\chi)^2
-\ft1{48}e\, e^{-\phi}\, F_4^2 + \ft1{48}e\, \chi\, F_4\cdot {}^*F_4 \ ,
\label{d8lag5}
\ee
which can easily be verified to give the same equations of motion as
the ones coming from (\ref{sllag}) together with the constraint
(\ref{trunc}).  Note that the truncation (\ref{trunc}) is $SL(2,\R)$
covariant, and thus the Bianchi identity and equation of motion
(\ref{sleq}) indeed inherit the global $SL(2,\R)$ symmetry that was
manifest in the Lagrangian (\ref{sllag}) prior to the truncation.
Before the truncation, $A_\3$ and $\tilde A_\3$ form an $SL(2,\R)$
doublet and the symmetry is realised in the Lagrangian; after the
truncation, $F_\4$ and $e^{-\phi}\, {*F}_\4$ form an $SL(2,\R)$
doublet and the symmetry is realised only in the equations of
motion. Note that prior to truncation, the Bianchi identities and
equations of motion can be written in the manifestly
$SL(2,\R)$-covariant forms
\be
dH_\4=0\ ,\qquad\qquad d\ast({\cal M} H_\4) =0\ .
\ee
The truncation (\ref{trunc}) then takes the manifestly 
$SL(2,\R)$-covariant form
\be
H_\4 = \Omega{\cal M}\, \ast H_\4\ ,
\ee
where $\Omega$ is the $SL(2,\R)$-invariant antisymmetric rank-two
tensor, which appears here in a (noncovariant) canonical form.
One can raise one index of this {\it a fortiori} $SO(2)$ invariant 
tensor and obtain the relation $\Omega^2=-1$.

\bigskip
\noindent{$D=6$:}
\bigskip

     The global symmetry of the scalar manifold is $E_5= O(5,5)$ in
six dimensions.  After dualising the 3-form potential to give an
additional vector potential $A_\1$, the vectors $A_{\1ij}$, ${\cal
A}_\1^{i}$ and $A_\1$ (after appropriate field redefinitions) form a
16-dimensional irreducible multiplet under $O(5,5)$, corresponding to
the weight vectors $(\vec a_{ij}, \vec b_i, -\vec a)$.  The
highest-weight vector is $-\vec a$.  The five 2-form potentials
$A_{\2i}$ cannot themselves form an $O(5,5)$ multiplet, but their
field strengths, together with the duals, form an irreducible
10-dimensional representation.  The associated dilaton vectors $(\vec
a_i, -\vec a_i)$ are the weight vectors of the 10, with $-\vec a_1$ as
the highest weight.  We may give an analogous discussion to the one in
$D=8$, and focus just on the sectors comprising the scalars and the
3-form field strengths.  (The vectors can be truncated consistently
 from the theory, thus simplifying the discussion.)  We may then
introduce a second set of five 2-form potentials $\tilde A_\2^{i}$, in
terms of which we define
\be
H_\3 = \pmatrix{dA_{\2i} \cr d\td A_{\2}^i}\ .
\ee
This set of field strengths transform as the 10-dimensional vector
representation under $O(5,5)$.  The Lagrangian for the kinetic terms
for the scalars and 2-form potentials can then be written in the
manifestly $O(5,5)$-invariant form
\be
{\cal L} =eR +\ft14 e\, {\rm tr}(\del_\mu {\cal M}^{-1}\, \del^\mu{\cal M})
-\ft1{24} e\, H_\3^{\rm T} {\cal M} H_\3 \ ,
\ee
where ${\cal M}$ is the $(O(5)\times O(5))\backslash O(5,5)$ coset
matrix defined in section 4.1.  Its explicit form is given by
(\ref{mmatrix}), with $G$ and $X$ as defined above (\ref{o55}) in Appendix C.
The Bianchi identities and equations of motion that follow from this
Lagrangian are
\be
dH_\3 =0\ ,\qquad \qquad d\ast({\cal M}\, H_\3)=0\ ,
\ee
from which we see that it is consistent to impose the $O(5,5)$-covariant
truncation
\be
H_\3 = \Omega\, {\cal M}\, \ast H_\3\ .
\ee

Here $\Omega$ is an $O(5,5)$-invariant metric that again can be
considered to be an invariant of $(O(5)\times O(5))$.  One of its
indices can be raised with the invariant (identity) metric of the
maximal compact subgroup to obtain $\Omega^2 = 1$.  The subsector of
the bosonic Lagrangian for six-dimensional maximal supergravity that
we have obtained here agrees with the results obtained in
\cite{tanii}, where the complete theory was obtained by direct
construction, rather than by dimensional reduction from $D=11$.

\bigskip
\noindent{$D=4$:}
\bigskip

     The fully-dualised scalar manifold in $D=4$ has an $E_7$ global
symmetry.  The only additional fields of higher degree are the 28
vectors, comprising 21 $A_{\1ij}$ and 7 ${\cal A}_\1^{i}$.  Their
associated field strengths, together with their duals, form the
56-dimensional irreducible representation of $E_7$ \cite{cj3}.  The
associated dilaton vectors, $(\vec a_{ij}, \vec b_i, -\vec a_{ij},
-\vec b_i)$ are the weight vectors of the 56, with $-\vec b_7$ as the
highest weight.

           The $D=4$, $N=8$ supergravity with manifest $E_7$ global
symmetry was obtained by first dimensionally reducing
eleven-dimensional supergravity, and then dualising the seven 2-form
potentials $A_{\2i}$ to give rise to additional scalars $\chi^i$
\cite{cj3}.  It is also necessary to dualise the twenty-one
pseudo-vectors $A_{\1ij}$ to give twenty-one vectors. Together with
the seven Kaluza-Klein vectors, they form a 28-dimensional
representation of $SL(8, \R)$.  The bosonic Lagrangian can then be
written as \cite{cj3}
\be
{\cal L}_1 = e\, R + \ft14 e\, \tr(\del_\mu {\cal M} \del^\mu {\cal M} )
+\ft18 e\, F^{ab}_{\mu\nu}\ {*G}^{\mu\nu}_{ab}\ ,\
\ee
where ${\cal M}$ parameterises the coset  
$SU(8)\backslash E_7$, constructed in
section 4, and $F^{ab}_{\mu\nu}$, with indices $a,b=(i,8)$, 
are the field strengths of the twenty-eight vectors.  $G^{ab}_{\mu\nu}$
is given by
\be
{*G}_{\mu\nu}^{ab} = - 4 \fft{\delta {\cal L}}{\delta F_{\mu\nu}^{ab}}\ ,
\ee
and is therefore a linear combination of $F_{\mu\nu}^{ab}$ and
${*F}_{\mu\nu}^{ab}$.  At the level of the Lagrangian, the global
symmetry is $SL(8,\R)$, which extends to $E_7$ at the level of the
equations of motion, where the twenty-eight $F_{\mu\nu}^{ab}$ and
twenty-eight
$G_{ab\mu\nu}$ form a 56-dimensional representation of $E_7$.
Writing
\be
H_\2 = \pmatrix{F\cr G}\ ,
\ee
it was shown that they in fact satisfy the duality relation $H_\2 =
\Omega {\cal M} {*H}_\2$, with \cite{cj3}
\be
\Omega = \pmatrix{0& I \cr -I & 0}\ .
\ee
This allows us to write down an $E_7$-invariant
Lagrangian ${\cal L}_2$, where the $G$ fields are regarded as
independent of $F$.  The Lagrangian is given by
\be
{\cal L}_2 = e\, R +\ft14 e\, \tr(\del_\mu {\cal M}
\del^\mu {\cal M}^{-1}) -\ft18 H_\2^{\rm T} {\cal M} H_\2\ .
\ee

\subsection{Doubled Lagrangians}

    We now present a general proof that all the equations of motion
following from the even-dimensional doubled Lagrangians that we have
been discussing here, with a doubled set of potentials for the field
strengths of degree $n=\ft12 D$, are indeed the same as the equations
of motion from the original Lagrangians after we impose a universal
twisted self-duality constraint.  We have already seen that this is
true for the equations of motion for the fields of degree $n$
themselves; it remains to be established that the equations of motion
for the other fields are the same, we shall now consider them.  The
structure of the ``doubled'' Lagrangians is
\be
{\cal L}_2 =- \ft1{4\, n!}\, H^{\rm T}{\cal M} H + L(\phi)
\ ,\label{lag111}
\ee
where $\phi$ denotes all the remaining fields, including the complete
set of scalar fields, the metric $g_{\mu\nu}$ {\it etc.}\ with
Lagrangian $L(\phi)$, and
\be
H=\pmatrix{ F \cr G}\ .\label{hfg}
\ee
Here $F=dA$ is written in terms of the original potentials $A$, while
$G=dB$ is written in terms of the ``doubled'' potentials $B$.  The
fields $H$ satisfy the Bianchi identity $dH=0$ and equations of motion
$d({\cal M} {*H})=0$.  Here the fields are real and the matrix $\cal M
=\v^{\rm T} \eta \v $ is symmetrical.

      We then impose the twisted self-duality constraint (coined some
time ago a silver rule of supergravity)
\be
H=\Omega{\cal M} {*H}\ .\label{con1111}
\ee
Acting with another $*$, and using the fact that ${**H}=(-1)^{n-1}\,
H$, this implies that it squares to a multiple of the unit matrix,
$(\Omega{\cal M})^2 =(-1)^{n-1}\, I$.  We also have
$(\Omega\eta)^2=(-1)^{n-1}$.  We may use the constraint
(\ref{con1111}) to solve for the field strengths $G$ in terms of $F$
and $*F$, giving $G=f(\phi)\, F + g(\phi) {*F}$.  We may then write a
Lagrangian purely in terms of the original fields in the form
\be
{\cal L}_1 = \ft1{2 \, n!}\, F\cdot {^*G} + L(\phi)\ ,\label{lag222}
\ee
where $G$ is expressed in terms of $F$ as above.  It is obvious that
the equations of motion for the $(D/2)$-form field strengths are the
same for the Lagrangians (\ref{lag222}) and (\ref{lag111}).  
Note that if the
solution for $G$ is substituted into $H$ given in (\ref{hfg}), it then has
the property that in any even dimension:
\be
H^{\rm T}{\cal M} H=0 \ .\label{zero}
\ee
This can be seen by using (\ref{con1111}) to re-express $H^{\rm T}{\cal M}
H$ as the manifestly-vanishing expression $H^{\rm T}\wedge \Omega H$.  

    With these preliminaries, we are in a position to show that the
equations of motion for the other fields $\phi$ that follow from
(\ref{lag111}), after imposing the constraint (\ref{con1111}), are the
same as the equations of motion that follow from (\ref{lag222}).  To
do this we vary (\ref{zero}) with respect to $\phi$, to get
\be
\ft12 H^{\rm T}\fft{\delta {\cal M}}{\delta\phi}\, H +\fft{\delta
H^{\rm T}}{\delta\phi}\, {\cal M}H =0\ .
\ee
Now, we can use (\ref{con1111}) in the second term, to give
\be
\ft12 H^{\rm T}\fft{\delta {\cal M}}{\delta\phi}\, H = -\fft{\delta
H^{\rm T}}{\delta\phi}\,\Omega {*H} = -\fft{\delta}{\delta\phi} (F\cdot
{*G})\ .
\ee
It is now evident that $\delta{\cal L}_1/\delta\phi = \delta{\cal
L}_2/\delta\phi$, and hence the scalar equations of motion from the
two Lagrangians agree.\footnote{It is possible to add a linear term
$H^{\rm T} X(\phi)$ to the Lagrangian (\ref{lag111}) provided that
$X(\phi)$ satisfies ${*X} {\cal M} \Omega X =0$ and we modify the
self-duality constraint (\ref{con1111}).  The proof is analogous, with
appropriate minor modifications.}
 
         To conclude this subsection, we present some results on
duality in even dimensions.  The maximal possible duality symmetry in
an even dimension for a given set of $N$ (prior to doubling) field
strengths with degree $D/2$ depends on whether $D=4k$ or $D=4k+2$;
they are $Sp(2N)$ and $O(N,N)$ respectively \cite{tanii}. (This
generalises the $D=4$ result in \cite{gz}.)  The maximal coset space
for the scalar fields also depends on the spacetime signature.  The
cosets for the maximal duality symmetries are summarised below in
Table 1.

\bigskip\bigskip

\centerline{
\begin{tabular}{|c|c|c|}\hline
 & Lorentzian & Euclidean \\ \hline
$D=4k$ & $U(N)\backslash Sp(2N)$ & $GL(N) \backslash Sp(2N)$\\ \hline
$D=4k+2$& $O(N)\times O(N)\backslash O(N,N)$  &
$O(N,C)\backslash O(N,N)$ \\ \hline
\end{tabular}}

\bigskip

\centerline{Table 1:  Maximal global symmetry and scalar coset for $N$
$D/2$-form field strengths}
\bigskip\bigskip

          In Lorentzian spacetime, the stability group for the scalars
is the maximal compact subgroup of the global symmetry group, whilst
in Euclidean space, the denominator group is non-compact \cite{gha}.
Also, in Euclidean space the kinetic terms for the axionic scalars,
which couple to the tensor fields, have the opposite sign.

          The introduction of the doubled formalism, where the
Lagrangian ${\cal L}_2$ is invariant under the full duality group,
allows us to define conserved Noether currents in the usual way.  If
we impose the self-duality constraint (\ref{con1111}) on the current,
it defines conserved currents for the original Lagrangian ${\cal
L}_1$.  In the case of $D=4$, this procedure gives rise to the same
currents as defined in \cite{gz}.  These currents are non-local
whenever the dual potentials appear explicitly in the expression.  It
is worth noting that even for the subgroup which leaves the original
Lagrangian ${\cal L}_1$ invariant, this procedure does not reproduce
the current defined directly from ${\cal L}_1$.  They differ by the
topological current
\be
J^{\mu_1} \sim \epsilon^{\mu_1\cdots\mu_D} \del_{\mu_2}
X_{\mu_3\cdots \mu_D}\ .
\ee

\subsection{Dualisations of higher-degree fields, and global symmetries}

     We have seen earlier in the paper that a convenient way to
discuss the global symmetries of the dimensionally-reduced
supergravities is to consider first the symmetries of the scalar
manifold.  The global symmetries of the scalar sector are themselves
dependent on the choice of dualisations, in the sense that one obtains
inequivalent symmetry groups if $(D-2)$-form potentials in the direct
reduction are not dualised to scalars, or if existing scalars are
``undualised'' to give $(D-2)$-form potentials.  These differences in
the global symmetry groups persist even at the level of the equations
of motion.

    One might be tempted to think that the global symmetries of the
scalar manifold would automatically extend to the entire theory, since
the higher-degree field strengths transform linearly.  However, this
is not in general true.  What is true is that in the toroidally
dimensionally reduced theories coming from eleven-dimensional
supergravity, there always exists a choice of dualisations for the
higher-degree fields such that the global symmetries of the scalar
sector do indeed extend to the full theory.  In some cases, this will
be true only at the level of the equations of motion, whilst in other
cases, the symmetry is realised also at the level of the Lagrangian.
What is perhaps more surprising is that there are examples where, even
at the level of the equations of motion, the global symmetry of the
scalar sector can be broken as a result of performing, or not
performing, certain dualisations of the higher-degree fields.  The
basic reason for this is that it is not in general the case that the
field equations and Bianchi identities involve only field strengths;
bare potentials can occur too.  In such cases the global symmetries
must clearly be realised on the potentials themselves, which is
possible in terms of local field transformations only if all the
potentials associated with a would-be irreducible multiplet have the
same degree.

     A simple example arises in six dimensions, where in the direct
reduction we have a total of $15=10+5$ vector potentials $A_{\1ij}$
and ${\cal A}_\1^{i}$, and one 3-form potential $A_\3$.  If the latter
is dualised to a 1-form, then the sixteen vector potentials can form
an irreducible 16 of $O(5,5)$, and the $O(5,5)$ global symmetry of the
scalar manifold can then be realised also in the entire theory.  Now
let us consider the situation where instead the 3-form potential is
not dualised.  The 4-form field strength and the rest of the fifteen
vector potentials could still form a 16-dimensional multiplet if it
were the case that the potentials were all covered by derivatives, so
that the symmetry transformations could be implemented on the sixteen
field strengths.  However, if the 3-form field strength is not
dualised, the corresponding 4-form field strength $\hat F_\4$ is given
by (\ref{interm}).  In these original variables, the ${\cal L}_{FFA}$
terms are trilinear and do not involve the Kaluza-Klein vectors ${\cal
A}_\1^{i}$.  Thus indeed there are no bare vector potentials in the
equation of motion for the field strength $\hat F_\4$.  However, the
Bianchi identity for $\hat F_\4$ does contain bare Kaluza-Klein
potentials.  On the other hand, by changing variables to the hatted
potentials, where $F_\4$ is given by (\ref{cjvar}), there will now be
no Kaluza-Klein vectors in the Bianchi identity.  Instead, the ${\cal
L}_{FFA}$ terms now become complicated and they will involve bare
Kaluza-Klein vector potentials.  Either way, the occurrence of bare
potentials in the equations of motion or Bianchi identities is
unavoidable, and in fact there is no possible redefinition of fields
that can circumvent the problem.  Since one is therefore forced to
realise the $O(5,5)$ on the sixteen potentials, rather than their
field strengths, it is necessary to dualise $A_\3$ to a vector in
order to be able to give a realisation of the global symmetry in terms
of local field transformations.  Another example of this kind is
discussed in section 9, where we observe that the $SL(2,\R)$ symmetry
of the type IIB theory is lost if one of its 3-form field strengths is
dualised to a 7-form.

     A contrasting example arises in $D=7$.  We saw in section 6.1
that the $SL(5,\R)$ symmetry can be realised in the full theory at the
level of the Lagrangian, provided that the 3-form potential $A_\3$
arising from the direct reduction from $D=11$ is dualised to yield a
2-form potential.  There are then five 2-form potentials in total,
which form an irreducible 5 of $SL(5,\R)$.  In fact one can make field
redefinitions such that these potentials appear in the Lagrangian only
via their field strengths.  This means that one can perform arbitrary
dualisations of these 3-form field strengths and the resulting
theories, at the level of the equations of motion, will still have the
unbroken global $SL(5,\R)$ symmetry.  (For the usual reasons, the
symmetry can only be realised as local field transformations on the
field strengths, in dualisation choices where 3-form and 4-form field
strengths are to be assembled into an irreducible multiplet.)

       In summary, we note that at the level of the Lagrangian the
global symmetries must necessarily be realised on the potentials,
since these are the fundamental fields.  Consequently, the global
symmetry will be broken if some members of an irreducible multiplet
are dualised to fields of the dual degree, since then a realisation of
the symmetry in terms of local field transformations becomes
impossible.  At the level of the equations of motion, on the other
hand, the symmetries can instead be realised on the field strengths,
provided that only the field strengths, and not their bare potentials,
appear in the equations of motion and the Bianchi identities.  The
global symmetry can then be preserved under dualisations, as long as
the dualisation of some members of an irreducible multiplet does not
result in the appearance of bare potentials for any fields in the rest
of the multiplet.  Otherwise, the global symmetry will again be
broken.

\section{R-R dualisation}

        In the previous sections, we studied the global symmetries of
toroidally-compactified eleven-dimensional supergravity.  We have seen
that the choice of whether or not to dualise certain field strengths
can affect the global symmetries of the theories.  At the level of the
Lagrangian, we saw in section 2 that the undualised theories have a
global $GL(11-D,\R)\semi \R^q$ symmetry, which can be extended to
$E_{11- D}$ if all field strengths of ranks $>\ft12D$ are dualised.
In all cases where the full dualisation is performed, the
$GL(11-D,\R)$ symmetry is preserved, and now forms a subgroup of the
enlarged $E_{11-\sst D}$ symmetry.

        Of course, if we do not insist on obtaining a formulation of
the theory with the $E_{11-\sst D}$ symmetry, then we can selectively
perform dualisations on only a subset of the higher-degree field
strengths.  In \cite{lptdual} the case was considered where only those
fields that correspond to type IIA string Ramond-Ramond fields were
dualised.  The motivation for this came from perturbative string
theory, where the NS-NS gauge potentials, namely the two-form and the
metric, couple to the string world-sheet directly, rather than through
their field strengths.  Thus in terms of a sigma model action we might
not wish to dualise these NS-NS fields, in order to have global
symmetries that act locally on the potentials that couple directly to
the world-sheet.  The R-R fields, on the other hand, couple in the
world-sheet string action only through their field strengths, and
hence these can be freely dualised without compromising the locality
of the global symmetry transformations.  In fact, the realisation of
the perturbative T-duality symmetry $O(10-D,10-D)$ of the
toroidally-compactified type IIA theory requires only that the R-R
fields be dualised, while the NS-NS fields can be left undualised
\cite{lptdual}.

          If we insist that the NS-NS gauge potentials are left
undualised, then the supergravity theories have the symmetries
$E_{11-D}$ for $D=9$, 8 and 7, but $O(10-D,10-D)\semi \R^q$ for $D\le
6$, where $q=2^{8-D}$ \cite{lptdual}.  The global symmetries in
$D\ge7$ are the same as in the fully dualised theories that we
discussed previously, since it is only the R-R potential $A_\3$ that
suffers dualisation in these dimensions.

     In $D=6$, the usual $E_5=O(5,5)$ symmetry of the fully-dualised
theory requires the dualisation of all of the five 3-form field
strengths, since they and their duals together comprise a
10-dimensional irreducible multiplet under $O(5,5)$.  Since one of
these 3-forms is an NS-NS field, the requirement that no NS-NS fields
be dualised will break the $O(5,5)$ symmetry to $O(4,4)$.  In fact the
full global symmetry is now $O(4,4)\semi \R^8$.  (The fact that the
NS-NS 2-form potential need no longer be covered by a derivative
allows a redefinition of fields in which all 8 R-R axions acquire
shift symmetries.)  This $O(4,4)\semi \R^8$ symmetry is actually a
subgroup of $E_{5}$.  This can be seen from the fact that the scalar
sector in $D=6$ is unaffected by the decision not to dualise the NS-NS
fields, and that the global symmetry of the full equations of motion
must be equal to or a subgroup of the global symmetry of the scalar
sector.

    In $D=5$, the story is similar since the 3-form gauge potential
that dualises to a scalar is an R-R field. Thus the scalar Lagrangian
with only R-R dualisation is the same as the that for full
dualisation, and hence the $O(5,5)\semi \R^{16}$ is a subgroup of
$E_6$.  In both $D=5$ and $D=6$, the $E_{11-D}$ symmetry of the scalar
sectors does not extend to the full theories with only R-R
dualisations, even at the level of the equations of motion.  Only the
$O(4,4)\semi \R^8$ and $O(5,5)\semi \R^{16}$ subgroups remain as
global symmetries.

     The story changes in $D=4$ and $D=3$, in that the $O(6,6)\semi
\R^{32}$ and $O(7,7)\semi \R^{64}$ symmetries for the R-R dualisations
are not subgroups of $E_7$ and $E_8$.  The reason for this can be seen
by looking at the scalar sectors of the theories.  In these cases, the
scalar manifolds with only R-R dualisations are different from the
scalar manifolds for the full dualisations, since there are NS-NS
$(D-2)$-form potentials that will no longer be dualised to give
additional scalars.  In fact already in the scalar sectors, one now
finds that the global symmetries are instead the $O(6,6)\semi \R^{32}$
and $O(7,7)\semi \R^{64}$ groups mentioned above.  It is easy to see
that these cannot be contained in $E_7$ and $E_8$, since these allow
only the smaller groups $\R^{27}$ and $\R^{36}$ as maximal abelian
subgroups.

    In the rest of this section, we shall concentrate on studying the
scalar sectors of the $D=4$ and $D=3$ theories obtained by performing
only R-R dualisations.  First, we show how to identify the R-R fields
and the NS-NS fields. In $D$ dimensions, the field content is given by
(\ref{dfields}).  Now in $D=10$ the metric, the dilaton and the 2-form
gauge potential $A_{\2 1}$ are NS-NS fields, while the 3-form gauge
potential $A_\3$ and the vector potential $A_{\1}^1$ are R-R fields.
This separation into NS-NS and R-R fields is preserved under the
subsequent steps of dimensional reduction.  It follows that in the
$D$-dimensional undualised theory the breakdown of the fields into
NS-NS and R-R is as follows:
\bea
{\rm NS-NS}:&& A_{\2 1} \quad A_{\1 1\a} \quad A_{\0 1\a\beta}
\quad {\cal A}_{\1}^\a \quad {\cal A}_\0^\a{}_\beta\quad \vec \phi
\quad g_{\mu\nu}\ ,\label{nsns}\\
{\rm R-R}:&& A_\3 \quad A_{\2\a} \quad A_{\1\a\beta} \quad
      A_{\0\a\beta\gamma} \quad {\cal A}_\1^{1} \quad
     {\cal A}_\0^1{}_\a\ ,\label{rr}
\eea
where we have decomposed the internal index $i$ as $i=(1,\a)$.

          In $D=4$, there are seven 2-form gauge potentials $A_{\2i}$
which could be dualised to scalars, of which $A_{\2 1}$ is an NS-NS
field while the six remaining potentials $A_{\2\a}$ are R-R fields.
Instead of dualising all seven, as we did in section 3.2, let us now
only dualise the six R-R potentials, and so instead of introducing
seven Lagrange multipliers for the Bianchi identities (\ref{f3bi}), we
now introduce only six multipliers $\chi^\a$, for the Bianchi
identities $d(\td\gamma^i{}_\a F_{\3i}) = 0$.  Thus we add Lagrange
multiplier terms
\be
{\cal L}_{\rm LM} = -d\chi^\a \wedge \td\gamma^i{}_\a F_{\3i}\label{rrlm}
\ee
to the Lagrangian.  We now repeat an analysis analogous to that in
section 3.2, except that now we treat only the six fields $F_{\3\a}$
as auxiliary.  Solving algebraically for these, we find that
$F_{\3\a}=e^{-\vec a_\a\cdot \vec \phi}\, G_\1^\a$, where
\be
G_\1^\a=\td\gamma^\a{}_\beta \, \Big(d \chi^\beta +\ft1{72}
       A_{\0k\ell m} \, d A_{\0npq}\, \epsilon^{\beta k\ell mnpq}
\Big)\ .\label{d4grr}
\ee
(In deriving this, we made use of the fact that $\tilde\gamma^\a{}_1
=0$.)  After substituting back into the Lagrangian, the resulting
theory is invariant under the transformations
\be
\delta A_{\0ijk} = c_{ijk}\ ,\qquad
\delta \chi^{\a} = k^{\a} - \ft1{72}\epsilon^{\a ijk\ell mn} 
\, c_{ijk} \, A_{\0\ell mn}\ ,
\ee
together with the usual $GL(6,\R)$ transformations described by
$\Lambda^\a{}_\beta$, and also those corresponding to the parameters
$\Lambda^1{}_{\a}$.  Note that the original $GL(7,\R)$ breaks down to
$GL(6,\R)$, since $\chi^{(\a)}$ is invariant under
$\delta_{\sst\Lambda^1{}_\a}$.  This invariance can be seen by
considering the variation of ${\cal L}_{\rm LM}$ in (\ref{rrlm}) under
the $\Lambda^1{}_\a$ transformations.  Noting that $F_{\3i}$ is
invariant, as it is under all $GL(7,\R)$ transformations (since $i$
here is a tangent-space index), and that from (\ref{gtgtran}), $\delta
\tilde\gamma^1{}_\a = \Lambda^1{}_\a$ and
$\delta\tilde\gamma^\a{}_\beta=0$, we see that if $\chi^\a$ is
invariant then ${\cal L}_{\rm LM}$ transforms by
\be
\delta{\cal L}_{\rm LM} = -d\chi^\a\, \wedge F_{\3 1}\, \Lambda^1{}_\a\ .
\label{dellag}
\ee
But $F_{\3i} = \gamma^j{}_i\, \hat F_{\3j}$, implying, since
$\gamma^j{}_1 = \delta^j_1$, that $F_{\3 1} = \hat F_{\3 1}$. Hence
from (\ref{interm}) we have that $F_{\3 1}= dA_{\2 1}$ when the
vectors are set to zero (as we are assuming here since we are
simplifying the discussion by focussing on the scalar sector), and
consequently, we see that (\ref{dellag}) is a total derivative.  This
shows that the fields $\chi^\a$ are indeed inert under
$\Lambda^1{}_\a$ transformations.

    The non-trivial commutation relations in this case are
\be
[\delta_c,\delta_{c'}]=\delta_k\ ,\qquad  
k^\a =\ft1{36} \epsilon^{\a ijk\ell mn}\, {c}_{ijk}\, {c'}_{\ell mn}\ .
\ee
Putting all this together, we find a  maximal abelian symmetry of
dimension 32, corresponding to the parameters
\be
\Lambda^1{}_\a\ , \quad k^\a\ , \quad c_{\a\beta\gamma}\ .
\ee
Note that this $\R^{32}$ symmetry corresponds precisely to the shift
symmetry of the full set of R-R scalars, namely six ${\cal
A}_0^1{}_\a$, six $\chi^{\a}$ and twenty $A_{\0\a\beta\gamma}$.  Note
also that although we simplified the discussion by setting the vectors
to zero, the result remains true in the full theory \cite{lptdual}.
This is because all the R-R fields can be covered by derivatives
simultaneously in $D=10$ already.

           From the point of view of the coset construction described
in sections 4 and 5, the generator $D_{1}$ in the positive-root
algebra of $E_7$, which was associated with the axion dual to the
NS-NS potential $A_{\2 1}$, can be rescaled and the scale factor sent
to zero.  After this contraction of the algebra, we can consistently
truncate the generator, resulting in a theory with an $O(6,6)\semi
\R^{32}$ global symmetry.  This procedure of dualisation can be also
understood from point of view of a double coset.  With respect to the
simple root $\vec r_2=\vec b_{12}$, the 63 positive roots of $E_7$ are
graded as 1 level-2 root, associated with $D_1$, 32 level-1 roots and
30 level-0 roots.  Thus the double coset can be denoted by
$SU(8)\backslash E_7 /N^{\vec r_2}_2$, implying that the generator
$D_1$ of the Borel group is contracted and truncated out.

    The analysis in $D=3$ is similar.  In the undualised theory we
have 28 $A_{\1 ij}$ and 8 ${\cal A}_\1^{i}$ potentials, which could be
dualised to axionic scalars.  If we instead dualise only the R-R
subset, namely the $A_{\1 \a\beta}$ and ${\cal A}_\1^{1}$ potentials,
then the theory will have an $O(7,7)\semi \R^{64}$ global symmetry,
with a maximal abelian subalgebra $\R^{64}$ corresponding to the shift
symmetries of the full set of 64 R-R axionic scalars.  In fact the
remaining undualised 14 NS-NS vectors correspond to the 14 level-2
generators of the Borel group of $E_8$, with respect to the simple
root $\vec r_2$, and hence the dualisation is equivalent to the double
coset $SO(16)\backslash E_{8}/ N^{\vec r_2}_2$.

           Note that in $D=4$ and $D=3$ the associated $O(6,6)\semi
\R^{32}$ and $O(7,7)\semi \R^{64}$ symmetries are perturbative in
nature, from the point of view of string theory.  In fact these
theories, where only R-R fields have been dualised to give additional
axions, do not have any non-perturbative symmetries.  Such symmetries
can arise in the scalar Lagrangian only when some of the NS-NS
$(D-2)$-form gauge potentials are dualised to scalars, in the process
of which the scalar manifold is changed.

         So far in this paper, we have considered the global
symmetries of the maximal supergravities coming directly from the
dimensional reduction of eleven-dimensional supergravity.  We have
also considered how these symmetries are modified when we dualise
either the full set of higher-rank field strengths, or alternatively
just the R-R subset. In particular, in $D=4$ or $D=3$ dimensions
various different choices can be made, depending on which set of
$(D-2)$-form gauge potentials are dualised to give additional axionic
scalars.  Each different choice can give rise to a new version of the
supergravity, with a different global symmetry, whose positive-root
algebra can be understood from the contraction and truncation of the
$E_{11-\sst D}$ symmetry that arises in the case of full
dualisation.  In fact a useful way to discuss the global symmetries in
the different dualisations is to begin by considering the
fully-dualised theories with the $E_{11-D}$ symmetries, and then pass
to the other cases by ``undualising'' certain fields, or in some
circumstances, deliberately raising the rank of field strengths by
further dualisations.  There are many more possibilities than the
no-dualisation, R-R dualisation and full-dualisation examples that we
have considered so far.  Another example is the following.  In $D=4$
dimensions all 28 vector potentials in the Lagrangian can be covered
simultaneously by derivatives, implying that there can be a commuting
$\R^{56}$ symmetry in $D=3$, realised by the the 56 scalars coming
from the dimensional reduction of the 28 vectors in $D=4$.  (28 arise
as scalars already, and the remaining 28 come from dualising the 28
vectors in $D=3$.)  In this case, the Kaluza-Klein vector arising from
the metric in the reduction from $D=4$ to $D=3$ can no longer be
dualised. In this version of $D=3$ supergravity we therefore have an
$E_7\semi \R^{56}$ global symmetry, since the dimensional reduction
and the dualisation preserve the $E_7$ symmetry that was already
present in $D=4$.  In $D=4$ the 28 vectors and their duals formed a 56
of $E_7$; in $D=3$ they have reduced to 56 axions that again form a 56
of $E_7$.

\section{Abelian symmetries in maximal supergravities}

     A global $\R$ symmetry in a toroidally-compactified supergravity
corresponds to a constant shift symmetry of an axion.  We can choose a
basis where this axion is covered by a derivative everywhere in the
Lagrangian or the equations of motion.  A set of abelian $\R$
symmetries arises when a set of axions can all be covered by
derivatives simultaneously.  It is of interest to look for the maximal
such sets of commuting $\R$ symmetries.  One application is for the
construction of massive supergravities in lower dimensions, which can
be obtained by performing a generalised Scherk-Schwarz reduction in
which an axion is allowed an additional linear dependence on the
compactification coordinate \cite{ss2,bdgpt,clpst,llp}.  Such
reductions can be simultaneously performed on each of a set of axions
that have commuting $\R$ symmetries, and thus it is useful to identify
the maximal such set.

     The dimensions of the maximal abelian subgroups for simple 
Lie groups are well understood by mathematicians.  
They are given by \cite{ma}
\bea
&& \hbox{Simply-laced}:\nn\\
A_n:&& \Big[\ft14 n(n+2) \Big]\ ,\qquad
D_n:\quad \ft12 n(n-1)\ ,\nn\\
E_6:&& 16\ ,\qquad E_7:\quad 27\ ,\qquad
E_8:\quad 36\ .\nn\\
\nn\\
&& \hbox{Non-simply-laced}:\nn\\
B_n:&&1+ \ft12 n(n-1)\ ,\quad n\ge 4, \quad \hbox{with 3 and 5 for
$B_2$ and $B_3$}\ ,\nn\\
C_n:&& \ft12 n(n+1)\ ,\qquad F_4:\quad 10\ ,\qquad
G_2:\quad 3\ . 
\eea

Thus it straightforward to obtain the maximal $\R^{q}$ symmetries for
the fully-dualised maximal supergravities which have $E_{11-D}$ global
symmetries, namely $q=\{1, 3, 6, 10, 16, 27, 36\}$ for dimensions
$D=\{9,8,7,6,5,4,3\}$.  The identification of the sets of axions that
realise these maximal abelian $\R$ symmetries was studied in
\cite{solv1,solv2}, using the method of solvable Lie algebras.  The
conclusion is that for $D\ge 4$, the maximal $\R$ symmetry can be
realised by all the ``new'' axions in $D$ dimensions that do not exist
in $(D+1)$ dimensions.  In other words, these are the axions coming
from the dimensional reduction of the vectors in $(D+1)$ dimensions.
This counting for maximal abelian $\R$ symmetries breaks down in
$D=3$, where the 36 axions are given by the dualisations of the 36
vectors, as we saw in section 3.

          In this section, we shall investigate maximal abelian $\R$
symmetries (but not necessarily those of maximal dimension)
 in the toroidal compactifications of eleven-dimensional
supergravity.  We shall prove that maximal abelian subalgebras of the
positive-root algebra correspond precisely to sets of axions that
can be simultaneously covered by derivatives everywhere in the
Lagrangian.  Then the abelian $\R$ symmetries in $D$-dimensional
supergravity can be analysed by studying the abelian subalgebras of
the associated positive-root algebra.

\subsection{$D\ge 6$}
             
          First let us consider $D\ge 6$, where there is no
complication from dualisations involving scalars.  The axions in these
cases are given by $A_{\0ijk}$ and ${\cal A}_0^i{}_j$.  The scalar
Lagrangian is given by (\ref{dgenlag}) with all the higher-forms set
to zero.  The non-linear Kaluza-Klein modifications for $F_{\1 ijk}$ and
${\cal F}_\1^i{}_j$ are given in (\ref{A.6}). To begin with, we work
to bilinear order in fields:
\bea
F_{\1 ijk} &=& dA_{\0 ijk} - {\cal A}_\0^\ell{}_i\, dA_{\0\ell jk}
-{\cal A}_\0^\ell{}_j\,  dA_{\0 i\ell k} - {\cal A}_\0^\ell{}_k\,
dA_{\0ij\ell} + \cdots\ ,\nn\\
{\cal F}_\1^i{}_j &=& d{\cal A}_\0^i{}_j - {\cal A}_\0^k{}_j\, 
d{\cal A}_0^i{}_k + \cdots\ .\label{blcs}
\eea
 From these bilinear terms, it is manifest that we cannot cover
$A_{\0\ell jk}$ and ${\cal A}_\0^\ell{}_i$ with derivatives
simultaneously, implying that their $\R$ symmetries are non-commuting.
Similarly, we cannot cover ${\cal A}_\0^i{}_k$ and ${\cal A}_\0^k{}_j$
with derivatives simultaneously.  

        The above observation is closely related to the positive-root
algebra of the theory.  As we saw in section 3, the positive-root
algebra can be translated via the Jacobi identities to a set of
summation rules for the dilaton vectors of the axions, which are the
positive roots of the global symmetry algebra.  If the sum of any two
dilaton vectors gives rise to a third one, then the associated
generators of the positive-root algebra do not commute; otherwise,
they do commute.  The dilaton vectors for $A_{\0ijk}$ and ${\cal
A}_\0^i{}_j$ are $\vec a_{ijk}$ and $\vec b_{ij}$ respectively.  Thus
we see that the non-commutativities of the shift symmetries of axions
in (\ref{blcs}) are exactly equivalent to the non-commutativity of the
corresponding root vectors, {\it i.e.}
\be
\vec b_{ij} +\vec b_{jk} = \vec b_{ik}\ ,\qquad
\vec a_{ijk} + \vec b_{i\ell} = \vec a_{\ell jk} \ ,\label{abcom1}
\ee
which is already given in (\ref{abcom}).  Thus in order to have axions
with commuting $\R$ symmetries, we must choose a subset whose
dilaton vectors correspond to positive roots that commute.  In other
words, since (\ref{abcom1}) defines the algebra of the root system, we
must choose a subset of the axions such that their dilaton vectors
cannot, using the summation rules in (\ref{abcom1}), generate any
other dilaton vectors for any axions.

       The above argument concentrated on the bilinear terms in the
non-linear Kaluza-Klein modifications.  However, it is clear from the
chain structure of the higher-order terms in $\gamma^i{}_j$ given in
(\ref{gam}) that any subset of axions that have commuting $\R$
symmetries at the bilinear level will continue to have commuting $\R$
symmetries when the higher-order terms are included as well.

       We shall now illustrate this with a few examples.  The first
non-trivial case occurs in $D=8$, where the global symmetry is given
by $E_3=SL(3,\R) \times SL(2,\R)$.  The association of the dilaton
vectors, positive roots and axions is given in the following table:

\bigskip\bigskip

\centerline{
\begin{tabular}{|c|c|c|}\hline
Dilaton Vectors & $\ \ \ \ \ell_1\ \ \ell_2\ \ \ell_3 \ \ \ $  &\ \ \
Axions\ \ \  \\ \hline\hline
 $\vec b_{13}$  & $0\quad 1\quad 1$ & ${\cal A}_\0^1{}_3$ \\ \hline
 $\vec b_{23}$  & $0\quad 0\quad 1$ & ${\cal A}_\0^2{}_3$ \\ \hline
 $\vec b_{12}$  & $0\quad 1\quad 0$ &  ${\cal A}_\0^1{}_2$ \\ \hline
 $\vec a_{123}$ & $1\quad 0\quad 0$ & $A_{\0 123}$ \\ \hline
\end{tabular}}

\bigskip

\centerline{Table 2:  Dilaton vectors, positive roots and axions in
$D=8$}
\bigskip\bigskip

\noindent
The entries in the second column denote the coefficients $\ell_i$ in
the expressions $\sum_i \ell_i\, \vec r_i$ for the positive roots. 
Thus there are two ways to get the maximal abelian commuting
$\R$ symmetry, which  has three commuting generators:
\be
\{ \vec b_{13}, \vec b_{23}, \vec a_{123} \} \leftrightarrow
\{ {\cal A}_\0^1{}_3, {\cal A}_\0^2{}_3, A_{\0 123} \}
\label{d8count1}
\ee
or
\be
\{ \vec b_{13},\vec b_{12}, \vec a_{123} \} \leftrightarrow
\{ {\cal A}_\0^1{}_3, {\cal A}_\0^1{}_2, A_{\0 123} \}
\ .\label{d8count2}
\ee
In each case, these sets of three axions can all be covered by
derivatives simultaneously in the Lagrangian.  In case 1, given by
(\ref{d8count1}), all the three axions carry an index 3, implying that
they are the new ones arising from the reduction from $D=9$.  In this
case, there is one R-R axion ${\cal A}_\0^1{}_3$, while the other two
are NS-NS fields.  This set of maximal abelian $\R$ symmetries was
also found in \cite{solv1,solv2}.  In case 2, given by
(\ref{d8count2}), we have only one NS-NS axion, namely $A_{\0 123}$,
while the other two are R-R fields.  This second $\R^3$ symmetry can
be understood more generally
from the fact that the theory has a global $\R\times
SL(3,\R)$ symmetry, which is a subgroup of $E_3$, where $\R$ is
realised on the axion $A_{\0 123}$, and hence commutes with
$SL(3,\R)$.

       Another example that can be presented in detail is in $D=7$.  
The global symmetry is $E_4= SL(5,\R)$; the dilaton
vectors, positive roots and axions are summarised in table 3:

\bigskip\bigskip

\centerline{
\begin{tabular}{|c|c|c|}\hline
Dilaton Vectors & $\ \ \ \ell_1\ \ \ell_2\ \ \ell_3 \ \ \ell_4\ \ \ $
& \ \ \ Axions\ \ \  \\ \hline\hline
$\vec a_{234}$ & $1\quad 1\quad 1\quad 1$ & $A_{\0 234}$ \\ \hline
$\vec a_{134}$ & $1\quad 0\quad 1\quad 1$ & $A_{\0 134}$ \\ \hline
$\vec b_{14}$  & $0\quad 1\quad 1\quad 1$ & ${\cal A}_\0^1{}_4$ \\ \hline
$\vec a_{124}$ & $1\quad 0\quad 0\quad 1$ & $A_{\0 124}$ \\ \hline
$\vec b_{24}$  & $0\quad 0\quad 1\quad 1$ & ${\cal A}_\0^2{}_4$ \\ \hline
$\vec b_{13}$  & $0\quad 1\quad 1\quad 0$ & ${\cal A}_\0^1{}_3$ \\ \hline
$\vec b_{34}$  & $0\quad 0\quad 0\quad 1$ & ${\cal A}_\0^3{}_4$ \\ \hline
$\vec b_{23}$  & $0\quad 0\quad 1\quad 0$ & ${\cal A}_\0^2{}_3$ \\ \hline
$\vec b_{12}$  & $0\quad 1\quad 0\quad 0$ & ${\cal A}_\0^1{}_2$ \\ \hline
$\vec a_{123}$ & $1\quad 0\quad 0\quad 0$ & $A_{\0 123}$       \\ \hline
\end{tabular}}

\bigskip

\centerline{Table 3: Dilaton vectors, positive roots and axions in 
$D=7$}
\bigskip\bigskip

     Sets of axions with abelian $\R$ symmetries can be associated
with sets of positive roots that all have a 1 entry in one of the
columns of coefficients $\ell_i$.  This is because there is no
coefficient $2=1+1$ for any of the positive roots.  The maximal
abelian symmetry is $\R^6$, which can be realised in two different
ways, namely by looking at the positive roots with a 1 in column 4 or
in column 3.  The six axions associated with commuting positive roots
determined by column 4 all have an index 4, implying that these are
the new axions appearing in the reduction from $D=8$ to $D=7$.  This
set comprises two R-R and four NS-NS axions.  This set was also
obtained in \cite{solv1,solv2}.  The column-3 commuting roots
represent a new, alternative, choice for realising the $\R^6$
symmetry.  In this case, there are three R-R axions and three NS-NS
axions.

    Looking instead at column 1, we see that the four axions
$A_{\0ijk}$ have a commuting $\R^4$ symmetry.  This is precisely the
$\R^4$ in $GL(4,\R)\semi \R^4$ discussed in section 2.  Note that once
all the four $A_{\0ijk}$ axions are covered by derivatives everywhere
in the Lagrangian, no further axions can be covered.  Although the
maximal abelian symmetry is $\R^6$, we cannot extend this $\R^4$ any
further.  This can be understood from the fact that $\R^4$ is the
maximal abelian subalgebra in $GL(4,\R)\semi \R^4$, which itself is a
maximal subalgebra of $E_4$.  Note that as we remarked in section 2,
the full $D=7$ Lagrangian, when the higher-form potentials are not
dualised, has a global $GL(4,\R)\semi \R^4$ symmetry.  The extension
to an $E_4$ symmetry can be done at the level of the equations of
motion, or at the level of the Lagrangian if the 4-form field strength
is dualised to a 3-form field strength.

     Finally, looking at column 2, we can see that the four R-R
axions, namely $A_{\0 234}$ and ${\cal A}_\0^1{}_\a$ have commuting
$\R^4$ symmetries, which also cannot be extended to $\R^6$.  This
corresponds to another maximal subalgebra of $E_4$, namely
$O(3,3)\semi \R^4$, and $\R^4$ is the maximal abelian subalgebra of
$O(3,3)\semi \R^4$.  This example follows the general pattern where
only the R-R fields are dualised \cite{lptdual}, which was discussed
in section 4.

        In $D=6$ there are a total of 20 axions, corresponding to the
20 positive roots of the $E_5=D_5$ group.  We shall not present the
details, because of the large number of axions that are involved, but
the analysis is analogous to the previous two examples.  The maximal
abelian symmetry for $D_5$ is $\R^{10}$, and can be realised in two
ways.  One is the set of axions that all carry an index 5, {\it i.e.}\
the new ones in $D=6$.  In this case, there are four R-R axions and
six NS-NS axions.  Another way to realise the $\R^{10}$ is by the set
of all the 10 axions $A_{\0 ijk}$, which is also a maximal abelian
symmetry of $GL(5,\R)\semi \R^{10}$.  In this case, they also comprise
four R-R axions and six NS-NS axions.  We can instead easily read off
an $\R^8$ symmetry, realised by all the eight R-R axions.  This $\R^8$
is the maximal abelian subalgebra of $O(4,4)\semi \R^8$.

   Another interesting possibility in $D=6$ is to dualise one of the
original axions, namely $A_{\0345}$, to a 4-form potential.  By
inspecting the list of positive roots and their association with
dilatons, analogous to Tables 2 and 3 in $D=8$ and $D=7$, one can see
that having removed $A_{\0 345}$ from the set, the remaining positive
roots now have an enlarged maximal abelian subalgebra, yielding an
$\R^{12}$ symmetry.  This is realised as shift symmetries on the 12
axions $A_{\0 1\a\beta}$, $A_{\0 2\a\beta}$, ${\cal A}_\0^1{}_\a$ and
${\cal A}_\0^2{}_\a$, where $\a,\beta =3,4,5$.  The abelian symmetry
in this case is larger than the $\R^8$ of the R-R dualistion or the
$\R^{10}$ of the no-dualisation or full-dualisation versions of the
theory.

\subsection{$3\le D\le 5$}

         In these lower dimensions, the theories contain $(D-2)$-form
gauge potentials that can be dualised to give additional axions.  We
shall show that the abelian $\R$ symmetries in these cases are also
governed by the algebras of the positive-root systems.  In $D=5$, let
us consider the scalar Lagrangian where the 3-form gauge potential is
dualised to a scalar.  It follows from (\ref{d5g}) that the axions
$A_{\0 ijk}$ and $A_{\0 \ell mn}$ cannot be covered by derivatives
simultaneously if $ijk\ell mn$ are all different, implying that the
corresponding shift symmetries are non-commuting.  This
non-commutativity is precisely equivalent to the non-commutativity
implied by the sum rules for their associated dilaton vectors, given
by (\ref{d5vecsum}).  In $D=4$, it follows from (\ref{d4g}) that the
non-commutativity of the $\R$ symmetries of the axions $\{A_{\0 ijk},
A_{\0 \ell mn}\}$ is described by the sum rules for their associated
positive roots, given by (\ref{d4g}).  The story is similar in $D=3$,
where the non-commutativity for the axions $\{A_{\0ijk}, A_{\0\ell
mn}\}$ and for $\{ A_{\0 ijk}, \lambda_{jk}\}$, which can be seen from
(\ref{g1eq}) and (\ref{gg1eq}), is also implied by the sum rules for
their dilaton vectors, given by (\ref{d3vecsum}).

        Having established the equivalence of the non-commutativity of
the axionic shift symmetries and the sum rules for the associated
positive roots in $3\le D\le 5$, the task of finding the maximal sets
of axions that can be simultaneously covered by derivatives becomes
equivalent to that of finding the maximal numbers of commuting
positive roots.  It is straightforward to verify that this leads to
exactly the same counting of commuting $\R$ symmetries as we obtained
in section 2, with an identical identification of the axions on which
the symmetries are realised.

        The above discussion was focussed on the versions of the
supergravities where all the $(D-2)$-form potentials are dualised to
scalars.  However, the same methods can also be applied to the
non-dualised or partially dualised cases.  The key point is that if
one of the $(D-2)$-forms is not dualised, then its associated dilaton
vector should not be included as a positive root.  For example in
$D=5$, if the 3-form gauge potential is not dualised, then $-\vec a$
is not a positive root in the system, and hence the sum rule $\vec
a_{ijk} + \vec a_{\ell mn}=-\vec a$ no longer implies that the sum of
$\vec a_{ijk}$ and $\vec a_{\ell mn}$ gives rise to another positive
root.  The consequent commutativity implies that there will be a
global $\R^{20}$ symmetry, which is the maximal abelian subalgebra of
$GL(6,\R)\semi \R^{20}$, the symmetry group of the undualised theory.
A similar analysis applies to $D=4$ and $D=3$, where the maximal
abelian $\R$ symmetries can be read off from the sum rules for the
dilaton vectors for any choice of dualisations, including for example
the no-dualisation or R-R dualisation possibilities.

     In summary, we have shown that we can read off the abelian
symmetries from the $E_{11-D}$ positive-root algebra of the
fully-dualised theories.  The maximal abelian symmetries (of 
largest dimension for each particular dualisation choice) of the
various versions, including non-dualised, fully-dualised and R-R
dualised cases, are given in Table 4 below.

\bigskip

\centerline{
\begin{tabular}{|c|cccccccc|}\hline
Dim. &10&9&8&7&6&5&4&3 \\ \hline\hline
No dual & 1&2&3&4&10&20&35&56 \\ \hline
R-R dual &1&2& 3& 6& 8& 16& 32& 64 \\ \hline
Full dual &1&2& 3& 6& 10& 16& 27& 36 \\ \hline
Max. &1&2& 3& 6& 12& 21& 35& 64 \\ \hline
\end{tabular}}

\bigskip

\centerline{Table 4: Maximal $\R$ symmetries}
\bigskip\bigskip

\noindent We have also listed in the row denoted ``Max'' the largest
maximal abelian symmetry that can be achieved in any of the various
versions of the theories.  The $\R^{12}$ and $\R^{21}$ cases in $D=6$
and $D=5$ occur in the direct dimensional reduction of type IIB
supergravity, which we shall discuss in the next section.

\section{Dimensional reduction  of type IIB supergravity}

     As is well known, the toroidal dimensional reductions of type IIB
supergravity are equivalent to those of the type IIA theory, and are
simply related by field redefinitions.  In fact there are two
different routes that one can follow in the descent from the type IIB
theory to $D=9$.  This is because there is a self-dual 5-form field
strength in type IIB, which, under Kaluza-Klein reduction, can give
either a 5-form field strength or a 4-form field strength (but not
both) in $D=9$, depending upon how the reduction is
performed.\footnote{More precisely, the self-duality of the 5-form
field strength in $D=10$ implies that its 5-form and 4-form reductions
are related by Hodge dualisation in $D=9$. The Lagrangian in $D=9$ can
be formulated using the potential for either one or the other of these
as the fundamental field.}  If the latter choice is made, the fields
in $D=9$ are precisely those of the $D=9$ reduction from $D=11$,
modulo simple local field redefinitions.  If, on the other hand, the
reduction to $D=9$ is expressed in terms of the 5-form field strength,
the theory is then related by non-local field redefinitions involving
the 4-form potential.  In fact, it is the theory one would get from
$D=11$ by choosing to perform an inverse dualisation of the 3-form
potential after reduction to $D=9$.  Subsequent direct dimensional
reductions without any dualisations will then continue to yield
versions of the lower-dimensional supergravities that can be
interpreted as specific inverse dualisations of the supergravities
that we discussed in the previous sections.

     The global symmetries of these direct type IIB reductions will be
$(SL(2,\R)\times GL(9-D,\R))\semi \R^r$, where the $SL(2,\R)$ factor
is directly inherited from the $SL(2,\R)$ symmetry of type IIB in
$D=10$, the $GL(9-D,\R)$ is the usual global symmetry from toroidal
reduction as discussed in section 2. (The symmetry is $GL(9-D,\R)$
rather than $GL(10-D,\R)$ because of a breaking of the symmetry by the
self-duality condition on the 5-form in $D=10$.)  The abelian factor
$\R^r$ describes the maximal commuting shift symmetries of the axions
coming from the antisymmetric tensors.  The values of $r$ in each
dimension are $r=\{0,1, 3, 6, 12,21,35,57\}$ in
$D=\{10,9,8,7,6,5,4,3\}$.  We shall discuss the origin of these $\R$
symmetries later in the section.  One reason for looking at these
reductions of the type IIB theory is that they can sometimes give rise
to maximal abelian shift symmetries that are larger than the ones
obtained via the type IIA route.  For example, as we mentioned in
section 8 there is a version of six-dimensional supergravity in which
there are 12 abelian symmetries, exceeding the 10, 8 and 10 abelian
symmetries of the non-dualised, R-R dualised and fully-dualised
versions from $D=11$.  This new version is in fact nothing but the
direct reduction of the type IIB theory, although it can of course
instead be understood as a specific inverse dualisation of the type
IIA reduction.

     The bosonic sector of type IIB supergravity comprises the metric,
a dilaton $\phi$, an axion $\chi$, two 2-form potentials
$A_{\2}^{(i)}$, and a 4-form potential whose associated field strength
is self dual.  The 4-form potential, $\chi$ and $A_{\2}^{(2)}$ are R-R
fields, and the remainder are NS-NS. Owing to the self-duality of the
5-form field strength, there is no simple way to write a covariant
Lagrangian for these fields alone.  However by adding extra degrees of
freedom, namely by removing the self-duality condition, one can write
a Lagrangian whose equation of motion yield the type IIB equations
after imposing the self-duality constraint as a consistent truncation
\cite{bbo}.  (A similar technique was used in section 6 when we
constructed Lagrangians for the higher-degree fields in even
dimensions.) Thus our starting point is the Lagrangian
\bea
{\cal L} &=& eR +\ft14 e\, {\rm tr}(\del_\mu{\cal M}^{-1}\, \del^\mu{\cal M})
-\ft1{12}e\, H_\3^T\, {\cal M}\, H_\3 -\ft1{240} e\, H_\5^2 \nonumber\\
&&- \ft1{2\sqrt2} \epsilon_{ij}\, B_\4\wedge 
dA_{\2}^{(i)}\wedge dA_{\2}^{(j)} \ ,\nn\\
&=& eR -\ft12 e\, (\del\phi)^2 -\ft12 e\, e^{2\phi}\, (\del\chi)^2 
-\ft1{12} e\, e^{-\phi}\, (F_\3^{(1)})^2 -\ft1{12} e\, e^\phi\, 
(F_\3^{(2)})^2 \nn\\
&& -\ft1{240} e\, H_\5^2 - \ft1{2\sqrt2} \epsilon_{ij}\, B_\4\wedge 
dA_\2^{(i)}\wedge dA_\2^{(j)}\ ,\label{2blag}
\eea
where 
\be
{\cal M}= \pmatrix{ e^{-\phi} + \chi^2\, e^\phi & -\chi\, e^{\phi} \cr
                    -\chi\, e^\phi & e^\phi }\ ,\qquad\qquad
H_\3 =\pmatrix{dA_\2^{(1)}\cr dA_\2^{(2)} }\ .
\ee
The field strengths appearing in (\ref{2blag}) are defined as follows:
\be
F_\3^{(1)}=dA_\2^{(1)}\ , \qquad F_\3^{(2)} = 
dA_\2^{(2)} -\chi\, dA_\2^{(1)}\ ,
\qquad H_\5 =dB_\4 + \ft1{2\sqrt2}\epsilon_{ij} A_\2^{(i)}\wedge
dA_\2^{(j)}\ .\label{2bcs}
\ee
The Lagrangian is manifestly $SL(2,\R)$-invariant.

     The equations of motion following from (\ref{2blag}) are
\bea
&&R_{\mu\nu} = \ft12 \del_\mu\phi\, \del_\nu\phi +\ft12 e^{2\phi}\, 
\del_\mu\chi\, \del_\nu\chi + \ft1{48}(H^2_{\mu\nu} -\ft1{10}
 H_\5^2\, g_{\mu\nu})\label{ein} \\
&&+ \ft14e^{-\phi} ((F_\3^{(1)})^2_{\mu\nu} -\ft1{12} 
(F_\3^{(1)})^2 g_{\mu\nu})
+\ft14e^{\phi} ((F_\3^{(2)})^2_{\mu\nu} -\ft1{12} 
(F_\3^{(2)})^2 g_{\mu\nu})\ ,
\nn\\
&&d\ast H_\5 = \ft1{2\sqrt2} \epsilon_{ij}\, F_\3^{(i)}\wedge F_\3^{(j)}
\ ,\label{5feq}\\
&&d\ast (e^{-\phi}\, F_\3^{(1)} -e^\phi\, \chi\, F_\3^{(2)} ) 
= -\ft1{\sqrt2}
H_\5\wedge (F_\3^{(2)} +\chi\, F_\3^{(1)})\ ,\label{f1eq}\\
&&d\ast (e^{\phi}\, F_\3^{(2)}) = \ft1{\sqrt2}
H_5\wedge F_\3^{(1)}\ ,\label{f2eq}\\
&&\nabla^\mu(e^{2\phi}\, \del_\mu \chi) =-\ft16\, e^\phi\,
F_{\mu\nu\rho}^{(1)}\,  F^{(2)\mu\nu\rho}\ ,
\label{chieq} \\
&&\square\phi =  e^{2\phi}\, (\del\chi)^2 +\ft1{12}\, e^{-\phi}\,
(F_\3^{(1)})^2 -\ft1{12}\, e^{\phi}\, (F_\3^{(2)})^2\ ,\label{phieq}
\eea
where $\square\phi\equiv \nabla^\mu\, \del_\mu\phi$. 

Note that the equation (\ref{5feq}) for $H_\5$ implies, since we also
have the Bianchi identity $dH_\5=\ft1{2\sqrt2}\epsilon_{ij}\,
F_\3^{(i)}\wedge F_\3^{(j)}$, that we can consistently impose the
self-duality condition $H_\5=\ast H_\5$.  After doing this, the
equations (\ref{ein})-(\ref{phieq}) become precisely the field
equations of type IIB supergravity \cite{sch}.  Note also that the
self-duality constraint preserves the $SL(2,\R)$ symmetry, since $H_5$
is a singlet under $SL(2,\R)$.  The equations of motion for the
3-forms $F_3^{(i)}$ can be written more elegantly as
\be
d\ast({\cal M}\, H_\3) = -\ft1{\sqrt2}
H_\5\wedge \Omega H_\3\ ,\qquad \qquad  \Omega=\pmatrix{0&1\cr -1&0}\ .  
\label{meq}
\ee

    Let us now consider some possible dual formulations of the type
IIB theory, and their global symmetries.  One possibility involves
replacing the axion $\chi$ by an 8-form potential.  Such a dualisation
is discussed in the context of a pure $(\phi,\chi)$ matter system in
Appendix B, where it is shown that the original global $SL(2,\R)$
symmetry of the scalar manifold is broken to a global $\R$ symmetry.
Another possibility is to consider dualising one or both of the 2-form
potentials $A_\2^{(i)}$.  These both enter the field equations and
Bianchi identities only via their field strengths $F_\3^{(i)}$.
However, as we discussed earlier in the paper, a set of fields cannot
necessarily be simultaneously dualised merely because they appear in
the equations of motion and Bianchi identities only through their
field strengths.  Rather, we use the (sufficient) property that their
bare potentials be absent also in the Lagrangian.  This criterion is
not satisfied simultaneously by the potentials $A_\2^{(i)}$, on
account of the ``Chern-Simons'' modification to the field strength
$H_\5$, given in (\ref{2bcs}).  Although the simultaneous dualisation
of the two potentials $A_\2^{(i)}$ is therefore not possible in this
way, we can nevertheless dualise either one or the other.  Having done
so, the bare potential of the undualised field will appear in the new
equations of motion and Bianchi identities.  Consequently, the
original $SL(2,\R)$ global symmetry will be broken, since it can no
longer be realised in terms of any local transformations on the new
set of fields.  In fact the unbroken global symmetry will be $\R$ if
the NS-NS 2-form $A_\2^{(1)}$ is dualised, or $\R\semi \R$ if instead
the R-R 2-form $A_\2^{(2)}$ is dualised.

     We now turn to the toroidal dimensional reduction of the type IIB
theory.  Normally, when a field strength of degree $n$ is
dimensionally reduced on a circle, it yields two lower-dimensional
field strengths, of degrees $n$ and $n-1$.  However, in a case such as
that of the 5-form $H_\5$ in the type IIB theory, its reduction on a
circle gives just one or the other of the lower-dimensional fields, on
account of the self-duality constraint.  (In other words, the 5-form
and 4-form field strengths that we would obtain in $D=9$ are not
independent, once the self-duality in $D=10$ is imposed, but are
related by nine-dimensional Hodge duality.  Either one or the other
can be chosen as the independent reduced field in $D=9$.)  We shall
choose the reduction scheme in which the 4-form potential for the
5-form field strength is taken as the fundamental field in $D=9$.  The
resulting axionic field strength content in each lower dimension is
given in Table 5 below, where $F_{\sst{(n)}}$ denotes an $n$-form
field strength coming from the field strengths already present in
$D=10$, while ${\cal F}_{\sst{(n)}}$ denotes one coming from the
dimensional reduction of the metric.

\bigskip
\bigskip
\centerline{
\begin{tabular}{|c|c|c|c|c|c|c|c|c|c|c|}\hline
$D$ &  \multicolumn{5}{c|}{NS-NS} &
\multicolumn{5}{c|}{R-R} \\ \hline
 & $F_\3$ & $F_\2$ & $F_\1$ & ${\cal F}_\2$ &
${\cal F}_\1$ & $F_\5$ & $F_\4$ & $F_\3$ &
$F_\2$ & $F_\1$ 
\\ \hline\hline
10 & 1 & 0& 0& 0 & 0&$\ft12$ & 0& 1& 0& 0 \\ \hline
9 & 1 & 1& 0& 1 & 0& 1 & 0& 1& 1& 1 \\ \hline
8 & 1 & 2& 1& 2 & 1& 1 & 1& 1& 2& 2 \\ \hline
7 & 1 & 3& 3& 3 & 3& 1 & 2& 2& 3& 4 \\ \hline
6 & 1 & 4& 6& 4 & 6& 1 & 3& 4& 5& 7 \\ \hline
5 & 1 & 5& 10& 5 & 10& -- & 4& 7& 9& 12 \\ \hline
4 & 1 & 6& 15& 6 & 15& -- & --& 11& 16& 21 \\ \hline
3 & -- & 7& 21& 7 & 21& -- & --& --& 27& 37 \\ \hline
\end{tabular}}
\bigskip
\centerline{Table 5. Field strengths in the type IIB reductions}
\bigskip    

     First of all, we note that if we fully dualise all higher-degree
field strengths to lower-degree ones, then in $D\le9$ the counting of
fields indeed reduces to that of the fully-dualised reductions of
$D=11$ supergravity.  In this section, we shall instead study the
global symmetries of the theories whose field contents are given in
the table.

    Following this reduction scheme, the scalar sectors in $D\ge 7$
are identical to those in the direct type IIA reductions, and thus the
scalar manifolds again have global $E_{11-D}$ symmetries in these
dimensions.  In $D=6$, the total number of scalars resulting from the
direct reduction of the type IIB theory is 24, comprising 5 dilatons
and 19 axions.  The Lagrangian has the global symmetry
$(SL(2,\R)\times GL(3,\R))\semi \R^{12}$, implying in particular that
there are 12 commuting axionic shift symmetries.  If we were to
dualise the 5-form field strength in $D=6$, we would get one further
axion, and the symmetry of the new scalar manifold would become
$E_5=O(5,5)$, which has an $\R^{10}$ maximal abelian symmetry.  Note
that the axion dual to the 4-form potential is $A_{\0 345}$ in the
type IIA notation.  It is easy to verify in the type IIA reductions of
the previous sections that if $A_{\0 345}$ is dualised in $D=6$, the
axions $A_{\0 1\a\beta}$, $A_{\0 2\a\beta}$, ${\cal A}_\0^1{}_\a$ and
${\cal A}_\0^2{}_\a$ can indeed all acquire commuting shift
symmetries.  (In fact in $D=6$ the scalar Lagrangian has an
$(SL(2,\R)\times GL(4,\R))\semi \R^{12}$ symmetry, since $H_\5$ is a
singlet under $GL(4,\R)$.  However, this $GL(4,\R)$ symmetry is broken
to $GL(3,\R)$ in the full theory, since the three 4-form field
strengths can form an irreducible representation under $GL(3,\R)$
only.  If we were to dualise these three 4-forms, while keeping the
5-form undualised, then we would instead have a global
$(SL(2,\R)\times GL(4,\R))\semi \R^{12}$ symmetry in the full theory.)

     In $D=5$, we see from Table 5 that there are four 4-form field
strengths that can be dualised to give additional scalars.  In the
notation of the $D=11$ reduction, these scalars would be $A_{\0
\a\beta\gamma}$, where the indices range over $3,4,5,6$.  The global
symmetry of the type IIB version can therefore be understood by taking
the fully-dualised $D=11$ version with its $E_6$ global symmetry, and
then inversely dualising the scalars $A_{\0 \a\beta\gamma}$.  The
maximal abelian symmetry is then $\R^{21}$, realised by the axions
${\cal A}_\0^1{}_\a$, ${\cal A}_\0^2{}_\a$, $A_{\0 1\a\beta}$, $A_{\0
2\a\beta}$ and $\chi$, where the last axion is the dual of $A_\3$ in
the $D=11$ reduction.

    A similar analysis can be given in $D=4$ and $D=3$.  In four
dimensions there are twelve 3-form field strengths, which can be
viewed as coming from the inverse dualisation of the axions
$A_{\0\a\beta\gamma}$, $\chi^1$ and $\chi^2$ of the fully-dualised
$D=11$ reduction.  (The axions $\chi^{i}$ were themselves the
dualisations of the 2-forms $A_{\2i}$ in the $D=11$ reduction; see
section 3.)  In the $D=11$ notation, the type IIB version then has a
maximal abelian $\R^{35}$ algebra that corresponds to simultaneous
shift symmetries for the axions ${\cal A}_\0^1{}_\a$, ${\cal
A}_\0^2{}_\a$, $A_{\0 1\a\beta}$, $A_{\0 2\a\beta}$ and $\chi^{\a}$,
where now the indices range over $3,4,5,6,7$.  Finally, in $D=3$ we
have 41 2-form field strengths, which can be viewed in the notation of
the $D=11$ fully-dualised reduction as the inverse dualisation of
$A_{\0 \a\beta\gamma}$, $\chi^{1\a}$, $\chi^{2\a}$, $\chi^{12}$,
$\chi_{1}$, $\chi_{2}$ and $\chi_{\a}$ (see section 3).  In the $D=11$
notation, the 57 maximal abelian $\R$ symmetries are associated with
simultaneous shifts for ${\cal A}_\0^1{}_\a$, ${\cal A}_\0^2{}_\a$,
$A_{\0 1\a\beta}$, $A_{\0 2\a\beta}$, $\chi^{\a\beta}$.

     Note that in all dimensions, in the direct reductions of the type
IIB theory, the original $SL(2,\R)$ symmetry acts on the 1 and 2
indices of the $D=11$ notation, and leaves invariant the $\a$ indices.
The $\a$ indices rotate under $GL(9-D,\R)$.  In all cases, the
symmetries of the type IIB reductions are described by deleting the
simple roots $\vec r_1=\vec a_{123}$ and $\vec r_3=\vec b_{23}$ in the
$E_{11-D}$ Dynkin diagram given in Table 1.

\section{Conclusions}

     In this paper, we have studied the global symmetries of the
maximal $D$-dimensional supergravities, obtained by toroidal
dimensional reduction from $D=11$.  If one simply performs a direct
dimensional reduction, without dualising any of the gauge fields, the
resulting theory in $D$ dimensions has a global $GL(11-D,\R)\semi
\R^q$ symmetry, where the first factor has its origin in the general
coordinate invariance in $D=11$, restricted to the internal
compactified dimensions.  The abelian factor $\R^q$ comes from the
original local gauge symmetry of the 4-form field strength in $D=11$,
and is realised on the $q=\ft16(11-D)(10-D)(9-D)$ axions that come
from the reduction of its 3-form potential.  In all dimensions,
$GL(11-D,\R)\semi \R^q$ acts on the various potentials in the theory,
and it is a global symmetry of the action.  In $D=8$ the equations of
motion actually have a larger global symmetry, namely $SL(3,\R)\times
SL(2,\R)$, where in the case of the 3-form potential, the symmetry is
now taken to act instead on its field strength $F_4$.  In fact the
$SL(2,\R)$ factor in the enlarged symmetry group acts as a kind of
electric/magnetic duality symmetry, under which $F_4$ and its dual
form a doublet.  In dimensions $D\le 7$, it is natural to consider
alternative formulations of the theories, in which all potentials
whose field strengths have degrees greater than $\ft12 D$ are dualised
to give fields of lower degrees.  If this is done, one obtains
theories that can also have enlarged global symmetries, namely
$E_{11-D}$ in its maximally non-compact version.  In odd dimensions
these are symmetries of the action, while in even dimensions they are
symmetries only of the equations of motion, owing to the need to put
the field strengths of degree $\ft12D$ and their duals into a single
irreducible multiplet under $E_{11-D}$.

     It is important to appreciate that the theories obtained by
performing dualisations of the kind described above are non-locally
related to their original undualised versions.  This is because the
potentials for the dualised fields are related to the potentials for
the undualised fields by non-local field transformations.  Thus at
least in cases where the global symmetry is realised at the level of
the action, where the potentials are the fundamental fields, it should
come as no surprise that the global symmetries can be affected by the
process of dualisation.  More precisely it is the choice of the
realisation of the symmetry that dictates its form; roughly speaking
the $E_{11-D}$ group can act more or less faithfully.  We discussed a
simple example of this in Appendix B.1, where the global $SL(2,\R)$
symmetry of a dilaton/axion system was shown to be broken if the axion
was dualised to a $(D-2)$-form potential; the $SL(2,\R)$ symmetry acts
via non-linear transformations on the dilaton and axion themselves,
and it cannot be realised in terms of local transformations on the
dilaton and $(D-2)$-form potential.  Similarly even for higher-degree
fields, where the global symmetries act linearly, at the level of the
action they must act on the potentials themselves, and so they cannot
be implemented in terms of local field transformations of the
fundamental variables if some members of a would-be irreducible
multiplet under the global symmetry are dualised to have the conjugate
degree.
 
     What is perhaps more surprising is that even at the level of the
equations of motion, the global symmetries can also be affected by the
choice of dualisation for the higher-degree field strengths.  In
simple situations one can become accustomed to the idea that only
field strengths, and not bare potentials, appear in equations of
motion and Bianchi identities.  In theories where such is the case
then indeed, provided one views the field strengths rather than their
potentials as the fundamental quantities on which the symmetries act,
the symmetry need not be affected provided that bare potentials are
still absent after the dualisation.  In more complicated situations,
however, including many of the supergravity theories that we have been
considering, the appearance of bare potentials in the equations of
motion or Bianchi identities can be unavoidable, for certain choices
of dualisation.  In cases where this happens, one would then again be
forced to realise the symmetry on the potentials themselves, and this
would not be possible in terms of local field transformations if some
of the potentials in a would-be irreducible multiplet had degrees dual
to the rest.  Thus in particular it is not always true that the
sectors of the theories involving the higher-degree field strengths,
on which the symmetries act linearly, will necessarily respect the
global symmetry of the scalar manifold. We discussed an example in
$D=6$, where the failure to dualise the 3-form potential to give
another vector leads to a breaking of the $O(5,5)$ global symmetry of
the scalar sector, even at the level of the equations of motion.

     The upshot of the above considerations is that in general there
is no unique answer for the global symmetry group of $D$-dimensional
maximal supergravity; it depends upon what choices of dualisations are
made.  Put another way, in such cases there is not a unique maximal
supergravity; rather, there exist inequivalent theories, related to
one another by non-local field transformations, which have different
global symmetries.  In this paper, we considered some specific choices
of dualisations that could, in some sense, be considered ``natural.''
As well as the non-dualised versions obtained by direct dimensional
reduction, and the fully-dualised versions with the $E_{11-D}$
symmetries, another natural choice is when only those fields which
from the ten-dimensional type II string point of view are
Ramond-Ramond fields are allowed to be dualised.  The motivation for
considering this is that in perturbative string theory the NS-NS
fields couple to the worldsheet via their potentials, whilst the R-R
fields have only field-strength couplings.  Thus if the global
symmetries are to act locally on the fundamental fields of
perturbative string theory, then no dualisations of NS-NS fields
should be permitted.  Under these circumstances, we saw that the
global symmetries can be different from either of the previous two
possibilities.  A final natural class of theories that we considered
were the ones obtained by the direct dimensional reduction of type IIB
supergravity.  One has a choice, in the reduction to nine dimensions,
as to whether the self-dual 5-form in $D=10$ will be described in
terms of a 3-form or a 4-form potential in $D=9$.  If the former is
chosen, the theory immediately coincides, after simple local field
transformations, with the reduction of eleven-dimensional supergravity
to $D=9$.  If instead the latter choice is made, the theory in $D=9$
is non-locally related to the direct reduction of eleven-dimensional
supergravity, and it provides a natural starting point for further
dimensional reduction.  Of course it, and its lower-dimensional
descendants, can all be viewed as certain specific ``inverse
dualisations'' of the usual $D=11$ reductions, but their natural type
IIB origin endows these versions of the supergravities with a
preferred status.

      It turns out that the fully-dualised versions of the
supergravities, which have $E_{11-D}$ global symmetries, enjoy a
preferred r\^ole and provide a convenient way of discussing the global
symmetries for all other possible versions, which can then all be
viewed as specific ``inverse dualisations.''  The reason for this
stems from the fact that it is these fully-dualised versions that
maximise the numbers of scalar fields, and while the higher-degree
fields in the theory need not necessarily respect the global symmetry
of the scalar manifold, the latter certainly sets an ``upper bound''
on the global symmetry of the entire theory.  We showed that in the
fully-dualised versions, the axionic scalars are in one-to-one
correspondence with the positive-root generators of $E_{11-D}$.  The
symmetries of the scalar manifolds in the various other dualisations
can then be understood in terms of the Coxeter-Dynkin diagrams for the
$E_{11-D}$ algebras.  Inverse dualisations of certain axions implies
the removal of the associated positive roots from the algebra.  In
particular, we showed that the non-dualised, R-R dualised and type IIB
dualised versions of the supergravities correspond respectively to the
deletion of the simple roots $\vec r_1$, $\vec r_2$ and $(\vec r_1,
\vec r_3)$ respectively, as described below Diagram 1. As well as
determining the global symmetries of the scalar manifolds, these
considerations also allowed us to find the dimensions of the maximal
abelian subalgebras in the various versions of the supergravities.
These determine the number of simultaneous commuting shift symmetries
of the axions in the theories.

     In the sense described above, one might think of the
fully-dualised versions of the supergravities as having the
``largest'' symmetries of all the possible versions.  However, this
viewpoint should be treated with some caution because versions
involving fewer dualisations can have global symmetry groups that are
neither contained in nor contain $E_{11-D}$, and so there is no one
version of the theory in a given dimension that could be said to
encompass the others.  Perhaps the lesson that should be drawn from
these considerations is that there is a richer variety of
supergravities than has sometimes been appreciated, and that sometimes
the apparently simple sounding question of ``what is the symmetry''
can have a rather involved answer.

\pagebreak
\section*{Acknowledgement}

         We are grateful to E. Bergshoeff, M.J. Duff, M. Duflo,
S. Ferrara, V. Kac, B. Kostant and K.S. Stelle for discussions, and were 
supported in part by EC contract ERBCHHBGT9220176.

\appendix
\section{Lagrangian of $D$-dimensional maximal supergravity}

    We begin by establishing some notation and conventions.  The
components of an $n$'th-rank antisymmetric tensor $F_{\sst{(n)}}$ are 
related to its expression as an $n$-form according to
\be
F_{\sst{(n)}}= \fft1{n!}\, F_{\mu_1\cdots \mu_n}\, dx^{\mu_1}\wedge \cdots
dx^{\mu_n}\ .
\ee
The Hodge dual operation $\ast$ in $D$ dimensions is defined by
\be
\ast(dx^{\mu_1}\wedge \cdots dx^{\mu_n}) =\fft1{m!}\, 
\varepsilon_{\nu_1\cdots\nu_m}{}^{\mu_1\cdots\mu_n}{}\,
dx^{\nu_1}\wedge \cdots dx^{\nu_m}\ ,
\ee
where $m=D-n$ and $\varepsilon_{\mu_1\cdots \mu_D}$ is the Levi-Civita 
tensor.  From these definitions, it follows that the components of the 
$m$'th rank antisymmetric tensor $G_{\sst{(m)}}$ dual to
$F_{\sst{(n)}}$ are given by
\be
G_{\mu_1\cdots \mu_m}=\fft1{n!}\, \varepsilon_{\mu_1\cdots \mu_m}{}^{
\nu_1\cdots \nu_n}\, F_{\nu_1\cdots\nu_n}\ .
\ee
         
         Following the discussion and notations of section 2.1,
the Lagrangian for the bosonic $D$-dimensional toroidal
compactification of eleven-dimensional supergravity (with any
dualisation) then takes the form \cite{lpsol}
\bea
{\cal L} &=& eR -\ft12 e\, (\del\vec\phi)^2 -\ft1{48}e\, e^{\vec a\cdot
\vec\phi}\, F_\4^2 -\ft{1}{12} e\sum_i
e^{\vec a_i\cdot \vec\phi}\, (F_{\3i})^2
-\ft14 e\, \sum_{i<j} e^{\vec a_{ij}\cdot \vec\phi}\, (F_{\2ij})^2
\label{dgenlag}\\
&& -\ft14e\, \sum_i e^{\vec b_i\cdot \vec\phi}\, ({\cal F}_\2^i)^2
-\ft12 e\, \sum_{i<j<k} e^{\vec a_{ijk} \cdot\vec \phi}\,
(F_{\1ijk})^2 -\ft12e\, \sum_{i<j} e^{\vec b_{ij}\cdot \vec\phi}\,
({\cal F}_\1^i{}_j)^2 + {\cal L}_{\sst{FFA}}\ ,\nn
\eea
where the ``dilaton vectors'' $\vec a$, $\vec a_i$, $\vec a_{ij}$,
$\vec a_{ijk}$,
$\vec b_i$, $\vec b_{ij}$ are constants that characterise the couplings of
the dilatonic scalars $\vec \phi$ to the various gauge fields.
They are given by \cite{lpsol}
\bea
&&F_{\sst{MNPQ}}\qquad\qquad\qquad\qquad\qquad\qquad\qquad\qquad
{\rm vielbein}\nonumber\\
{\rm 4-form:}&&\vec a = -\vec g\ ,\nonumber\\
{\rm 3-forms:}&&\vec a_i = \vec f_i -\vec g \ ,\nonumber\\
{\rm 2-forms:}&& \vec a_{ij} = \vec f_i + \vec f_j - \vec g\ ,
\qquad\qquad\qquad\qquad\qquad \,\,\, \,\vec b_i = -\vec f_i \ ,
\label{dilatonvec}\\
{\rm 1-forms:}&&\vec a_{ijk} = \vec f_i + \vec f_j + \vec f_k -\vec g
\ ,\qquad\qquad\qquad\qquad\vec b_{ij} = -\vec f_i + \vec f_j\ ,\nonumber
\eea
where the vectors $\vec g$ and $\vec f_i$ have $(11-D)$ components
in $D$ dimensions, and are given by
\bea
\vec g &=&3 (s_1, s_2, \ldots, s_{11-\sst D})\ ,\nonumber\\
\vec f_i &=& \Big(\underbrace{0,0,\ldots, 0}_{i-1}, (10-i) s_i, s_{i+1},
s_{i+2}, \ldots, s_{11-\sst D}\Big)\ ,\label{gfdef}
\eea
where $s_i = \sqrt{2/((10-i)(9-i))}$.  It is easy to see that they satisfy
\be
\vec g \cdot \vec g = \ft{2(11-D)}{D-2}, \qquad
\vec g \cdot \vec f_i = \ft{6}{D-2}\ ,\qquad
\vec f_i \cdot \vec f_j = 2\delta_{ij} + \ft2{D-2}\ .\label{gfdot}
\ee
Note also that
\be
\sum_i \vec f_i = 3\, \vec g\ .\label{fsum}
\ee
In fact the vectors $\vec f_i$ can be written as $\vec f_i = \sqrt2
\vec e_i + \a \vec g$, where $\a = (3- \sqrt{D-2})/(11-D)$ and $\vec
e_i$ are orthonormal vectors, {\it i.e.}\ $\vec e_i \cdot \vec e_j
=\delta_{ij}$.  It follows from (\ref{fsum}) that we have $\vec g = 
\sqrt{2/(D-2)} \sum_i \vec e_i$.  A useful identity that follows from
(\ref{gfdot}) is that
\be
\sum_i (\vec f_i\cdot \vec x)^2 = 2\vec x^2 + (\vec g\cdot\vec x)^2 
\ ,\label{ecid}
\ee
where $\vec x$ is an arbitrary vector.  Note that the $D$-dimensional
metric is related to the eleven-dimensional one by
\be
ds_{11}^2 = e^{\ft13 \vec g\cdot\vec\phi} \, ds_{\sst D}^2 +
\sum_i e^{2\vec\gamma_i\cdot\vec\phi}\, (h^i)^2\ ,\label{met}
\ee
where $\vec \gamma_i=\ft16\vec g -\ft12\vec f_i$, and
\be
h^i=dz^i + {\cal A}_1^i + {\cal A}_0^i{}_j\, dz^j\ .
\ee

     The naive field strengths are associated with the gauge
potentials in the obvious way; for example $F_\4$ is the field
strength for $A_\3$, $F_{\3i}$ is the field strength for $A_{\2i}$,
{\it etc}.  In general, the field strengths appearing in the kinetic
terms are not simply the exterior derivatives of their associated
potentials, but have non-linear Kaluza-Klein modifications as well.
On the other hand the terms included in ${\cal L}_{\sst{FFA}}$, which
denotes the dimensional reduction of the $F_\4\wedge F_\4\wedge A_\3$
term in $D=11$, are best expressed purely in terms of the potentials
and their exterior derivatives.  The complete details of all the field
strengths appearing in this Appendix, in the notation we are using
here, were obtained in \cite{lpsol}. (We correct some sign errors
here.) The field strengths are given by
\bea
F_\4 &=& \td F_\4 - \gamma^i{}_j\, \td F_{\3i}\wedge {\cal A}_\1^{j} +\ft12
\gamma^i{}_k\, \gamma^j{}_\ell \, \td F_{\2ij} \wedge {\cal A}_\1^{k}\wedge
{\cal A}_\1^{\ell}\nn\\
&& - \ft16 \gamma^i{}_\ell\, \gamma^j{}_m\, \gamma^k{}_n\,
\td F_{\1ijk}\wedge {\cal A}_\1^{\ell} \wedge {\cal A}_\1^{m} \wedge
{\cal A}_\1^{n}\ ,\nonumber\\
F_{\3i} &=& \gamma^j{}_i\, \td F_{\3j} + \gamma^j{}_i\, \gamma^k{}_\ell 
\, \td F_{\2jk}
\wedge {\cal A}_\1^{\ell} + \ft12 \gamma^j{}_i\, \gamma^k{}_m\, 
\gamma^\ell{}_n\, 
\td F_{\1jk\ell}\wedge {\cal A}_\1^m \wedge {\cal A}_\1^{n}
\ ,\nonumber\\
F_{\2ij} &=& \gamma^k{}_i\, \gamma^\ell{}_j \, \td F_{\2k\ell} -
\gamma^k{}_i\, \gamma^\ell{}_j \, \gamma^m{}_n \,\td F_{\1k\ell m}\wedge
{\cal A}_\1^{n}\ ,\label{A.6}\\
F_{\1ijk} &=& \gamma^\ell{}_i\,  \gamma^m{}_j\, \gamma^n{}_k \,
\td F_{\1\ell mn}\ ,\nonumber\\
{\cal F}_\2^i &=& \td {\cal F}_\2^{i} - \gamma^j{}_k\,  
\td {\cal F}_\1^i{}_j \wedge
{\cal A}_\1^{k}\ ,\nonumber\\
{\cal F}_\1^i{}_j &=& \gamma^k{}_j\, \td {\cal F}_\1^i{}_k\ ,\nn
\eea
where the tilded quantities represent the unmodified pure exterior
derivatives of the corresponding potentials, and $\gamma^i{}_j$ is
defined by
\be
\gamma^i{}_j=[(1+{\cal A}_0)^{-1}]^i{}_j=\delta^i_j 
-{\cal A}_\0^i{}_j +
{\cal A}_\0^i{}_k\, {\cal A}_\0^k{}_j +\cdots\ .\label{gam}
\ee
Recalling that ${\cal A}_\0^i{}_j$ is defined only for $j>i$ (and vanishes
if $j\le i$), we see that the series terminates after a finite number of
terms.  We also define here the inverse of $\gamma^i{}_j$, namely $\tilde
\gamma^i{}_j$ given by
\be
\tilde \gamma^i{}_j = \delta^i_j +{\cal A}_\0^i{}_j\ .\label{gaminv}
\ee
Note that the upper index on $\tilde \gamma^i{}_j$ is a tangent-space index,
while the lower is a world index.  Conversely, the upper index on 
$\gamma^i{}_j$ is a world index, while the lower is a tangent-space index.
These characteristics reflect themselves in their $GL(11-D,\R)$ 
transformations, since these act only on world indices.  Thus from
(\ref{glnax}) we have that
\be
\delta \gamma^i{}_j =-\Lambda^i{}_k\, \gamma^k{}_j \ ,\qquad
\delta \tilde\gamma^i{}_j = \Lambda^k{}_j\, \tilde\gamma^i{}_k\ .
\label{gtgtran}
\ee

    The term ${\cal L}_{\sst{FFA}}$ in (\ref{dgenlag}) is the dimensional
reduction of the $\td F_\4\wedge\td F_\4\wedge A_\3$ term in $D=11$, 
and is given in lower dimensions by \cite{lpsol}
\bea
D=10:&&\ft12 \td F_\4\wedge \td F_\4 \wedge A_\2\ ,\nonumber\\
D=9: &&\Big(\ft14 \td F_\4 \wedge \td F_\4 \wedge A_{\1ij}-\ft12 
\td F_{\3i}
\wedge \td F_{\3j} \wedge A_\3\Big)\epsilon^{ij}\ ,\nonumber\\
D=8: && \Big(\ft1{12} \td F_\4\wedge \td F_\4 A_{\0ijk} -\ft16 
\td F_{\3i}\wedge
\td F_{\3j} \wedge A_{\2k} -\ft12 \td F_\4\wedge \td F_{\3i} \wedge 
 A_{\1jk}\Big) \epsilon^{ijk}\ ,\nonumber\\
D=7:&& \Big(\ft16 \td F_\4\wedge \td F_{\3i} A_{\0jkl} -\ft14 
\td F_{\3i}\wedge
\td F_{\3j} \wedge A_{\1kl} +\ft18 \td F_{\2ij}\wedge \td F_{\2kl}
\wedge A_\3\Big)\epsilon^{ijkl}\ ,\nn\\
D=6: && \Big(\ft1{12} \td F_\4\wedge \td F_{\2ij} 
A_{\0klm} \!-\!\ft1{12}
\td F_{\3i}\wedge \td F_{\3j} A_{\0klm} +\ft18 \td F_{\2ij}\wedge
\td F_{\2kl} \wedge A_{\2m}\Big) \epsilon^{ijklm}\ ,\nonumber\\
D=5: && \Big(\ft1{12} \td F_{\3i}\wedge \td F_{\2jk}  
A_{\0lmn} +\ft1{48}
\td F_{\2ij}  \wedge \td F_{\2kl}\wedge A_{\1mn} \label{ffaterms}\\
&&\qquad -\ft1{72}
\td F_{\1ijk}\wedge \td F_{\1lmn} \wedge A_\3\Big)
\epsilon^{ijklmn}\ ,\nonumber\\
D=4:&& \Big(\ft1{48} \td F_{\2ij}\wedge \td F_{\2kl} A_{\0mnp} 
-\ft1{72} \td F_{\1ijk}\wedge \td F_{\1lmn} \wedge A_{\2p}\Big)
\epsilon^{ijklmnp}\ ,\nn\\
D=3:&& -\ft1{144}\, \td F_{\1ijk}\wedge 
\td F_{\1lmn}\wedge A_{\1pq} \, \epsilon^{ijklmnpq}\ ,\nonumber\\
D=2: && -\ft1{1296}\, \td F_{\1ijk}\wedge \td F_{\1lmn} A_{\0pqr}
\, \epsilon^{ijklmnpqr}\ .\nonumber
\eea
Here, and elsewhere in the paper, we commonly omit the Hodge $\ast$ symbol
when writing the Wess-Zumino terms in the Lagrangian.  It is then understood
that $dA_{\sst{(m)}}\wedge dB_{\sst{(n)}}\wedge C_{\sst{(p)}}$ represents
a contribution 
\be
{\cal L}_{\rm WZ} =\fft1{m!\, n!\, p!}  
\epsilon^{\mu_1\cdots\mu_{m+1}\, \nu_1\cdots
\nu_{n+1} \,\rho_1\cdots\rho_p}\, \del_{\mu_1}A_{\mu_2\cdots \m_{m+1}}\,
\del_{\nu_1} B_{\nu_2\cdots \nu_{n+1}}\, C_{\rho_1\cdots\rho_p}\ .
\ee

     The expressions for the non-linear Kaluza-Klein modified field 
strengths can be simplified considerably by introducing redefined 
potentials $\hat{\cal A}_\1^{i}$, $\hat A_{\1ij}$, $\hat A_{\2i}$ 
and $\hat A_\3$, given by solving:
\bea
&&\hat{\cal A}_\1^{i} = 
\gamma^i{}_j\, {\cal A}_\1^{j}\ ,\qquad A_{\1ij} = \hat A_{\1ij} +
A_{\0ijk}\, \hat{\cal A}_\1^{k}\ ,\nn\\
&&A_{\2i}= \hat A_{\2i} - \hat A_{\1ij}\wedge \hat{\cal
A}_\1^{j}+ \ft12 A_{\0ijk}\, \hat{\cal A}_\1^{j}
\wedge \hat{\cal A}_\1^{k} \ ,\label{aredef}\\
&&A_\3=\hat A_\3 +\hat A_{\2i}\wedge \hat{\cal A}_\1^i +
\ft12 \hat A_{\1ij} \wedge \hat{\cal
A}_\1^{i}\wedge \hat{\cal A}_\1^{j}+\ft16 A_{\0ijk}\, 
\hat{\cal A}_\1^{i}\wedge \hat{\cal A}_\1^{j}\wedge 
\hat{\cal A}_\1^{k}\ .\nn
\eea
The various Kaluza-Klein modified field strengths are given by
\bea
{\cal F}_\2^{i} &=& \tilde\gamma^i{}_j\, \hat{\cal F}_\2^{j}\ ,\qquad
F_{\2ij} =  \gamma^k{}_i\, \gamma^\ell{}_j\, \hat F_{\2k\ell}\ ,\nn\\
F_{\3i} &=& \gamma^j{}_i\, \hat F_{\3j}\ ,\qquad F_\4 = \hat F_\4\ ,
\label{hatfieldstr}
\eea
where $\tilde\gamma^i{}_j$ is the inverse of $\gamma^i{}_j$, given by 
(\ref{gaminv}), and
\bea
\hat{\cal F}_\2^i &=& d\hat{\cal A}_\1^{i}\ ,\qquad
\hat F_{\2ij} = d\hat A_{\1ij} +A_{\0ijk}\,  \hat{\cal F}_\2^{k} \ ,\nn\\
\hat F_{\3i} &=& d\hat A_{\2i} + \hat A_{\1ij}\wedge \hat{\cal F}_\2^j \ ,
\qquad \hat F_\4 = d\hat A_\3 + \hat A_{\2i}\wedge \hat{\cal F}_\2^i\ .
\label{cjvar}
\eea

     The non-dualised $D$-dimensional Lagrangian obtained by the
direct reduction of eleven-dimensional supergravity is then given by
\be
{\cal L}= e\, R + {\cal L}_{\rm scalar} + {\cal L}_\2 +{\cal L}_\3 +
{\cal L}_\4 +{\cal L}_{FFA}\ ,\label{laglag}
\ee
where ${\cal L}_{\rm scalar}$ is the kinetic Lagrangian for the scalar
sector, ${\cal L}_\2$, ${\cal L}_\3$ and ${\cal L}_\4$ are the kinetic
Lagrangians for the 2-form, 3-form and 4-form field strengths, and
${\cal L}_{FFA}$ represents the remaining Wess-Zumino terms.  The
kinetic terms are given by \cite{cj3}
\bea
{\cal L}_{\rm scalar} &=&-\ft14 e\, g^{ik}\, g^{j\ell}\, 
\del_\mu g_{ij} \,
\del^\mu g_{k\ell} + \ft14 e\, \Big(\fft{\del\omega}{\omega}\Big)^2 \nn\\ 
&&-\ft1{12}e\, \omega g^{i\ell}\, g^{jm}\, g^{kn}\,
\partial_\mu A_{\0ijk}\, \partial^\mu A_{\0\ell mn}\ ,\label{scall}\\
{\cal L}_\2 &=& -\ft18 e\, \omega\, g^{ik}\, g^{j\ell}\, \hat
F_{\mu\nu ij}\, \hat F^{\mu\nu}{}_{k\ell} -\ft14 e\, g_{ij}\,
\hat{\cal F}_{\mu\nu}^{i}\, \hat{\cal F}^{\mu\nu j} \ , 
\label{2l}\\
{\cal L}_\3 &=& -\ft1{12} e\, \omega\, g^{ij}\, \hat
F_{\mu\nu\rho i} \, \hat F^{\mu\nu\rho}_j \ ,\label{3l}\\
{\cal L}_\4 &=& -\ft1{48} e\, \omega\, \hat F_{\mu\nu\rho\sigma}\,
\hat F^{\mu\nu\rho\sigma}\ .\label{4l}
\eea
The metrics $g^{ij}$ and $g_{ij}$ are defined by
\be 
g^{ij} \equiv \sum_k \gamma^i{}_k\, \gamma^j{}_k\, 
e^{\vec f_k\cdot\vec\phi} \ ,\qquad
g_{ij} \equiv \sum_k \tilde\gamma^k{}_i\, \tilde\gamma^k{}_j\, 
e^{-\vec f_k\cdot\vec\phi} \ ,\label{gggg}
\ee
and 
\be
\omega\equiv ({\rm det}\, (g_{ij}))^{\ft13} = e^{-\vec
g\cdot\vec\phi}\ .
\ee
Note that the summation index $k$ in both of the equations in
(\ref{gggg}) is a tangent-space index.  The first two terms in ${\cal
L}_{\rm scalar}$ come from the dimensional reduction of the
eleven-dimensional metric, and are equal to $-\ft12(\del\vec\phi)^2 
-\ft12\sum_{i<j} e^{\vec b_{ij}\cdot\vec\phi}\, ({\cal F}_\1^i{}_j)^2$.
The derivation of ${\cal L}_{\rm scalar}$ involves the use of the
identity (\ref{ecid}), and the fact that $\gamma^i{}_j$ and
$\td\gamma^i{}_j$ vanish when $i>j$.  As we shall show in Appendix C,
the first two terms in ${\cal L}_{\rm scalar}$ describe an
$O(11-D,\R) \backslash GL(11-D,\R)$ coset, which is the scalar
manifold of the dimensional reduction of the pure gravity sector of
the eleven-dimensional theory. 

     A general remark is in order here about the distinction between
internal world indices and tangent-space indices.  We have used the
same kind of index, $i,j,k\ldots,$ for both of these, because in many
of the discussions it would have been inconvenient to have to make the
distinction.  However, it is useful to take note of which indices are
of which kind.  Let us therefore, just within the confines of this
paragraph, introduce the notation that $i,j,k\ldots$ represent world
indices and $a,b,c\ldots$ represent tangent-space indices.  Then the
indices on the various fields we have been using in this paper are as
follows:
\bea
&& A_{\2 i}\ , \qquad A_{\1ij}\ ,\qquad A_{\0ijk}\ ,
\qquad {\cal A}_\1^a\ ,\qquad
{\cal A}_\0^a{}_i\ ,\nn\\
&&\tilde F_{\3 i}\ ,\qquad  \tilde F_{\2ij}\ ,\qquad \tilde F_{\1 ijk}
\ ,\qquad \tilde {\cal F}_\2^a\ ,\qquad \tilde {\cal F}_\1^a{}_i\ ,\nn\\
&&F_{\3 a}\ ,\qquad F_{\2ab}\ ,\qquad F_{\1abc}\ ,\qquad 
{\cal F}_\2^a\ ,\qquad {\cal F}_\1^a{}_b\ .
\eea
In addition, we have $\gamma^i{}_a$ and $\tilde\gamma^a{}_i$.

       In terms of the hatted potentials, we saw that the non-linear
Kaluza-Klein modifications (\ref{cjvar}) for the higher-degree field
strengths became very simple.  However, this is achieved at the price
of making the ${\cal F}_{FFA}$ terms in (\ref{ffaterms}) very much
more complicated, with up to seventh powers of fields arising.  In low
dimensions, this is not a problem in the fully-dualised formulations,
since in fact these terms conspire to cancel against others arising
from the dualisations.  In higher dimensions an intermediate
redefinition of fields seems to be more useful, which does not change
the structure of the ${\cal L}_{FFA}$ terms, but which nonetheless
considerably simplifies the non-linear Kaluza-Klein modifications to
the various field strengths.  This can be done by again introducing
$\hat{\cal A}_\1^{i}$ as in (\ref{aredef}), but now writing the hatted
field strengths defined in (\ref{hatfieldstr}) in terms of the
original gauge potentials $A_\3$, $A_{\2i}$, $A_{\1ij}$ and
$A_{\0ijk}$, rather than the hatted potentials defined in
(\ref{aredef}).  In terms of these potentials we have
\bea
\hat {\cal F}_\2^{i} &=& d \hat{\cal A}_\1^{j}\ ,\qquad
\hat F_{\2ij} = dA_{\1ij} - dA_{\0ijk}
\wedge\hat {\cal A}_\1^{k}\ ,\nn\\
\hat F_{\3i} &=& dA_{\2i} + 
dA_{\1ij}\wedge \hat {\cal A}_\1^{j}
+\ft12 dA_{\0ijk}\wedge  \hat {\cal A}_\1^{j}\wedge  
\hat {\cal A}_\1^{k}\ ,\label{interm}\\
\hat F_\4 &=& dA_\3 - dA_{\2i} \wedge\hat {\cal A}_\1^{i} +\ft12
dA_{\1ij}\wedge  \hat {\cal A}_\1^{i}\wedge  \hat {\cal A}_\1^{j}
-\ft16 dA_{\0ijk}\wedge  \hat {\cal A}_\1^{i}\wedge 
\hat{\cal A}_\1^{j}
\wedge  \hat {\cal A}_\1^{k}\ .\nn
\eea
The kinetic terms in the Lagrangian are still given by (\ref{laglag}),
while the ${\cal L}_{FFA}$ terms are given by (\ref{ffaterms}).

\section{$O(2)\backslash SL(2,\R)$ coset manifold}

   In this appendix, we illustrate the details of the coset construction
for the scalar manifolds in section 4 by considering the example of
the two-dimensional $SL(2,\R)/O(2)$ manifold. We beging by introducing the
Cartan generator $H$, and the raising and lowering operators $E_\pm$,
which may be taken to be 
\be
H=\tau_3=\pmatrix{1&0\cr 0&-1}\ ,\qquad E_+ = \pmatrix{0&1\cr
0&0}\ ,
\qquad E_- =\pmatrix{0&0\cr 1&0} \ .
\ee
Then we can parameterise the coset as 
\be
\v= e^{\ft12\phi\, H}\, e^{\chi\, E_+}\,
=\pmatrix{e^{\ft12\phi} & \chi\, e^{\ft12\phi}\cr
             0 & e^{-\ft12\phi} }\ .\label{triang}
\ee
Then we see that 
\be
d\v\, \v^{-1} = \pmatrix{\ft12 d\phi & e^\phi\, d\chi\cr
             0 & -\ft12 d\phi}
    = \ft12 d\phi\, H + e^\phi\, d\chi\, E_+
\ee
Clearly we can now write the Lagrangian as
\be
{\cal L} =\ft14{\rm tr}\Big(\del {\cal M}^{-1}\, \del {\cal M})
\Big)
= -\ft12 (\del\phi)^2 -\ft12 e^{2\phi}\, (\del \chi)^2\ ,\label{lag}
\ee
where ${\cal M}=\v^{\rm T}\, \v$. This Lagrangian is obviously
invariant under $SL(2,\R)$ transformations $\v \rightarrow \v'' =\v \,
U$, where $U$ is a constant $SL(2,\R)$ matrix of the form
\be
U=\pmatrix{a & b\cr c & d}\ ,\qquad\qquad ad-bc=1\ .
\ee
However, it is clear that $\v''$ is no longer in the upper triangular
gauge (\ref{triang}).  We can perform a {\it local} compensating
transformation, however, to get a $\v'$ which {\it is} back in
upper-triangular gauge:
\be
\v' = {\cal O}\, \v'' = {\cal O} \, \v\, U\ ,
\ee
where ${\cal O}$ is a field-dependent $SL(2,\R)$ matrix in the $O(2)$
subgroup, satisfying ${\cal O}^{\rm T}\, {\cal O}=1$.  After a little
algebra, we find that ${\cal O}$ is given by
\be
{\cal O}= (c^2+ e^{2\phi}\, (c\, \chi+a)^2)^{-1/2}\, 
\pmatrix{(c\, \chi +a)\, e^\phi & c\cr c & (c\, \chi+a)\, e^\phi}\ .
\ee
Thus for a given global $U$, with constant components $a,b,c,d$, there is
a local ${\cal O}$, with field-dependent and $U$-dependent components,
which restores the upper-triangular gauge.  This means that
$\v$ defined in (\ref{triang}) parameterises elements in the coset
$O(2)\backslash SL(2,\R)$. 

\subsection{Dualisation of scalars: an $SL(2,\R)$ example}

      In section 3, we show the dualisation of the $(D-2)$-form gauge
potentials to scalars in $D=5$, 4, and 3 has the effect of changing
the global symmetry $GL(11-D,\R)\semi \R^q$ of the undualised theory
to $E_6$, $E_7$ and $E_8$ respectively. These examples, however, are
rather complicated, and it is difficult to study exactly how the
symmetry alters under the dualisation.  We focussed principally on how
the maximal abelian $\R$ symmetries are altered under the
dualisations.  Here, we shall present a simpler example, namely a
$D$-dimensional scalar Lagrangian with a global $SL(2,\R)$ symmetry.
It contains the metric, a dilaton $\phi$ and an axion $\chi$.  The
Lagrangian is given by
\be
e^{-1}{\cal L} = R-\ft12(\del\phi)^2 -\ft12 e^{2\phi}\, (\del\chi)^2\ .
\label{sl2rlag}
\ee
This is precisely of the form of the scalar Lagrangian for type IIB 
supergravity when $D=10$.  The theory has the global $SL(2,\R)$ symmetry
\be
\tau \longrightarrow \tau'=\fft{a\tau + b}{c\tau + d}
\ ,\label{sl2ntr0}
\ee
where $\tau = \chi + {\rm i} e^{-\phi}$, and $ad-bc=1$.

       Since the axion $\chi$ appears in the Lagrangian only through
its derivative, it follows that it can be obtained by dualising a
$(D-2)$-form gauge potential $A_{\sst D-2}$.  Consider a theory with
the metric, a dilaton and such a potential $A_{\sst D-2}$, with
Lagrangian given by
\be
e^{-1}{\cal L} = R -\ft12 (\del \phi)^2 -\ft1{2(D-1)!} F_{\sst{(D-1)}}^2
e^{-2\phi}\ ,\label{sl2rdual}
\ee
where $F_{\sst{(D-1)}} = d A_{\sst{(D-2)}}$.  If we dualise the gauge
potential $A_{\sst{(D-2)}}$, we recover the scalar Lagrangian
(\ref{sl2rlag}).  However this original undualised Lagrangian
(\ref{sl2rdual}) has just one scalar, which has an $\R$ global
symmetry, namely
\be
\phi\longrightarrow \phi'= \phi + c\ ,\qquad A_{\sst{(D-2)}}
\longrightarrow A_{\sst{(D-2)}}' = e^{2c} A_{\sst{(D-2)}}\ .
\ee
There is also a local gauge symmetry for the gauge field $A_{\sst{(D-2)}}$.

         The $SL(2,\R)$ symmetry of the two-scalar system
(\ref{sl2rlag}) involves three nontrivial transformations, namely the
shift symmetry $\chi' = \chi + {\rm const.}$ of the axion; the
inversion $\tau' = - 1/\tau$; and a shift symmetry $\phi' = \phi +
{\rm const.}$ of the dilaton, together with the necessary rescaling of
the axion.  In the original undualised theory (\ref{sl2rdual}), only
the last of these symmetries is preserved: The constant shift symmetry
of $\chi$ simply becomes obsolete, since there is no field on which to
realise it.  In fact, the global shift symmetry is replaced by the
local gauge symmetry of $F_{\sst{(D-1)}}$ in (\ref{sl2rdual}).  The
$\tau\rightarrow - 1/\tau$ symmetry in $SL(2,\R)$ also breaks down,
owing to its non-linearity; it can, however, be viewed as an on-shell
symmetry in the original theory (\ref{sl2rdual}).  It should be
emphasised here that the $SL(2,\R)$ symmetry is absent not only in the
Lagrangian (\ref{sl2rdual}), but also in its equations of motion.

     It is of interest to follow the various symmetries in detail in
the dualisation procedure, in order to see when and how they are
altered.  To dualise the potential in the Lagrangian (\ref{sl2rdual}),
we introduce the Lagrange multiplier $\chi$ for the $(D-1)$-form field
strength $F_{\sst D-1}=dA_{\sst D-2}$, to enforce the Bianchi identity
$dF_{\sst D-1}=0$.  This leads to the first-order Lagrangian
\be
e^{-1} {\cal L} = R - \ft12 (\del \phi)^2 -
\ft1{2(D-1)!} F_{\sst{(D-1)}}^2 e^{-2\phi} + \chi\, *d F_{\sst{(D-1)}}\ .
\label{sl2rfo}
\ee
The relevant equations of motion that follow from this are given by
\be
\square \phi + \fft{1}{(D-1)!} e^{-2\phi} F^2_{D-1} =0\ ,\quad 
dF_{\sst{(D-1)}} = 0\ ,\quad *F_{\sst{(D-1)}} = e^{2\phi}d\chi\ .
\label{eomsl2r}
\ee
Note that in this first-order formalism, the $SL(2,\R)$ symmetry is still
present, with the three transformation rules given by 
\bea
&&\chi \rightarrow \chi + c\ ,\quad e^{\phi} \rightarrow e^{\phi}
\ ,\quad F_{\sst{(D-1)}} \rightarrow F_{\sst{(D-1)}}\ ,\label{sl2rtr1}\\
&&e^{\phi}\rightarrow \lambda^2 e^{\phi}\ ,\quad
  F_{\sst{(D-1)}} \rightarrow \lambda^{2} F_{\sst{(D-1)}}\ ,\quad
  \chi \rightarrow \lambda^{-2} \chi\ ,\label{sl2rtr2}\\
&& \tau \rightarrow -\fft{\a^2}{\tau}\ ,\quad
{*F}_{\sst{(D-1)}} \rightarrow \fft{1}{\a^2} \Big((\chi^2 + e^{-2\phi})
\, {*F}_{\sst{(D-1)}} - 2 \chi d\phi - 2 d\chi\Big)\ .\label{sl2rtr3}
\eea
It is a matter of straightforward computation to verify that the
Lagrangian (\ref{sl2rfo}) is invariant under these.  At the level of
the equations of motion, the invariance under (\ref{sl2rtr1}) and
(\ref{sl2rtr2}) is manifest.  The transformation (\ref{sl2rtr3}) is
more complicated, sending the third first-order equation in (\ref{eomsl2r})
into itself, while transforming the other two equations into two
independent combinations of the original equations of motion.

       If we integrate out the auxiliary field $F_{\sst D-1}$, it
gives rise to the two-scalar system $(\phi, \chi)$ with Lagrangian
(\ref{sl2rlag}), and the $SL(2,\R)$ symmetry is preserved.  However,
if we instead integrate $\chi$, the resulting theory, whose Lagrangian
is given by (\ref{sl2rdual}), no longer has an $SL(2,\R)$ global
symmetry; but instead only a global $\R$ symmetry.  It does, however,
have an additional local gauge symmetry for $F_{\sst{(D-1)}}$.  The
reason for the loss of the global $SL(2,\R)$ symmetry can be seen from
(\ref{sl2rtr3}).  This transformation involves the undifferentiated
$\chi$ field, and thus cannot be implemented in terms of local
transformations either for the field strength $F_{\sst{(D-1)}}$ or for
its gauge potential.  Such a reduction of the global symmetry is not
surprising, since in the first-order formalism there is an auxiliary
field, which can allow the symmetry to be ``artificially''
enlarged. Integrating out the auxiliary fields then leads to a
reduction of the symmetry.  The Lagrangian (\ref{sl2rlag}) maintains
the full global symmetry, while losing the local gauge invariance. The
Lagrangian (\ref{sl2rdual}), on the other hand, maintains the local
gauge symmetry but loses part of the global symmetry.

        It should be noted that there is more than one way to write
down a first-order Lagrangian that can lead to (\ref{sl2rlag}) and 
(\ref{sl2rdual}).  In the first-order formulation (\ref{sl2rfo}) we
chose $\chi$ and $F_{\sst{(D-1)}}$ as the fundamental fields.  We could 
instead choose $B_{\sst{(D-2)}}$ and $F_\1$ as the fundamental
fields, with the Lagrangian now talking the form
\be
e^{-1} {\cal L} = R -\ft12 (\del \phi)^2 -
\ft12 F_\1^2 e^{2\phi} + B_{\sst{(D-2)}} \wedge d F_\1 \ .
\ee
In this case, even the first-order Lagrangian has only an $\R$ global 
symmetry, corresponding to a shift of the dilaton $\phi$.

         Note that the global $\R$ symmetry of the original theory
(\ref{sl2rdual}) is a subgroup of the $SL(2,\R)$ symmetry of the
two-scalar system.  This feature will in general be the case if the
Lagrangian has no ${\cal L}_{FFA}$ type topological terms.  In a
supergravity theory, however, the presence of such an ${\cal L}_{FFA}$
term can have the effect that the global symmetry group for the theory
where an axion is dualised may no longer be contained in the global
symmetry group of the theory where it is left undualised.

\section{$(O(n)\times O(n))\backslash O(n,n)$ coset manifolds}

     Consider the $2n\times 2n$ matrix
\be
\Omega=\pmatrix{0 & \oneone \cr \oneone & 0}\longrightarrow 
\pmatrix{ 0 & \delta_B^A \cr \delta_A^B & 0}\ ,
\ee
which is left invariant by the $O(n,n)$ infinitesimal
transformations generated by\footnote{This formulation is related by a
simple change of basis to the more familiar one where the metric is 
$\pmatrix{I & 0\cr 0 & -I}$ and the $O(N)\times O(N)$ subgroup is
manifest.} 
\be
L= \pmatrix{U & V \cr \wtd V & -U^T}\longrightarrow 
\pmatrix{U_A{}^B & V_{AB} \cr \wtd V^{AB} & -U_B{}^A}\ ,\label{onn}
\ee
where $U$ is an arbitrary real matrix, and $V=-V^T$, \ $\wtd V = -
\wtd V^T$.  There is a manifest $GL(n,\R)$ subgroup of $O(n,n)$
generated by matrices (\ref{onn}) with $V=\wtd V=0$.  Also, the
maximal compact subgroup $O(n)\times O(n)$ is generated by matrices
(\ref{onn}) with $U=-U^T$ and $\wtd V = V$.  We can parameterise the
coset $(O(n)\times O(n))\backslash O(n,n)$ in terms of upper
triangular matrices of the form
\be
\v=\pmatrix{S & R \cr 0 & (S^{-1})^T } \ ,
\ee
where, in order to satisfy the condition $\v \, \Omega\, \v^T=\Omega$, we
must have $R\, S^T + S\, R^T =0$.

    From $\v$, we may construct the matrix ${\cal M}=\v^T\, \v$, giving
\be
{\cal M} = \pmatrix{ S^T\, S & S^T\, R \cr R^T\, S & 
(S^T\, S)^{-1} + R^T\, R } \longrightarrow 
\pmatrix{ (S^T\, S)_{AB} & (S^T\, R)_A{}^B \cr (R^T\, S)^A{}_B & 
((S^T\, S)^{-1})^{AB}+ (R^T\, R)^{AB} }  \ .\label{mmatrix}
\ee
Defining $(S^T\, S)_{AB} = G_{AB}$ and $(S^{-1}\, R)^{AB}=
-X^{AB}$, and noting that $G_{AB}=G_{BA}$ and $X^{AB}=-X^{BA}$, we can
write ${\cal M}$ as
\be
{\cal M}= \pmatrix{G & -G\, X \cr X\, G & G^{-1} - X\, G\, X} 
\longrightarrow \pmatrix{G_{AB} & -G_{AC}\, X^{CB} \cr
                 X^{AC}\, G_{CB} & G^{AB} -X^{AC}\, G_{CD}\, X^{DB} }
\ .
\ee
The $(O(n)\times O(n))\backslash O(n,n)$ coset Lagrangian can then be 
written as
\bea
{\cal L} &=& \ft18 e\, {\rm tr}\, (\del_\mu {\cal M}^{-1} \, 
\del^\mu {\cal M})
    = \ft18e\, {\rm tr}\, (\Omega\, \del_\mu{\cal M}\, 
\Omega\, \del^\mu {\cal M}) \ ,\nn\\
&=& -\ft14 e\, G_{AC}\, G_{BD}\, \Big( \del_\mu G^{AB}\, \del^\mu G^{CD} 
+ \del_\mu X^{AB}\, \del^\mu X^{CD}\Big)\ .\label{onnlag}
\eea
Note that if we set the fields $X^{AB}$ to zero, we get the coset
Lagrangian for $O(n)\backslash GL(n,\R)$.
    
   The above formalism can be used to describe the scalar Lagrangians
of various supergravity theories.  For example, the global symmetry of
the supergravity describing the low-energy limit of the heterotic
string in $D>4$ dimensions is $O(10-D,10-D)$, which is the
perturbative T-duality group.  (This supergravity theory is obtained
from the type IIA supergravity by setting all the R-R fields to zero.)
Another example, which we shall study in detail, is the maximal
supergravity in $D=6$, which has $E_5=O(5,5)$ global symmetry. The
complete scalar Lagrangian coming from the dimensional reduction of
eleven-dimensional supergravity is given in (\ref{scall}).  Defining 
$G_{ij}$ and $X^{ij}$ by $g_{ij}= \omega\, G_{ij}$ and
$A_{\0ijk}=\ft12\epsilon_{ijk\ell m}\, X^{\ell m}$, we find that
(\ref{scall}) in $D=6$ reduces to
\be 
{\cal L}_{\rm scalar}= -\ft14 e\, G_{ik}\, G_{j\ell }\, 
\Big( \del_\mu G^{ij}\, \del^\mu G^{k\ell} 
+ \del_\mu X^{ij}\, \del^\mu X^{k\ell}\Big)\ ,\label{o55}
\ee
which is precisely of the form (\ref{onnlag}) with $n=5$.

The first two terms in (\ref{scall}) correspond to a scalar coset
manifold with a $GL(11-D,\R)$ symmetry, and could be cast into the
form of the $GL(n,\R)$-invariant Lagrangian (\ref{onnlag}) with
$X^{AB}$ set to zero, by making an appropriate Weyl rescaling of
$g_{ij}$. The $GL(11-D,\R)$ symmetry is augmented by the inclusion of
the third term in (\ref{scall}), describing the scalar Lagrangian for
the axions coming from the dimensional reduction of the antisymmetric
tensor in $D=11$.  We have already seen that in $D=6$, the inclusion
of the third term enlarges $GL(5,\R)$ to $O(5,5)$.  In $D=7$, the
enlarged symmetry group is $SL(5,\R)$.  We can see this from
(\ref{scall}) by making the redefinition $A_{\0ijk}=
\epsilon_{ijk\ell} \, X^\ell$ and defining a $5\times 5$ metric
$G_{AB}$ by
\bea
G_{AB}&=& \pmatrix{\fft1\omega \, g_{ij} & \fft1\omega \,g_{ik}\, X^k
\cr \fft1\omega \,g_{jk}\, X^k & \omega +\fft1\omega \, g_{k\ell}\, X^k
\, X^\ell} \ ,\nn\\
G^{AB}&=& \pmatrix{\omega \, g^{ij} +\fft1\omega \, X^i\, X^j 
 & -\fft1\omega\, X^i
\cr -\fft1\omega\, X^j &\fft1\omega} \ .\label{gsl5r}
\eea
Note that $G_{AB}$ has unit determinant.  Substituting this into the
general $GL(n,\R)$-invariant Lagrangian (\ref{onnlag}) where $X^{AB}$
is set to zero, we find that it precisely gives the scalar Lagrangian
(\ref{scall}) for $D=7$.  Thus the $SL(5,\R)$ symmetry is made
manifest.

    In $D=8$, we define $A_{\0ijk}= \epsilon_{ijk}\, \chi$ and $g_{ij}
= \omega\, G_{ij}$, and substitute these into (\ref{scall}).
Noting that $G_{ij}$ has unit determinant here, and making the further
redefinition $\omega= e^{-\phi}$, we find that the complete scalar
Lagrangian becomes
\be
{\cal L}_{\rm scalar} = -\ft14 e\, G_{ik}\, G_{j\ell }\, 
 \del_\mu G^{ij}\, \del^\mu G^{k\ell}  -\ft12 e\, (\del\phi)^2 -\ft12
e\, e^{2\phi}\, (\del\chi)^2\ .
\ee
Owing to the fact that $G_{ij}$ has unit determinant we see that the
first term describes the coset $O(3)\backslash SL(3,\R)$, and the
remaining two terms describe the coset $O(2)\backslash SL(2,\R)$ as
discussed in Appendix B.


\end{document}